\documentclass[a4paper,fleqn]{cas-sc}
\pdfoutput=1 
\usepackage[square,sort,comma,numbers]{natbib}

\def\tsc#1{\csdef{#1}{\textsc{\lowercase{#1}}\xspace}}
\tsc{WGM}
\tsc{QE}
\tsc{EP}
\tsc{PMS}
\tsc{BEC}
\tsc{DE}
\usepackage{graphicx}
\graphicspath{ {./images/} }
\usepackage{natbib}
\usepackage{hyperref}
\usepackage{picins}

\begin{document}
\let\WriteBookmarks\relax
\def\floatpagepagefraction{1}
\def\textpagefraction{.001}

\shorttitle{AI for Next Generation Computing: Emerging Trends and Future Directions}

\shortauthors{Sukhpal Singh Gill et~al.}

\title [mode = title]{AI for Next Generation Computing: Emerging Trends and Future Directions}

\author[1]{Sukhpal Singh Gill*} [orcid=0000-0002-3913-0369]

\ead{s.s.gill@qmul.ac.uk}

\affiliation[1]{organization={School of Electronic Engineering and Computer Science, Queen Mary University of London, London, UK}}

\author[2]{Minxian Xu} 

\ead{mx.xu@siat.ac.cn}

\affiliation[2]{organization={Shenzhen Institute of Advanced Technology, Chinese Academy of Sciences, Shenzhen, China}}

\author[3]{Carlo Ottaviani} 

\ead{carlo.ottaviani@york.ac.uk}

\affiliation[3]{organization={Department of Computer Science and York Centre for Quantum Technologies, University of York, York, UK}}

\author[4]{Panos Patros} 

\ead{panos.patros@waikato.ac.nz}

\affiliation[4]{organization={Department of Software Engineering, University of Waikato, Hamilton, Aotearoa New Zealand}}

\author[5]{Rami Bahsoon} 

\ead{r.bahsoon@cs.bham.ac.uk}

\affiliation[5]{organization={School of Computer Science, University of Birmingham, Birmingham, UK}}

\author[6]{Arash Shaghaghi} 

\ead{arash.shaghaghi@rmit.edu.au}

\affiliation[6]{organization={Department of Information Systems and Business Analytics, RMIT University, Melbourne, Australia}}

\author[1]{Muhammed Golec} 

\ead{m.golec@qmul.ac.uk}

\author[7]{Vlado Stankovski}

\ead{vlado.stankovski@fri.uni-lj.si}

\affiliation[7]{organization={Faculty of Computer and Information Science, University of Ljubljana, Ljubljana, Slovenia}} 

\author[8]{Huaming Wu} 

\ead{whming@tju.edu.cn} 

\affiliation[8]{organization={Center for Applied Mathematics, Tianjin University, Tianjin, China}}

\author[9, 10]{Ajith Abraham} 

\ead{ajith.abraham@ieee.org}

\affiliation[9]{organization={Machine Intelligence Research Labs, Auburn, WA, USA}}

\affiliation[10]{organization={Center for Artificial Intelligence, Innopolis University, Innopolis, Russia}}

\author[11,12]{Manmeet Singh} 

\ead{manmeet.singh@utexas.edu}

\affiliation[11]{organization={Jackson School of Geosciences, University of Texas at Austin, Austin, Texas, USA}}

\affiliation[12]{organization={Centre for Climate Change Research, Indian Institute of Tropical Meteorology, Pune, India}}

\author[13,14]{Harshit Mehta} 

\ead{harshit.mehta@utexas.edu}

\affiliation[13]{organization={Walker Department of Mechanical Engineering, Cockrell School of Engineering, The University of Texas at Austin, Texas, USA }}

\affiliation[14]{organization={Dell Technologies, Austin, Texas, USA}}

\author[15]{Soumya K. Ghosh} 

\ead{skg@cse.iitkgp.ac.in}

\affiliation[15]{organization={Department of Computer Science and Engineering, Indian Institute of Technology, Kharagpur, India}}

\author[16]{Thar Baker} 

\ead{tshamsa@sharjah.ac.ae}

\affiliation[16]{organization={Department of Computer Science, University of Sharjah, Sharjah, UAE}}

\author[17]{Ajith Kumar Parlikad}

\ead{aknp2@cam.ac.uk}

\affiliation[17]{organization={Institute for Manufacturing, Department of Engineering, University of Cambridge, Cambridge, UK}}

\author[18]{Hanan Lutfiyya} 

\ead{hanan@csd.uwo.ca}

\affiliation[18]{organization={Department of Computer Science, University of Western Ontario, London, Canada}}

\author[19]{Salil S. Kanhere} 

\ead{salil.kanhere@unsw.edu.au}

\affiliation[19]{organization={School of Computer Science and Engineering, The University of New South Wales (UNSW), Sydney, Australia}}

\author[20]{Rizos Sakellariou} 

\ead{rizos@manchester.ac.uk}

\affiliation[20]{organization={Department of Computer Science, University of Manchester, Oxford Road, Manchester, UK}}

\author[21]{Schahram Dustdar} 

\ead{dustdar@dsg.tuwien.ac.at}

\affiliation[21]{organization={Distributed Systems Group, Vienna University of Technology, Vienna, Austria}}

\author[22]{Omer Rana} 

\ead{ranaof@cardiff.ac.uk}

\affiliation[22]{organization={School of Computer Science and Informatics, Cardiff University, Cardiff, UK}}

\author[23]{Ivona Brandic} 

\ead{ivona.brandic@tuwien.ac.at}

\affiliation[23]{organization={Faculty of Informatics, Vienna University of Technology, Vienna, Austria}}

\author[1]{Steve Uhlig} 

\ead{steve.uhlig@qmul.ac.uk}

\cortext[cor1]{Corresponding author at: School of Electronic Engineering and Computer Science, Queen Mary University of London, London, E1 4NS, UK.}

% \section*{Highlights}
% \begin{itemize}
%   \item We investigate the prospects of AI/ML-based next generation computing systems.
% \item For successful computing services, we propose integrating emerging technologies and computing paradigms through AI-enhanced computing systems.
% \item For cloud, fog, edge, serverless and quantum computing environments, we outline trends and open challenges based on the use of AI/ML.
% \item We discuss the new research developments related to next generation computing with embedded intelligence and Explainable AI (XAI).
% \item We identify benefits and potential risks of computing approaches that make use of AI/ML algorithms.
% \end{itemize}

\begin{abstract}
Autonomic computing investigates how systems can achieve (user) specified ``control'' outcomes on their own, without the intervention of a human operator. Autonomic computing fundamentals have been substantially influenced by those of control theory for closed and open-loop systems. In practice, complex systems may exhibit a number of concurrent and inter-dependent control loops. Despite research into autonomic models for managing computer resources, ranging from individual resources (e.g., web servers) to a resource ensemble (e.g., multiple resources within a data center), research into integrating Artificial Intelligence (AI) and Machine Learning (ML) to improve resource autonomy and performance at scale continues to be a fundamental challenge. The integration of AI/ML to achieve such autonomic and self-management of systems can be achieved at different levels of granularity, from full to human-in-the-loop automation. In this article, leading academics, researchers, practitioners, engineers, and scientists in the fields of cloud computing, AI/ML, and quantum computing join to discuss current research and potential future directions for these fields. Further, we discuss challenges and opportunities for leveraging AI and ML in next generation computing for emerging computing paradigms, including cloud, fog, edge, serverless and quantum computing environments. 

\end{abstract}

 \begin{NoHyper}
\maketitle
\end{NoHyper}

\section{Introduction}
Autonomic Computing Initiative (ACI) from IBM were among the first industry-wide initiatives for the design of computer systems that require limited human interaction to achieve performance targets~\cite{kephart2003vision}. The Tivoli systems division at IBM focused initially at performance tuning of the DB2 database system using autonomic computing principles. 
The initiative was heavily inspired by observations from the functioning and coordination of the human nervous system and human cognition---i.e., the autonomic nervous system acts and reacts to stimuli independent of an individual's conscious input; an autonomic computing environment functions with a high level of Artificial Intelligence (AI), while remaining invisible to users~\cite{singh2017star}. Additionally, a human nervous system achieves multiple outcomes concurrently and seamlessly (e.g., internal temperature changes, breathing rates fluctuate, and glands secrete hormones as a response to stimulus) adhering to pre-defined/evolved ``limits'' and norms, and acting on impulses sensed or learned from the body itself or the environment. As for the human body, an autonomic computing environment is expected to work in response to the data it collects, sensed or learned, without an individual directly controlling functions used to manage a system~\cite{parashar2018autonomic}. 

Autonomic computing---also referred to as self-adaptive systems---is a field of investigation that studies how systems can achieve {\it desirable} behaviours on their own~\cite{puviani2013self}. It is common for these systems to be referred to as ``self-*'' systems, where ``*'' stands for the behaviour type~\cite{huebscher2008survey}, such as: self-configuration, self-optimization, self-protection and self-healing~\cite{elmroth2011self}.
 
An autonomic system's capacity to adapt to environmental changes is referred to as ``self-configuring''~\cite{kephart2007achieving}. The system automatically upgrades missing or obsolete components depending on error messages/alerts generated by a monitoring system~\cite{singh2015qos}. A self-optimizing autonomic system is one that can enhance its own performance by successfully completing computational jobs submitted to it, reducing resource overload and under-utilization~\cite{gill2018chopper}. Self-protection is an autonomic system's capacity to defend itself against potential cyber-attacks and intrusions. The system should also be detecting and preventing harmful assaults on the autonomic coordinator managing the overall system~\cite{gill2019resource}. Self-healing is a system's ability to discover, evaluate and recover from errors on its own, without the need for human intervention~\cite{singh2017star}. By decreasing or eliminating the effect of errors on execution, this self-* property improves performance through fault tolerance~\cite{derbel2009anema}. 

The ultimate vision is that neither self-managed systems nor self-healing systems need to be configured or updated manually~\cite{herrmann2005self}. In a broader sense, self-managed systems should be capable of controlling all of the aforementioned behaviours~\cite{kephart2015symbiotic}.  

\begin{figure}[t]
    \centering
    \includegraphics[width=1\linewidth,trim=0in 0.3in 0in 0.4in]{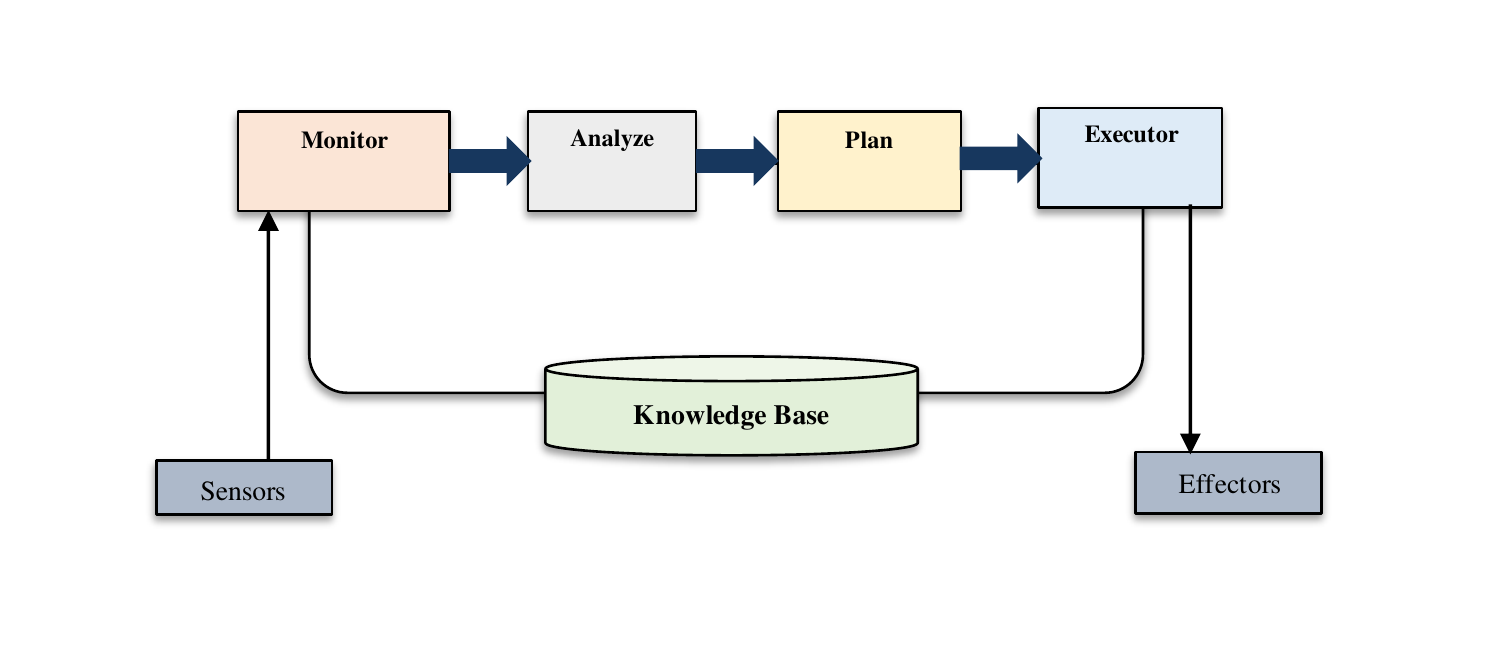}
    \caption{MAPE-K loop for Autonomic Computing}
    \label{fig:model1}
\end{figure}

Different practical systems realise these outcomes to varying levels of granularity and success. Also, the level of human intervention and control can vary.
As part of IBM's Autonomic Computing paradigm, the Autonomic Manager (AM) is a smart entity that interacts with the environment via management interfaces (Sensors and Effectors) and performs actions based on the information received from sensors and rules established in a low-level knowledge base. The AM is set up by an administrator using high-level warnings and acts. \color{black}Figure~\ref{fig:model1} illustrates IBM's autonomic approach in operation~\cite{kephart2003vision}. \color{black}Initial monitors acquire sensor data for regular inspection of Quality of Service (QoS) metrics whilst engaging with external hardware and send this data to the next component for further evaluation. In the Analyze and Plan modules, data collected from the monitoring module is analysed and appropriate action plans are drawn up in response to system warnings. Using the results of the data analysis, this autonomic system takes appropriate actions in response to the generated warnings. After a thorough review\color{black}, which includes verification and validation to provide guarantees that the adaptation will indeed work\color{black}, the plan is put into action by the Executor, whose primary goal is to ensure that the QoS of an executing application is maintained. An Executor monitors changes in the knowledge base and acts based on the results of the analysis.

\subsection{AI/ML for Next Generation Computing: A Vision}
AI and ML can be used to support and develop autonomic behaviours based on data collected about systems operations. ML techniques, for example, can be used to discover patterns in the workload, where these patterns can be used to optimise resource management~\cite{kephart2004artificial}. Additionally, to mitigate model uncertainty, ML-based dynamical system identification methods, such as recurrent neural networks, could be adaptively invoked by the autonomic manager to achieve self-learning. Thus, black- and gray-box models of the managed system can be generated during a concept drift and subsequently verified to check their sanity or even, detect mission-critical alterations of the system's operation~\cite{anderson2021self}. Further, AI may be employed in the analysis and planning stages of autonomic systems that are often arranged as monitor-analyze-plan and execute (MAPE) cycles~\cite{rutten2017feedback}, in addition to the use of techniques from control theory. It is the combination of feedback control with data-driven model construction using ML that offers key benefits in support autonomic self-management.  

Among the notable types of autonomous computing solutions: feedback-based control is one common solution. The use of self-organizing systems, such as particle swarm optimisation, cellular automata and genetic algorithms, are others. In the first category of solutions, systematic techniques for designing closed-loop systems capable of tracking system performance and altering control parameters are provided by autonomic computing~\cite{donepudi2018application}. There is a vast corpus of control theory literature and design tools that are used in these techniques. When it comes to the second type of solution, a variety of newly developing peer-to-peer approaches are now being employed to create massively scaled self-managing networks~\cite{ullah2020applications}.

\subsection{Motivation and Aim}
Autonomic computing has been integrated in computing paradigms such as cloud, fog, edge, serverless and quantum computing using AI/ML techniques~\cite{gill2019transformative}. The use of autonomic computing techniques is particularly significant when there is a large number of potential configuration options for a system. The greater the potential parameter space over which configuration options can vary, the greater the potential to optimise search over this space of possible options. Autonomic computing techniques are most useful {\it under the hood}, i.e. as a programmatic interface that can be invoked directly~\cite{buyya2018manifesto} from an application.  

There are many applications that can manage node failures, network setup/updates and a limited ability to carry out performance optimization on their own since most peer-to-peer networks are fundamentally autonomous. AI- and ML-based self-managing capabilities are becoming increasingly common in web services and data center management software, allowing these systems to automatically adapt to shifting workloads~\cite{kettimuthu2018towards}. However, autonomic features are not always included in schedulers and workflow managers, as such systems frequently lack the ability to monitor system condition and provide real-time feedback, making it difficult for these systems to be fully autonomous~\cite{harman2012role}. Integrating ``tuning'' capability that makes use of AI/ML techniques can extend the capability of such systems. For instance, self-managed computing platforms, such as Hadoop/MapReduce, provide self-healing and self-organizing capabilities that enable the use of a large number of resources~\cite{lopez2019self}. 

AI- and ML-based autonomic computing will become prevalent with increasing scale and interconnectivity of our systems, making manual administration and adaptation of such systems challenging and expensive. We expect AI- and ML-based autonomic computing will be the norm in the future---with human users still able to influence the behaviour of these systems through the use of judiciously integrated interfaces. Crucially, with the advent of cyberphysical systems and digital twins, quality-assured and mission-critical adaptations will become mandatory because the self-adaptive software will be responsible for physical assets, such as the unit operations of a processing plant.

But how should self-adaptive systems and AI/ML be combined? According to IBM, an autonomic system must meet the following eight criteria for computing systems using AI and ML techniques~\cite{salehie2005autonomic, nami2007survey, tarbell2019ai, singh2017star, singh2015qos, gill2018chopper, gill2019resource, psaier2011survey}:
\begin{itemize}
\item The resources that are available to the AI-powered system, as well as the capabilities and limits of the system, must be known by the system.
\item As the computing environment changes, e.g., because of a concept drift, the system must be able to adapt and reconfigure autonomously.
\item An efficient computer process requires a system that can maximise its performance via AI- and ML-based prediction.
\item When an error occurs, the system should be able to fix itself or redirect processes away from the source of the issue.
\item To ensure overall system security and integrity, the system must be able to detect, identify, and respond to numerous forms of threats automatically.
\item As the environment changes, the system must be able to interact with and develop communication protocols with other systems.

\end{itemize}

Despite the system's transparency, it must be able to predict demand on its resources, which can be forecasted with AI/ML techniques. Small, even inconspicuous computers will be able to communicate with each other across more linked networks, leading to the notion of ``The Internet of Everything (IoE)'', thanks in part to the emergence of ubiquitous computing and autonomic computing~\cite{lynn2018toward}. Crucially, AI-powered self-adaptive systems promise to cost-effectively and sustainably meet changing requirements in a changing environment and in the presence of uncertainty---vs., just adding more and more resources. Hence, in conjunction with the latest AI and ML techniques, autonomic computing is being studied and applied by a number of industry giants.

\subsection{Benefits of AI/ML-integrated Next Generation Computing}
AI-based Autonomic computing's primary advantage is lower total cost of ownership~\cite{ganek2004response}. As a result, maintenance expenditures will be significantly reduced. There will also be a reduction in the number of people needed to maintain the systems. AI-powered automated IT systems will save deployment and maintenance costs, time, and boost IT system stability. Companies will be able to better manage their business using IT systems that can adopt and implement directives based on business strategy and can make alterations in response to changing surroundings, according to the higher-order advantages. \color{black}Server consolidation is another benefit of using AI-based autonomic computing, since it reduces the cost and human labour required to maintain huge server farms~\cite{chaurasia2021comprehensive}. \color{black}Management of computer systems should be made easier using AI for autonomous computing. As a result, computing systems will be significantly improved. Another example of an application is server load distribution, which may be accomplished by distributing work across several servers~\cite{zhou2019distributing}. Further, cost-effective and sustainable power supply policies can be accomplished by continuously monitoring the power supply.

As a consequence of AI, the following changes have occurred in autonomic computing:
\begin{itemize}
\item Cost-effective: Using computer systems instead of on-site data centres has its advantages. Despite the high initial costs, organisations may easily acquire AI technology via a monthly charge in the cloud. Systems using AI may analyse data without involving a human being.
\item Autonomic: Enterprises may become more efficient, strategic, and insight-driven through the use of AI cloud computing. AI has the potential to boost productivity by automating tedious and repetitive tasks, as well as doing data analysis without the use of operator interaction.
\item Data Organization: Real-time personalisation, anomaly detection, and management scenario prediction may be achieved by integrating AI technology with Google Cloud Stream analytics.
\item Making Intelligent Decisions: Intelligence-based data security is critical as more cloud-based apps are deployed. Network traffic tracing and analysis made possible by AI-powered network security technologies. As soon as an abnormality is discovered, AI-powered systems can raise a red signal. Such strategy safeguards crucial information.
\end{itemize}

\subsection{\textcolor{black}{Related Surveys and Our Contributions}}
\textcolor{black}{As the area of computing continues to expand, there is a need for a fresh visionary work to review, upgrade and consolidate the current evidence and discuss potential trends and future perspectives in the field of computing. Varghese and Buyya \cite{varghese2018next} introduced an innovative survey on next generation cloud computing, which does not consider AI/ML. Abdulkareem et al. \cite{abdulkareem2019review} presented a review on AI for fog computing only. Massimo et al. \cite{merenda2020edge} explored literature for AI-based edge computing. Gill et al. \cite{gill2019transformative} presented a review on AI for cloud computing. The surveys from Kumar el al. \cite{kumar2021survey} and Li et al. \cite{li2020quantum}  highlighted the potential role of AI in quantum computing.  The suitability of AI for serverless computing is described in Hassan et al. \cite{hassan2021survey}.}

By combining AI/ML with cloud, fog, edge, serverless, and quantum computing, we've created the first review of its kind. Adding to the previous surveys, this new research gives a new imaginative approach to assessing and identifying the most current research challenges. Table~\ref{table:comparison_table} compares our review with existing surveys based on different criteria. 

\begin{table*}[]
\caption{\small \textcolor{black}{Comparison of Our Survey with Other Survey Articles. $\times$:= method supports the property.}} 
\label{table:comparison_table}
\begin{center}
\footnotesize
\begin{tabular}{|c|c|c|c|c|c|c|c|c|c|c|c|c|}
\hline
\textbf{Works} & \textbf{1} & \textbf{2} & \textbf{3} & \textbf{4} & \textbf{5} & \textbf{6} & \textbf{7} & \textbf{8} & \textbf{9} & \textbf{10} & \textbf{11} & \textbf{Publication Year} \\ \hline
           Varghese and Buyya \cite{varghese2018next}    &            &          &        x    &            &            &            &            &            &            &             &             & 2018    \\ \hline

            Abdulkareem et al. \cite{abdulkareem2019review}   &            &       x       &            &         x     &            &            &            &            &            &             &             &  2019  \\ \hline
               
           Gill et al.  \cite{gill2019transformative}   &            & x           &    x        &            &            &            &            &            &            &             &             &   2019 \\ \hline
           
                Massimo et al. \cite{merenda2020edge}   &            &       x     &            &            &        x    &            &            &            &            &             &             &  2020   \\ \hline
                
                Li et al. \cite{li2020quantum}   &            &        x    &            &            &            &            &       x     &           &            &             &             &  2020   \\ \hline
                
                Kumar el al. \cite{kumar2021survey}   &            &       x     &            &            &            &            &      x      &            &            &             &             &  2021   \\ \hline

                Hassan et al. \cite{hassan2021survey}   &            &       x     &            &            &            &  x          &            &            &            &             &             &  2021   \\ \hline

               Our Survey (This Paper)               &   x         &  x          &        x    &       x     &      x      &    x        &      x      &        x    &      x      &    x         &        x     &  2022  \\ \hline
\end{tabular}\\
\textbf{Abbreviations:} 1: Prospective Model, 2: AI, 3: Cloud Computing, 4: Fog Computing, 5: Edge Computing, 6: Serverless Computing, 7: Quantum Computing, 8:  Explainable AI (XAI), 9: Risks and Benefits of AI-integrated Next Generation Computing, 10: Hype Cycle, and 11: Intelligent Edge. 
\end{center}
\end{table*}

\subsubsection{Our Focus} 
\textcolor{black}{This paper leverages the expanding domain of Internet of Things (IoT), edge computing and the computing continuum as an exemplar application for AI-powered adaptation. There is a tremendous growth on applications that leverage such technologies, such as smart agriculture, environmental monitoring, industrial digital twins, smart cities, management of renewable energy generation/storage, etc. Nevertheless, our discussion can be expanded to other fields as well. }

\subsection{Article Organization}
\begin{figure}[t]
    \centering
    \includegraphics[width=1\linewidth]{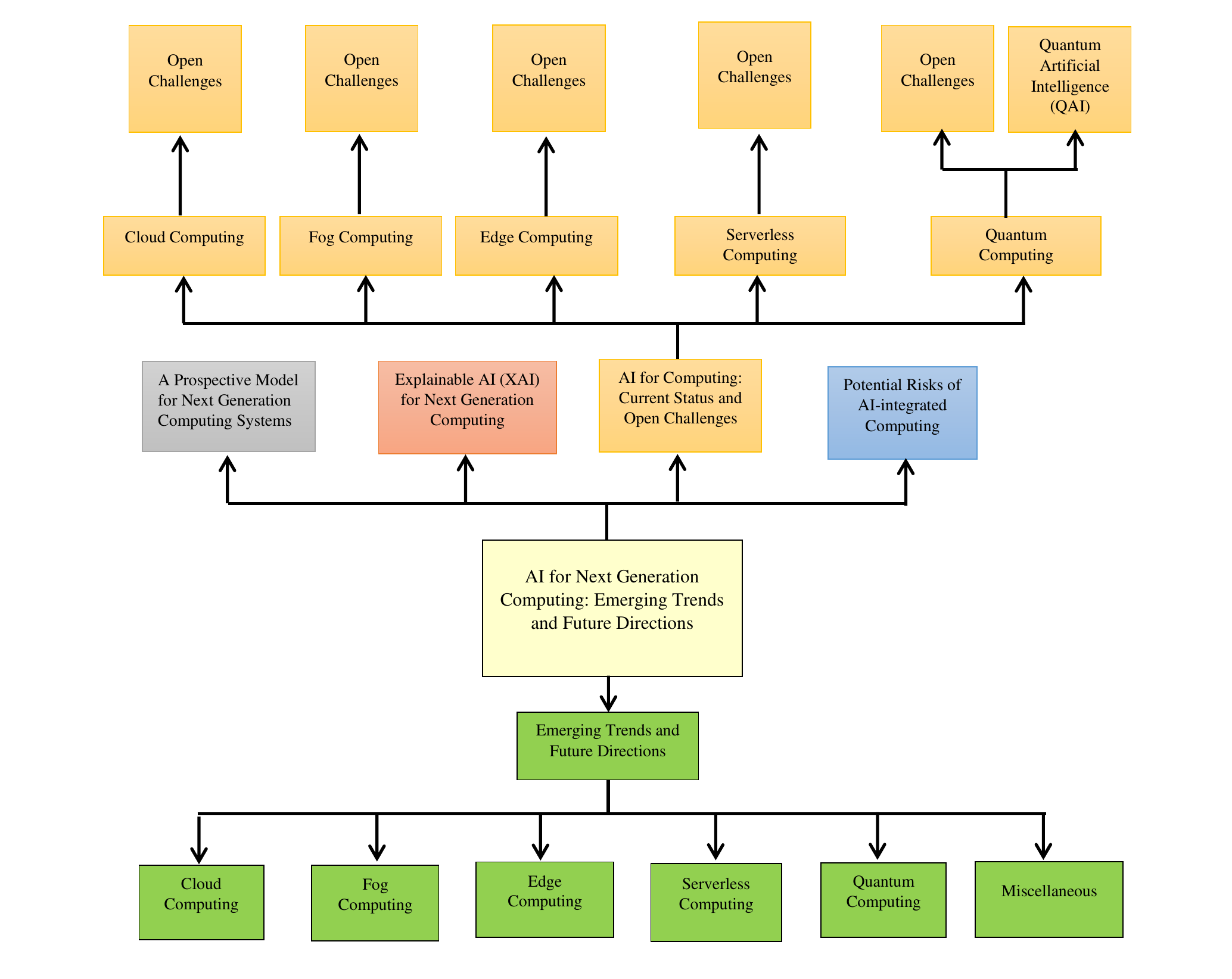}
    \caption{The organization of this survey}
    \label{fig:model2}
\end{figure}

The rest of this article is organized as illustrated in Figure~\ref{fig:model2}. Section~2 proposes a conceptual model. Section~3 is presenting vision and discussing various emerging trends in AI for cloud, fog, edge, serverless and quantum computing. Section~4 discusses the new research developments related to autonomic computing with embedded intelligence. Section~5 discusses the use of Explainable AI (XAI) for next-generation computing. Section~6 presents the potential risks of autonomic computing approaches that make use of AI/ML algorithms. Section~7 gives the hype cycle for autonomic computing and highlights the future directions. Section~8 concludes and summarizes the paper. 

\section{A Prospective Model for Next Generation Computing Systems} 
\begin{figure}[t]
    \centering
    \includegraphics[width=.85\linewidth]{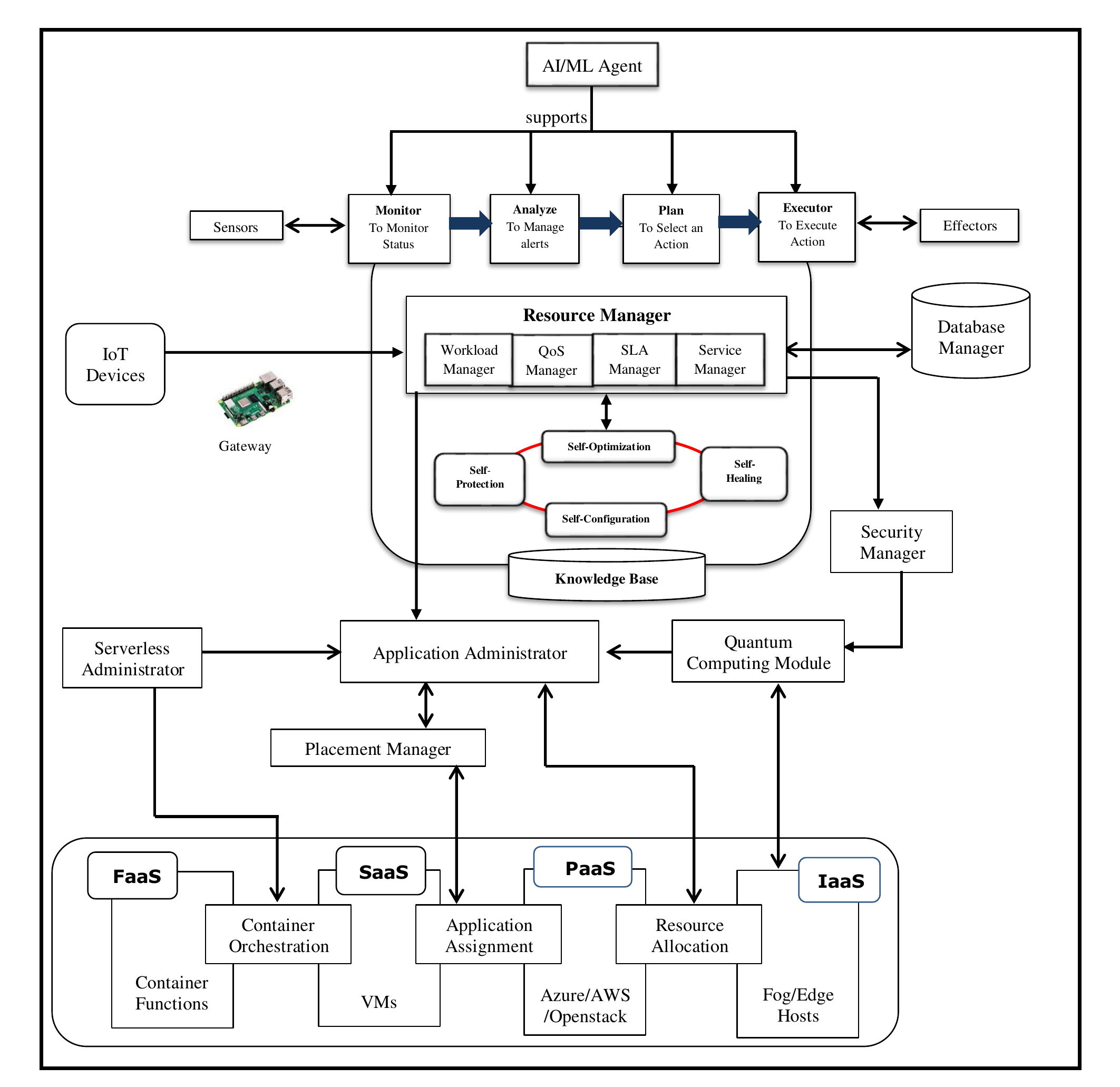}
    \caption{A Prospective Model for AI-integrated Next Generation Computing}
    \label{fig:model3}
\end{figure}

To show the relationship between AI/ML and autonomous computing systems, we propose a prospective software architecture model as shown in Figure~\ref{fig:model3}. Our proposal integrates advanced technologies to offer effective computing services that fulfill the demand for a variety of IoT applications.

\subsection{IoT Applications}
Gateway devices will be used by IoT/edge devices and end users to communicate with computer systems, abstracting away the interactions with sensors and actuators/effectors located on the edge \cite{goel2021review}. The system will communicate with various and multiple instances of IoT applications (such as healthcare, smart city, farming, and weather monitoring) \color{black} or their digital twins \color{black} to efficiently provide \color{black} AI and other autonomic \color{black} services~\cite{desai2021healthcloud}. 

\subsection{Resource Manager}
Distributed systems, including IoT edge platforms, require adaptive and fault-tolerant management of resources and scheduling of tasks. The proposed resource management module maintains the set of available and reserved resources (the number of CPUs utilised, the amount of memory, the price of resources, the kind of resources, and how many resources there are) as well as the desired resources, constraints (e.g., placement) and QoS per deployed task. Further, the module incorporates data supplied by the provider on the accessible and scheduled resources, as well as the resource specification (resource identity, resource category, configuration, data, use information, and pricing of resource).

When evaluating QoS, the QoS manager figures out how long it will take to complete a given workload. Priority queues (workloads with an urgent deadline in execution state) are created for critical cloud workloads based on Service Level Agreement (SLA) details, which includes details about the highest and lowest violation probability and penalty rate in the case of SLA violation. The service manager is responsible for overseeing all aspects of the system's operation. With the use of SLA and QoS information, a mapper may assign workloads to adequate resources, taking into consideration both SLA and QoS. After allocating the workloads to the available resources, the resource manager creates a workload schedule by predicting it using AI. In order to complete tasks within a certain budget and timeframe, the resource scheduler makes efficient use of the system's resources, which are predicted via AI/ML techniques. Finally, wherever possible, the resource manager will be providing explainable guarantees under uncertainty---potentially using explainable AI methods---that the proposed adaptation will indeed meet the desired QoS.

\subsection{Autonomic Model}

This future model employs IBM's autonomic computing model~\cite{kephart2003vision}, which emphasises self-healing, self-configuring, self-protecting, and self-optimizing features.

\begin{itemize}
\item Self-healing is aimed at making all required modifications to recover from defects in order to keep the system running without interruption~\cite{psaier2011survey}. Software, network, and hardware errors must not impair the efficiency of the algorithm or workload regardless of their severity~\cite{gill2019resource}. Any unintended exception in high resource-intensive applications can cause a software, hardware or network failure. AI-based systems can leverage a variety of data sources and sensor data to generate fault models and enable predictive---instead of reactive---fault detection and maintenance.

\item The primary goal of self-protection is to keep the system secure from hostile purposeful acts by keeping track of suspicious activity and responding appropriately in order to keep the system running smoothly~\cite{gill2018chopper}. To prevent an attack, the system must be able to tell the difference between what is lawful and what is not. AI-based prediction systems can be used to achieve this: for instance, the system could be trained to detect vulnerabilities in the communications configurations/policies or identify code smells in the user-submitted functions/lambdas. 

\item Installing missing or obsolete parts without requiring any human interaction is the primary goal of self-configuration. Depending on the situation, a developer may need to reinstall specific components or perform software upgrades~\cite{singh2017star}. Self-configuration takes care of the cost of resources and penalties for SLA violations, which can be predicted in advance through AI/ML.

\item Dynamic scheduling approaches are used to match jobs and workloads to the best available resources in the self-optimizing aspect~\cite{psaier2011survey}. The autonomic element's input is used to constantly enhance the system's performance through dynamic scheduling. AI/ML based adaptive scheduling can be used for data-intensive applications because it is flexible and can be adjusted to a changing environment with ease. Further, the impact of different QoS characteristics on system performance can be measured automatically~\cite{singh2015qos}. 

\end{itemize}

Models for complex distributed systems that can self-heal, self-configure, self-optimise and self-protect have been developed using this idea. Autonomic elements (AEs) are primarily in charge of managing resources on their own~\cite{kephart2003vision}. Figure~1 shows a schematic representation of the many components that make up an AE system. Interaction between all the AEs is necessary for the sharing of messages on system performance. AEs complete a necessary sub-task to maintain the system's performance based on interaction. There are four stages to the IBM model of an autonomic system~\cite{kephart2003vision}: Monitor, analyse, plan, and execute are the four steps in the process, which will be supported by AI/ML models to improve the monitoring, analysing, planning and execution. Further, AI-powered techniques could also improve the efficiency of the persistence (knowledge) component of the MAPE-K loop, especially in effectively resolving state synchronization in a highly-distributed and potentially unreliable environment.

\subsubsection{Sensors}
Sensors gather data on the QoS metrics of the present state nodes' performance~\cite{singh2017star, singh2015qos, gill2018chopper, gill2019resource}. Input from computation elements is first sent to the manager, which subsequently sends this information to Monitors through the manager node. Faults (software, network, and hardware), fresh updates on component status (outdated or missing), and security threats are all included in the recent developments (intrusion detection rate).

\subsubsection{Monitor}
Initially monitors data from the resource manager node to continually check performance variances by contrasting AI-based predicted and real outcomes~\cite{singh2017star, singh2015qos, gill2018chopper, gill2019resource}. The threshold value of QoS metrics, which also contains the highest value of SLA violation, is already recorded in the knowledge base. The faults (network, software, and hardware), fresh upgrades of resources (obsolete or lost), security assaults, variation in QoS parameters, and SLA violations are noted, and this data is transmitted to the next module for more investigation. Each node has a QoS agent deployed to monitor and predict the performance of the above-mentioned QoS parameters for self-optimization. Self-protection is achieved by installing security agents on all processing nodes, which are then utilised to track down both undiscovered and recognized attacks. After analysing the system's current database, additional abnormalities can be predicted using AI/ML. System invasions and system abuse are detected and classified as either normal or abnormal utilising its monitor and the system's attributes are compared with metadata.
Hardening agents for software, networks, and hardware will be reducing attack surfaces by identifying corresponding flaws to achieve self-healing and self-protection. When a new node is introduced to the cloud, the hardware hardening agent scans the drivers and validates the replica of the original drivers. The new node is inserted when the device driver has verified it. This node will create a warning if it is still present in the system.
The performance of the software and hardware components is monitored by agents for self-configuration. The software component agent retrieves the active component condition for all software components that are employed on separate processing nodes.

\subsubsection{Analyze and Plan}
When the monitoring module sends data, the Analyze and Plan unit evaluates it and identifies a strategy for reacting to the alarm~\cite{singh2017star, singh2015qos, gill2018chopper, gill2019resource}.
After a QoS agent generates an alert, the analysis unit begins predicting QoS metrics associated with a specific node. `DOWN' status is reported for that unit, and the unit is restarted, and the state of that node is measured. Alternatively, new resources are added if the node state goes to `ACTIVE'.
After an alarm is sent by a hardware or software agent, the analysis unit begins examining the behaviour of the node's hardware and software (self-healing). Node `N' should be set to ``DOWN'' if an alert is produced during workload execution and restarted, to measure the state of that node. The execution of the node's state switches to `ACTIVE' if execution is continued, or alternatively another reliable node is chosen. Self-protection begins by examining attack logs once an alarm is produced by the security agent and a signature is created by the analysing component.
After an alarm is issued by a hardware or software component, the analysis unit begins studying the behaviour of a node's hardware and software components. It is necessary to designate a hardware component as ``DOWN'', reset the failed component, and then start it again in order to predict whether or not it is ``CRITICAL'' or ``ERROR''. Again when the data has been processed, this framework takes care of implementing the alert-related actions on its own.

Further, before any adaptation does take place, the modules will first provide evidence that the proposed plan will indeed complete successful. This is achieved using a combination of formal guarantees, which can be derived from the use of control theory. An AI-model can also be used to predict when the users might issue a goal update--based on external information or other types of operational data---and prepare/assess an adaptation plan ahead of time.

\subsubsection{Executor}

A plan is put into action by the executor~\cite{singh2017star, singh2015qos, gill2018chopper, gill2019resource}, whose primary purpose in self-optimization to enhance QoS and execute tasks within a pre-defined deadline. Using the data from the analyzer, the executor may quickly, cheaply and efficiently add a new node to the pool of resources. If the resources are not already in the pool of available resources, then notify the user and negotiate a SLA before adding a new node from the backup pool of resources with the least amount of workload, price and power usage requirements. These aspects can be predicted in advance using AI/ML. A node that is not reliable should be replaced with a node that is the most stable amongst those available. To relaunch the node, the current status of a node is stored (checkpointed). The node is then restarted. If the problem persists, an alert is subsequently generated.

For self-healing, whenever a new component is introduced, it should be linked to other components and restarted. 

\subsubsection{Effector}
New policies, regulations, and notifications are sent to other computing nodes via the effector~\cite{singh2017star, singh2015qos, gill2018chopper, gill2019resource}, which serves as an interface between the various computing nodes. Through the effector, the computing nodes can work together to form a more powerful system. It is worthwhile mentioning that a system-of-systems approach is likely to be leveraged for such applications; hence, effectors of a top system might be triggering adaptation of a bottom system and so on.

\subsubsection{Knowledge Base}
\color{black}
The main aspects of information stored in the knowledge base are the following: (a) The current and previous states of the system (including deployed applications, available computing resources, etc.), whose values are read via the system's monitors. (b) The desired state of the system, which is driven by specifications set by the user/admin/operator of the system; they include both functional requirements, such as the microservices network of deployed applications, as well as nonfunctional requirements, such as QoS Service Level Objectives (SLOs) on desired response time, tail latency, target resource utilization, etc. (c) Current, past and predicted models---as well as meta models and surrogate models---of the system and its environment generated via AI/ML as well as the efficacy of the various AI/ML methods used for their training. (d) Current and past execution plans that are devised by the planner module and implemented by the executor module. (e) The actual code of the various interfaces the system provides to enable informed self-adaptation by autonomically incorporating improved methods for various operating aspects. For example, new AI/ML, scheduling and resource management algorithms, could be selected by the self-adaptation algorithm and added in, which would eliminate the need for the software engineering team to have to patch the system. (f) Further, the Knowledge Base will maintain pre-stored policies with predefined configurations to support system management. It is the responsibility of the system administrator to periodically update the policies stored in the Knowledge Base to reflect changes in resource scheduling regulations. A system admin will be replaced by an AI-based autonomic agent to handle the execution automatically.

Crucially, the knowledge module needs to provide a centralized location for the various running tasks, which could be executed as threads, processes and of course across multiple nodes of the distributed cluster, to safely store and exchange information. This kind of architecture is required for highly distributed systems; otherwise, direct communication between the various units will result in dramatic slowdowns due to locking and contention, increase the attack surface or even worse, into system failure due to synchronization issues, such as race conditions, that can invalidate information manipulated by multiple actors. Finally, the knowledge base needs to be replicated, potentially across multiple reliability zones, to assure business continuity in cases of hardware and communication failures or even a catastrophe that knocks down a whole datacenter. As such, distributed consensus and adaptive data recovery algorithms are required to maintain data validity.

\color{black}

\subsection{Service Management Layer}

There is a database manager in this position (which manages the data of IoT applications effectively). AI-based systems can be used by Security Manager to predict and guard against external threats on task execution~\cite{gill2021quantum}. Application data may be securely sent during task execution with the help of a blockchain service. At runtime, the serverless manager controls the cloud resources that IoT applications are consuming. With the integration of Serverless data pipelines with quantum computers, efficient load balancing and dynamic provisioning may be achieved for the edge computing paradigm. It is the responsibility of the application manager to control the deployment of IoT applications and to provide data for the allocation of resources in advance, which can be achieved using AI/ML. The placement module serves as a bridge between the application manager and the application placement module.

Four categories of services are included at the bottom layer~\cite{gill2021quantum}: function (FaaS), software (SaaS), platform (PaaS), and infrastructure (IaaS). Function containers are used to provide a virtual environment for computer systems that can be dynamically scaled up and down. SaaS uses the notion of virtualization based on VMs to deliver cloud-based services. Platform as a service can be provided via Microsoft Azure, Amazon Web Service (AWS), or OpenStack. By lowering latency and reaction time at the edge devices, fog and edge computing may be used to deliver the infrastructure service. Orchestrating containers using orchestration is an intermediary step between deploying containers as a service and deploying them as software as a service (SaaS). The placement of IoT applications for dynamic provisioning and management is handled by application placement, which is a bridge between SaaS and PaaS. Machine learning and artificial intelligence-based approaches are used to schedule the cloud resources of PaaS and IaaS~\cite{XuTSUSC2021}. Using quantum technology, the system is able to perform nonce or Proof-of-work (PoW) computations in a fraction of the time.

\section{AI for Computing: Current Status and Open Challenges}
\begin{figure}[t]
    \centering
    \includegraphics[width=1\linewidth]{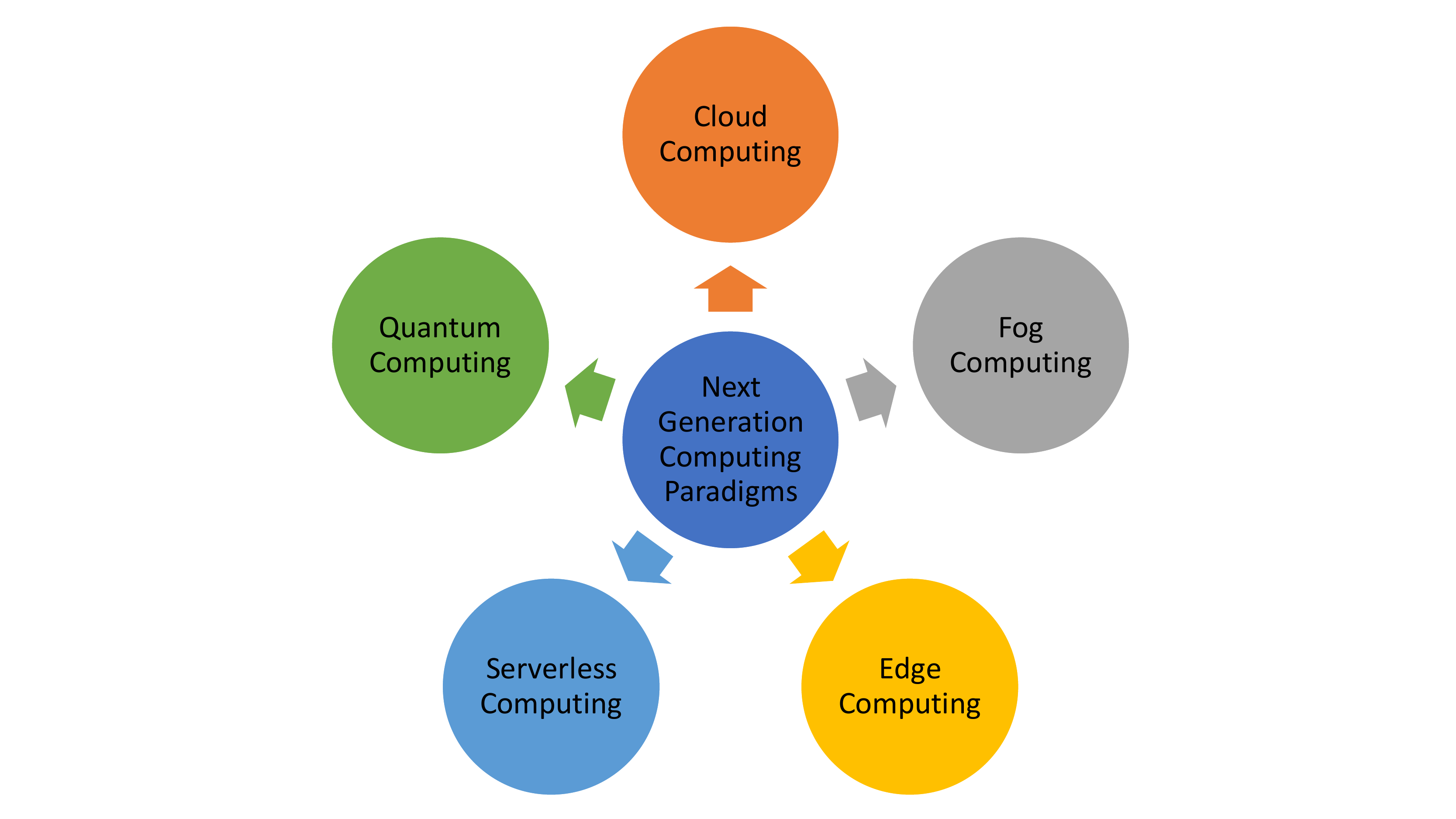}
    \caption{Emerging computing paradigms}
    \label{fig:model4}
\end{figure}

 It is very important to identify the research opportunities for leveraging AI and ML in next generation computing for emerging computing paradigms, including cloud, fog, edge, serverless and quantum computing environments as shown in Figure~\ref{fig:model4}. This section discusses various new trends and open challenges in AI-integrated next generation computing.

\subsection{Cloud Computing}
It is becoming increasingly evident that the rise of cloud computing and the rise of AI are mutually reinforcing. As a result, using AI in the cloud can improve the cloud's performance, efficiency, and digital transformation~\cite{pusztai2021slo}. AI in the cloud computing environment is a crucial key to enabling organisations to become more efficient, strategic and insight-driven, while at the same time providing greater flexibility, agility and cost savings~\cite{tuli2022hunter}. As a result, we turned to industry insiders for their insights about the expanding importance of AI in cloud computing. 

AI and cloud computing may be combined in a variety of ways to enhance cloud computing. AI tools and software are synched with the power of cloud computing in order to provide an enhanced value to the existing cloud computing environments~\cite{abdelaziz2018machine}. This combination makes enterprises efficient, strategic, and insightful. Data and applications hosted on the cloud allow businesses to be more responsive and adaptable, while also saving money for the company as a whole~\cite{abdelaziz2018machine}. Existing capabilities gain intelligence, and customers receive an excellent experience, thanks to the addition of this additional layer of AI that aids in the generation of insights from data~\cite{abdelaziz2018machine}. As a result, businesses may profit from a tremendously distinctive mix. Cloud is like a video game, which emits an enormous quantity of operating data and telemetry, much like a Tesla electric vehicle~\cite{ulrich2020top}. As a result, AI-based cloud computing is basically AI Ops, which uses algorithms to make sense of all this data rather than relying on humans~\cite{masood2019aiops, dang2019aiops}. In the post-COVID future, cloud-computing investment increased by 37 percent to \$29 billion in the first quarter of 2020 compared to the first quarter of 2019~\cite{pwc}. Integrating AI and cloud computing can therefore help businesses get closer to their consumers while also increasing their operational efficiency~\cite{nastic2020sloc}.

Cloud computing environments and solutions are helping businesses to become more agile, adaptable, and cost-effective because this significantly cuts infrastructure administration expenses for corporations~\cite{younas2020elicitation}. As a way to handle enormous data repositories, simplify data, improve workflows, and create real-time insights for day-to-day operations, AI gives companies more freedom. The operational weight may be shifted from processes and people to engineering and data~\cite{breiter2009life}. That is why AI is boosting cloud computing in a variety of ways. The Software as a Service (SaaS) paradigm is currently being used to successfully employ cloud-based AI~\cite{pop2016machine}. SaaS companies are incorporating AI into their solutions, which provides clients and end-users with enhanced capabilities. Another method businesses are adopting AI to enhance their present cloud infrastructure is through AI as a service~\cite{elger2020ai}. The use of AI makes applications more flexible and efficient, reducing mistakes and increasing production.

The cloud native paradigm derived from cloud computing has shifted the traditional monolithic cloud application into light-weight, loose-coupled and fine-grained microservices~\cite{XuContainerSurvey}. This paradigm can support the applications to be updated in a much more efficient manner. However, due to the increased number and time-sensitive features of microservices, their efficient management can be challenging. AI/ML based solutions can address some of the challenges, for instance, neural network based approach can be applied to predict workloads of microservices, and ML based techniques can be utilized to analyze the dependency of microservices. 

The following are various advantages to deploying AI in the cloud: 
\begin{itemize}
\item Enhanced data management: Data is king in today's data-driven world, which necessitates better ways to handle it. An enterprise's ability to keep track of that data is a major hurdle~\cite{chaudhary2018optimized}. Cloud-based AI tools and apps that recognise, update, catalogue, and provide real-time data insights to clients. AI techniques may also be used to detect fraudulent activity and identify anomalous system trends~\cite{rajeswari2022ai}. Banks and other financial institutions rely heavily on this technology to remain competitive and safe in today's high-risk climate.

\item Automation: Intelligent automation can now be implemented throughout a whole business thanks to the combination of AI and the cloud, which removes the last remaining roadblocks~\cite{surya2018streamlining}. Predictiveness is enhanced by AI since algorithmic models draw on historical data and other patterns to deliver in-the-moment insights~\cite{marshall2018cloud}. AI and cloud computing solutions can help businesses go from semi-structured to unstructured documents cognitively automated while also pushing the frontiers of effective infrastructure management, resulting in little downtime and impact~\cite{jha2021combining}. As a result, the cost of doing business is transformed, and the customer experience is transformed as well.

\item Cost Savings: Cloud computing allows businesses to just pay for the resources they utilise. 
This saves a significant amount of money compared to the typical infrastructure expenditures of building and maintaining massive data centres~\cite{robertson2021cloud}. Saved money may be utilised to build more strategic AI tools and accelerators, which can then be used to increase revenue and save the company money at the core~\cite{horn2019cost}. This will lead to better operational quality and cheaper expenses.

\end{itemize}
 
\subsubsection{Open Challenges}
The following issues that may arise when these two technologies are combined: 
\begin{itemize}

\item Integration: It is never easy to get started with a seamless integration of two different technologies. In order to accomplish this integration, companies must first move all of their apps and technologies to the cloud~\cite{surya2018streamlining}. This is no small feat for many companies. Businesses may only begin to consider cloud-based AI after undergoing such a seismic shift. Thus, the technological sync is excessively reliant on companies that are implementing tangible digital transformations of their systems.

\item Inadequate data: Large datasets with high-quality data are ideal for AI technologies. Businesses must make sure that their data is both accessible and clean in order for AI to be of any use~\cite{gonzalez2019biomedical}. Because data is often unorganised or missing, this is a big difficulty. It is critical that the solution's value be derived from high-quality data.

\item Security and privacy issues: To prevent data breaches, businesses must be vigilant about protecting their sensitive and financial information from adversaries, who are likely to target them~\cite{chatterjee2020adoption}.

\end{itemize}
 
This synchronisation of AI with the cloud necessitates tremendous knowledge, resources, and financial investment if it is to be worthwhile for businesses. It is only when cloud computing and AI systems are properly integrated that companies will be able to utilise a wide range of powerful machine learning capabilities, such as image recognition and natural language processing~\cite{carvalho2019off, blohm2019towards}. As a result, additional businesses will follow suit in the future. Businesses will require an AI cloud in order to keep up with the rapid advancements in cloud computing. After successful implementation of it, AI operations will eventually become the standard approach for cloud management~\cite{dang2019aiops}. The cloud is already a powerful technology, but they believe that AI will make it even more so. With this combination, data analysis and management will undergo a radical shift. The marriage of AI with the cloud is a game-changer and will bring unparalleled value to end-users in a world flooded with vast volumes of data~\cite{gonzalez2019biomedical}. Now that cloud computing and AI are more widely available, they are producing upheaval in a wide range of industries throughout the globe. It is clear that technology has moved from being merely operational to one of strategic importance. AI is expected to help the company tackle new and more visible challenges, as well as open up a new universe for its potential clients.

\subsection{Fog Computing}
Fog computing was established to supplement cloud computing services because of the rising use of the IoT and the necessity to handle and store massive amounts of produced data~\cite{singh2021fog}. IoT applications with minimal reaction time requirements can be supported by fog computing, which provides basic network services~\cite{gubbi2013internet}. It is difficult to distribute IoT application activities efficiently inside fog nodes in order to fulfil QoS and quality of experience (QoE) restrictions due to fogs' scattered, heterogeneous, and restricted resource nature~\cite{mahmud2019quality}. Vehicle-to-Everything (V2X), health monitoring and industrial automation employ fog computing because it provides computing capabilities near to the user to meet reaction time expectations for these applications~\cite{tong2019artificial}. As a result, these apps create enormous amounts of information from the widespread use of IoT devices. Because of delays in long-distance data transmission and network congestion, cloud computing is unable to meet latency requirements~\cite{singh2021fog}. It provides a network of gateways, routers, and compute nodes between the source of data and cloud computing centres. Because of the low latency and energy efficiency, as well as the reduction in bandwidth required for data transport, fog computing extends cloud computing \cite{gill2020thermosim}. Fog nodes can be used to process sensitive data instead of sending it to the cloud, which improves security~\cite{maroof2019plar}. Using the data generated from various IoT devices, these applications aim to provide helpful information while also addressing latency concerns~\cite{gubbi2013internet}. In recent years, researchers have increasingly turned to AI to help them analyse large amounts of data for the aforementioned uses. AI's Machine Learning (ML) and Deep Learning (DL) subfields give useful data insights and decision help~\citep{zou2019edge, firouzi2021convergence}. Following that, we are discussing some of the AI-enabled fog computing technologies that make these applications possible.

For the IoT, 5G signifies more than just a new era of wireless innovation. More than trillions of sensors, gadgets, and machines are powered by AI and run autonomously from the data centre to the edge of the network~\citep{zou2019edge, firouzi2021convergence}. In terms of speeding up data analysis and decision making, fog computing and edge computing are the two best technologies. Many ``fog devices'' will be networked and co-located as part of a distributed computing system known as fog computing~\cite{singh2021fog}. Edge management, data collection, monitoring, analytics, and streaming all take place at the edge of the network in the fog computing nodes \cite{teoh2021iot}. While fog computing is capable of connecting a limited number of devices, this technology has a far greater capability to handle real-time requests and to aggregate data from a much larger number of sources. Input-response delay is therefore greatly reduced. We have been able to access resources of all types, have scalable architectures with the press of a button, and utilise them from anywhere since Amazon's cloud was launched in 2006~\cite{czentye2019optimizing}. Cisco claimed in 2008 that IoT, is the one of the technologies that will benefit from the cloud, but its roots date back to 1999~\cite{satyanarayanan2015brief}. For example, we can save sensor data and act on it, automate processes using AI, and react in real time to circumstances that previously necessitated direct involvement. When IoT was first introduced, it promised to extend to both professional and personal areas, and how sensorization and communication protocols had to change to meet these new demands~\cite{tuli2020ithermofog}. New paradigms have emerged as a result of the integration of sensor data and the application of AI to it. New terms like ``Smart home'' are being used to describe new technologies that make it easier and more convenient to manage our homes' energy use and other aspects of our daily lives~\cite{skouby2014smart}. On a broader scale, the term ``Smart City'' is used to describe cities, while ``Smart Factory'' is used to describe manufacturing and processing facilities. One thing they all have in common is the utilisation of data and automated decision-making in combination with automation, which can be easily achieved using AI and ML techniques.

An example of this is altering the configuration of a computer or railway, putting the brakes on an autonomous automobile, or sending a warning for a preventative maintenance. It is evident from the examples that decision-making and action-taking cannot be done on the cloud, but rather on devices that are closer to the sensors that collect the data. In contrast to cloud computing, fog computing offers a variety of advantages for IoT applications \cite{gill2022manifesto}. First and foremost, quicker and real-time processing are possible because local processing is used rather than relying on the cloud. As a result of the large number of IoT devices already in use and expected in the future, less network traffic means better communication. Additionally, more apps may be developed and operated everywhere there is an Internet connection. We need to think how AI could help in the automation?

\subsubsection{Open Challenges}
Research on application deployment has already been done in several domains, such as industry and manufacturing, but there are still a number of issues that need to be addressed.
\begin{itemize}

\item Execution Time: For both service providers and customers, time is the most pressing issue. One of the key motivations for putting software in the fog is to speed up user reaction time. The time parameter was one of the performance indicators investigated in the literature~\cite{bonomi2012fog, naha2018fog}. When there are more demands, the QoS suffers. This difficulty has been partially alleviated by the presented techniques, but it remains a concern~\cite{goudarzi2020application}. In the application placement problem, applied techniques for optimising time performance metrics in the category of deep learning algorithms~\cite{luong2020machine, naveen2018search, lin2020fog}. We may be able to achieve better outcomes if we use different machine learning algorithms and evolutionary algorithms or novel combinatorial techniques.

\item Mobility-awareness: Fog computing's lack of mobility support may be noticed when dealing with a large number of mobile users with varying application needs~\cite{wang2019mobility}. Consequently, migration methods and architectures that can handle a wide range of mobility activities are required. Migrating VMs or containers is something that has been discussed in a few publications~\cite{waqas2018mobility}. Moving to a new location may be expensive, as well. Reinforcement learning~(RL) approaches like Q-learning and State–Action–Reward–State–Action~(SARSA) have been used to study this topic~\cite{zhao2017reinforcement}, but it remains a challenge in practical contexts where there are many requests~\cite{wang2022optimization, wang2020reinforcement}. 

\item Resource Scheduling: Another problem the authors encounter is managing resources in a dynamic environment like fog, which has a limited amount of resources and a short reaction time for the user. The Fog environment is less flexible than the cloud when it comes to resource sharing~\cite{rusman2019fog}. Therefore, the issue of efficient resource utilisation must still be addressed. Resource allocation was based on a survey of existing research and the use of neural networks, support vector machines, and k-nearest neighbours (KNNs)~\cite{majumdar2018kyasanur}.

\item Energy-efficiency: The amount of energy used if supplied policies and algorithms are improved, idle fog nodes may be turned-off and energy consumption can be avoided in combination with QoS and QoE since the application modules are situated in the dispersed fog nodes~\cite{bonomi2012fog, naha2018fog}. \color{black}Energy consumption and cost are influenced by memory, CPU, and bandwidth use which can be predicted using machine learning methods include K-means, KNNs, logistic regressions, branch and bounds and Deep Q-Network (DQN) and SARSA~\cite{ray2018compassionate}.%\cite{hussain2020machine, lv2020advanced, ray2018compassionate}.
\color{black}

\item Security and privacy: Fog infrastructure is critical to determining the security of the applications because of security concerns such as information degradation, identity disclosure, replay, and denial of service assaults~\cite{hussain2020machine}. Authentication, encryption, and data integration all need to be implemented in dynamic computing settings due to the lack of control that users have over their information~\cite{hussain2020machine}.

\item Fault-tolerance and availability: One of the primary reasons for the development of fog computing was to improve dependability. When it comes to fog computing, difficulties like sensor failure, a lack of access network coverage in a particular region or the entire network, service platform failure, and a broken user interface system connection are all part of the equation~\cite{zou2019edge}. Another challenge in the fog environment is to increase the availability of apps. A heuristic approach to improving service availability and QoS is to map applications to fog communities and then transitively put their services on the fog devices' community, according to the service placement problem~\cite{singh2021fog}.

\end{itemize}

Images, video, natural language processing (NLP), and robotics are some of the more recent fog computing applications that are only starting to emerge~\cite{carvalho2019off, blohm2019towards}. Fog computing's picture placement and processing is one of the most widely utilised sectors of AI in research and industry, with the goal of differentiating objects or people from one another and the capacity to classify and discriminate photos based on image processing algorithms~\cite{queralta2019edge}. The use of fog computing for image processing-based applications decreases response time and improves QoS. Placement in fog settings with effective scheduling algorithms might be beneficial in circumstances linked to medical applications that demand precision in image processing and fast processing of medical data~\cite{abdulkareem2019review}. According to a literature, deep learning algorithms, such as Convolution Neural Networks (CNN) and Generative Adversarial Network (GAN), can be used in the image processing area in fog~\cite{rihan2020deep}. 

Another area of interest in the sector is NLP~\cite{rihan2020deep}. For sound processing and recognition, cloud and fog environments are needed to store data. For security reasons, deep learning approaches with sound imitation might be useful. For example, scenarios for smart homes and processing and identifying the homeowner's speech from outsiders should be done with care and speed, wherein effective scheduling methods for placement of NLP applications in fog will be proposed. Techniques from the field of deep learning may be beneficial here. Industry, trade, agriculture, and health all benefit greatly from robotics, making it an essential issue for discussion~\cite{abdulkareem2019review}. In circumstances when quick judgments must be made, the usage of fog environments that employ machine learning to process and make decisions, may be acceptable. In order for robots to communicate data, they require an environment like fog that responds quickly to their commands~\cite{tanwani2019fog}. IoT defines each robot as an item capable of interacting with other IoT things and other robots~\cite{tian2019fog}. Literature reported~\cite{tian2019fog, tanwani2019fog, li2018deep} that deep learning approaches for placement robot tasks in fog, however further research is needed. In the future, methods and scheduling algorithms for fog computing's application placement problem will need to be established depending on the types and categories of request applications, according to research and evaluations of literature~\cite{ng2020anomaly}. An application placement issue in fog for robotics or simple image, video, audio processing is an example of this type of difficulty~\cite{pusztai2021pogonip}. As a result, the QoS and QoE will be enhanced using AI.

\subsection{Edge Computing}
Distributed computing has evolved from content delivery networks to become a generally accepted and commonly used edge computing paradigm that brings processing and data storage closer to the end user's location~\cite{nastic2021polaris}\cite{XuSPE2022}. Instant data that is created by the user, and only for the user, requires compute and storage on the edge, but big data always requires cloud-based storage~\cite{yu2017survey}. As customers spend more time on mobile devices, businesses have recognised they need to move key computation to the device in order to service more customers. The edge computing market has a chance to develop as a result of this. By the year 2023, it is expected to reach \$1.12 trillion~\cite{edgeai}. 74 percent of all data will need to be handled on the edge by 2022, according to Gartner, compared to 91 percent of all data currently being processed in centralised systems~\cite{edgeai}. 

Customers are more concerned about their privacy and want to know how and where their data is acquired and maintained. After completing the app's authentication procedure, a slew of businesses serve their clients by offering applications with AI-enabled tailored features~\cite{pusztai2021novel}. These aid users in protecting their personal information. Customers often utilise speakers, phones, tablets, and robots to access AI-enabled gadgets~\cite{sun2019ai}. Multiple levels of encryption and a dynamic encryption process are required due to the sensitive and personal nature of the data. Edge nodes facilitate the construction of a highly distributed architecture and help establish the appropriate security strategy for each device~\cite{lv2021intelligent}. There are worries about latency when data is sent across networks and devices since services are dispersed at both the network and device levels. Due to this delay, the work must be done on the fly. Having several endpoints of load balancing is a need when an application has to be end-to-end resilient and have a widely spread architecture. Resiliency at the device level is increased by the fact that data computing services are closer to the mobile device or on the edge (referred to as a ``cloudlet'')~\cite{hao2018edge}. We have to think, how these challenges can be overcame using AI?

Edge computing is a major enabler for AI, giving high-quality performance at a low cost. This is the best way to understand the link between AI and edge computing. We can benefit from the marriage of AI with cutting-edge computers~\cite{calo2017edge}. Edge Computing helps AI-enabled applications overcome the technical problems of AI-enabled applications because of the data- and compute-intensive nature of AI. AI and machine learning systems absorb vast volumes of data to spot patterns and deliver reliable suggestions~\cite{shakarami2020survey}. Cloud-based streaming of high-definition video data results in latency issues and increased costs, since huge bandwidth is utilised, in AI use cases requiring video analysis~\cite{naveen2018search}. When ML triggers, decisions and actions must be made in real time, the latency and dependence on central processing in the cloud are detrimental. Processing and decision making may be done at the source of data, which means that actions can be taken at the edge and backhaul expenses can be avoided, making the edge an ideal location for data processing~\cite{deng2020edge}. Rather of storing sensitive data on the cloud, the edge stores client location data. Streaming data to the cloud only includes the most important information and data sets, leaving the rest of the data behind~\cite{yang2019mobile}.

Due to their scattered and complicated nature, edge computing networks have brought several issues when it comes to infrastructure management. There are a number of activities that must be completed in order to effectively manage resources. These include workload estimation and task scheduling as well as VM consolidation, resource optimization, and energy conservation~\cite{zhu2020toward}. In dynamic, fast changing settings and in real-time scenarios, traditional pre-defined rules, largely based on operation research approaches, have been used for resource management in the past. AI-based technologies are increasingly being employed to address these concerns, particularly when decisions must be made. Approaches including AI, ML, and DL have become widespread in recent years. On the other hand, deciding where to carry out a work on the edge is a difficult choice that takes into account aspects such as the amount of traffic on edge servers and the mobility of users~\cite{zhou2019edge}. In order to further on the element of user mobility, the cache must be able to forecast where the user will go. For the sake of reducing expenses and energy consumption, it is located at an appropriate edge server. Reinforcement learning, neural network models, and genetic algorithms are some of the approaches that are employed~\cite{huh2019understanding}. 

In the commercial and industrial sectors, the advantages of Edge have quickly spread. Specifically, the reduction of IT equipment's growing expenditures on cloud and network bandwidth. All of the company's activities take place in different parts of the world. Only an estimated 1\% of the monitoring data is relevant for business insights like anomaly identification or future event prediction, despite the fact that the cloud and big data centres are overflowing with data~\cite{ranjan2014streaming}. Edge delivers high-quality business services through local processing, analytics, and local devices. This is the operational efficiency and significance of the edge, since it prevents the transfer of terabytes of unnecessary data to the cloud/data centres and only communicates pertinent actionable data to the end user~\cite{chen2019artificial}. Every day, new and creative applications for edge's capabilities emerge. Edge computing still has a problem in moving to the last-mile of dispersed networks, but new use cases in industrial applications show a strong convergence with AI in particular, offering substantial value to businesses~\cite{xu2020artificial}. In the automotive, construction, process, and manufacturing industries, augmented reality, virtual reality, and mixed reality are becoming increasingly popular~\cite{chen2019artificial}. This necessitates a scalable, highly adaptable, and quick-to-respond computing infrastructure that is always available. Provides a low-latency experience and application instances that are near to the end user AI and machine learning have a plethora of uses in Edge Computing. NLP and CNN are two developing technologies that are used in a wide range of everyday applications~\cite{rihan2020deep}. Smart retail, contact centres, security, and legal assistantship all benefit from NLP's ability to parse human voice, recognise handwriting and classify text. Use cases such as quality control, facial identification, healthcare, and industrial safety can benefit from CNN capabilities in visualisation algorithms, which enable to detect faces and other visual data~\cite{rihan2020deep}.

\subsubsection{Open Challenges}
As compared to cloud, fog, or serverless computing, the problems of edge computing environments are markedly different. As a result, the edge environment is plagued by issues related to scalability and performance, especially when dealing with mission-critical data and applications~\cite{shakarami2020survey}. It is tough to keep track of the health and state of each IT component, especially when there are so many remote edge locations to keep track of, much alone visualise and analyse their influence on other linked equipment, when considering the scale issue \cite{tuli2020healthfog}. Highly dispersed and diverse networks define edge settings. Because of the disparate nature of the infrastructure's components and the high costs associated with acquiring the various skills and resources required, this creates ``edge silos'', which only serves to complicate matters further~\cite{naveen2018search}. To handle the extremely dispersed and heterogeneous edge environment, AI-based intelligent software is crucial. It helps to collect and unify data from many sources and provides a highly abstracted ``low touch'' monitoring and administration, which eliminates human involvement. It is also possible to have entire client security without the client's involvement thanks to automated security and responses. You may pick from a variety of suppliers, avoiding vendor ``lock-in'', and switch out equipment with no negative impact on your business or cutting-edge efficiency~\cite{deng2020edge}.

Additionally, real-time performance management between end points, such as consumers and cloud/data centres, is a critical issue~\cite{yang2019mobile}. End-to-end views and data repositories at all sites are supported by technology tools that continually monitor performance metrics and data flow. For situations when edge equipment fails or is unavailable, edge infrastructure has built-in redundancy to isolate, repair and sustain acceptable levels of functioning~\cite{zhu2020toward}. If you run an edge data centre where multiple teams are in charge of different portions of the infrastructure, you'll run into certain inefficiencies. Advanced correlation and analytics, based on AI, are quite helpful in this situation for examining, consolidating, and unifying data from many sources, transforming data into information, and communicating that information with the concerned roles in the team. Information that can be used to help automated processes is supplied~\cite{zhou2019edge}.

The followings are the key open issues of adopting AI in edge computing: 
\begin{itemize}

\item With tremendous advantages, edge computing has a number of challenges to overcome. Edge computing adoption is being stifled by some of the causes listed above. Edge computing has no legal, societal, or ethical framework for the use of AI. We need to do more study on the present benchmark tools and techniques. Newer technologies have a hard time being integrated into existing legacy industrial systems since they aren't modular. Information security issues and a lack of integration testing with new entrants further limit technology use~\cite{huh2019understanding}.

\item Small and inexpensive, most edge devices do not require third-party API authentication, leaving them open to exploitation~\cite{yu2017survey}. They are designed with simplicity of use and low cost in mind, not security in mind. Edge devices that gather personal information, such as email addresses, phone numbers, health information, and credit card data, are on the rise as a result of specialised apps \cite{dhillon2020iotpulse}. The necessity for an AI based security framework before beginning large-scale and sensitive edge initiatives naturally causes IT and network administrators to be apprehensive~\cite{ranjan2014streaming}.

\item For AI-integrated edge workflows, new software frameworks and toolkits are needed~\cite{shakarami2020survey}. Working with heterogeneous hardware and platforms, as well as the resources available in a workflow, will be supported by these software frameworks~\cite{rihan2020deep}.

\item When it comes to the administration of edge devices, there are no set standards and regulations that apply universally. The complexity of the IoT network architecture increases as more IoT devices are added to the edge~\cite{chen2019artificial}. There is still much to learn about the local consequences of IoT standards for different companies across different geographies, despite the fact that the US and UK governments have produced them. Organizations require a framework of regulations and criteria before they can make the decision to shift their data and application assets to AI-integrated edge computing environments, according to their perspective~\cite{zhou2019edge}.

\item Only by fully integrating edge computing into existing cloud architecture can its full potential be realised using AI, making edge computing the crucial missing connection between data sources (the devices) and cloud computing (core network)~\cite{sun2019ai} because edge nodes have a limited storage and computing capacity~\cite{calo2017edge}.

\item The main strength of this system is its tight interaction with the cloud. As more businesses move to a multi-cloud environment, it becomes more difficult for the cloud to set up a redundant AI-integrated edge network to accommodate incoming data traffic from numerous nodes~\cite{lv2021intelligent}. High bandwidth needs and redundant data reporting and routing requirements for IT managers have increased the search for cloud suppliers capable of meeting these demands~\cite{shakarami2020survey}.

\item Additionally, establishing an AI-integrated edge computing environment requires an initial investment in edge-enabled software frameworks and hardware~\cite{hao2018edge}. The fact that this investment typically has to compete with other company objectives makes it a bottleneck. In developing countries, this problem is exacerbated.

\item Many small and medium-sized businesses (SMEs), IT managers, and government decision-makers are unaware of the possibilities and applications of AI-integrated edge computing~\cite{naveen2018search}. It may take some time for nations in Asia-Pacific that are still learning about cloud computing to adopt its capabilities for edge computing. Micro data centres, rather than edge workloads, are preferred by most service providers in emerging economies, as they are more cost-effective~\cite{deng2020edge}.

\item With the arrival of 5G, billions of devices will be able to communicate with each other, and the network will see a rise in the rate at which connected devices are added and removed~\cite{edgeai}. IT managers' judgement in implementing edge computing will always be questioned and adoption will never achieve its full potential without a standard, proven, and recognised edge monitoring technology~\cite{yang2019mobile}.

\item Faster R\&D interventions and innovations in security, governance and standards/frameworks are the road to alleviate the difficulties for the adoption of AI-integrated edge computing~\cite{zhu2020toward}.

\end{itemize}

Satellites in low-Earth orbit (LEO) are being built by private firms including SpaceX and Amazon to give worldwide broadband internet access \cite{bermbach2021future}. It will be vital to determine if and how edge computing principles could be implemented in LEO satellite networks as the group of subscribers to such a access network increases.

\subsection{Serverless Computing}
When it comes to designing cloud-native apps, serverless computing is becoming increasingly popular. Serverless is a cloud computing paradigm that abstracts away the management of operational aspects~\cite{kounev2021toward}. Because developers no longer have to worry about maintaining infrastructure, serverless computing is likely to expand considerably quicker~\cite{castro2019rise}. Because of this, cloud service providers may more easily manage infrastructure and automated provisioning with serverless computing. The time and resources required for infrastructure management are also reduced as a result of~\cite{fox2017status}. It is the purpose of serverless computing to guarantee that the finest serverless technologies are used so that the investment is minimised and the return is highest~\cite{akkus2018sand}. Serverless computing and infrastructure are characterised by the following terms:
\begin{itemize}

\item Functions: Using event-driven models, serverless functions are implemented in serverless computing. Because the code is automatically executed as events occur, they are able to speed up the development process~\cite{lee2018evaluation}. As a result, numerous services can be linked to the present application. Using these features, you may effectively create the pay-per-execution model~\cite{jangda2019formal}. It is billed for the time and resources used on executing code under this paradigm.

\item Kubernetes: Developers have the option of bringing their own containers to Kubernetes through Serverless Kubernetes~\cite{lloyd2018serverless}. In Kubernetes-managed clusters, these containers may be automatically scaled up or scaled down. In order to deal with exceptional traffic situations and fluctuating workloads, this automatic scaling function is activated.

\item Workflows: A low-code or no-code approach is used with serverless workflows~\cite{al2018making}. It is an aim of this method to reduce the planning overheads associated with several activities at once. With these processes, developers may connect various cloud and on-premises services. Serverless computing has the capability to learn new APIs or standards, so interactions do not need to be coded~\cite{mohanty2018evaluation}.

\item Application environments: Both the back-end and front-end of a serverless application environment are hosted on a dedicated server service. Fully-managed services via dedicated servers assume responsibility for the application's scalability, security, and compliance monitoring~\cite{feng2018exploring}. As a result, running an AI-based application on a serverless computing platform is significantly simpler, as serverless computing meets the dynamic scalability and security requirements of applications while still conforming to industry standards~\cite{perez2018serverless}.

\item API Gateway: An API gateway that is both centrally controlled and entirely managed is achievable with a serverless API gateway~\cite{hassan2021survey}. This application makes it feasible to administer, secure, and analyse APIs on a global scale. The management of authorisation and other services (such as content and user services) is therefore made simpler for a serverless API gateway. Serverless computing infrastructure enables automated API support using AI and database connectivity for every service that requires it, as stated previously~\cite{baldini2017serverless}.

\end{itemize}

Platforms across the board are embracing AI because it is the future of technology. We've been able to make better, faster judgments because of these AI-powered platforms. They've changed the way businesses do business, the way customers interact with them, and the way we gather and analyse business data. Complicated machine learning systems have a significant impact on the productivity and efficiency of developers~\cite{castro2019rise}. A serverless design, on the other hand, addresses most of the issues that developers experience. Using a serverless architecture, the machine learning models are properly handled and the resources are effectively managed~\cite{golec2021ifaasbus}. As a result of this design, developers may spend more time and resources working on AI model training rather than server infrastructure management.

Complex challenges typically need the development of machine learning systems. They analyse and preprocess data, train models, and fine-tune AI models, among other things. As a result, APIs should be able to run smoothly~\cite{eivy2017wary}. Serverless computing and AI should be used to ensure that data storage and message delivery are uninterrupted. Machine learning models may benefit greatly from serverless architecture, which offers a wide range of options and advantages~\cite{wang2019distributed}. Virtually little administration is required to run any form of application or back-end service. As incoming requests of any traffic volume come in, the infrastructure provider allocates its own computing execution power accurately.

AI/ML integrated serverless architecture will have following merits:

\begin{itemize}

\item Fair pricing: Serverless design makes execution-based pricing possible, so you only pay for services that are really being used~\cite{gupta2020utility}. As a result, the pricing model is more flexible and the cost is significantly reduced.

\item Independent work: Serverless computing enables the development teams to operate autonomously with little intervention and delays. Models are viewed as distinct functions because of this. Invoking this function has no impact on the rest of the system and can be done at any time~\cite{hassan2021survey}.

\item Autoscaling: This feature frees up the developer to work on other projects while the system adjusts itself to the changing scope~\cite{schuler2021ai}. Storage prediction is no longer necessary when using autoscaling because developers may make changes on the fly.

\item Pay-as-per-usage: Using a new model called ``pay-per-use'', customers only pay for resources when they really use them. You don't pay for the amount of servers with serverless computing, but rather for the use of services~\cite{tu2018pay}. Combined with the scale-to-zero feature of serverless, one just has to pay for the number of executions and the length of time resources are utilised for.

\item Hassle-free server management: Serverless computing provides backend services that may be accessed only when they are needed, freeing users from the burden of managing servers~\cite{bilal2021great}. A serverless service eliminates the need for the user to be concerned with the infrastructure that underpins the service. With serverless backends, service providers don't have to modify their setups if they want to raise or reduce the amount of bandwidth they are reserving or paying for~\cite{schuler2021ai}. It was difficult and expensive for web developers to own the hardware necessary to run a server before the introduction of the Internet.

\item High availability: Serverless programmes have become more popular due to their built-in availability and fault tolerance. No need to construct services that will deliver these features to your application, therefore you don't have to. Your company doesn't need to invest in new capabilities because they are constantly available~\cite{douceur2001optimizing}. 

\end{itemize}

By reinventing automation and enhancing the corporate environment, AI has taken over today's life and made it easier. Machine learning algorithms on serverless architecture may be used in a variety of ways to make jobs easier and data more accurate~\cite{schuler2021ai}: 
\begin{itemize}
\item Applications that employ GPS gather user data, such as their location and their purchasing habits, to provide suggestions about their preferences or the next thing they should buy. AI assesses the frequency of alerts and suggests a number of options that the app users may be able to bear and enjoy before turning off the notifications. Using this method ensures that clients find the material useful and enriches the user experience~\cite{tu2018pay}. 

\item Using AI models, it is possible to examine a customer's financial viability before recommending an increase in purchasing power. Prior to requesting any further information, the system will run a credit check to determine their creditworthiness. As soon as all of the prior invoices have been paid, the system decides if the transaction should go through or be put on hold. 

\item As part of logistics, it is important to keep an eye on the routes and determine how traffic overloads influence customers. In order to help businesses make better decisions and enhance customer service, AI analyses the routes and proposes alternate routes~\cite{eivy2017wary}.
\end{itemize}

\subsubsection{Open Challenges}

The following is a quick description of the open issues and challenges that serverless computing presents for AI applications:

\begin{itemize}
\item Vendor locking: If a company has committed to a cloud-based provider that provides technological implementation, switching suppliers will be difficult. The lack of industry-wide standards is responsible for around half of the challenges to cloud computing adoption~\cite{nastic2017serverless}. To understand the consequences of vendor lock-in, no amount of investigation or study can be relied upon. Those who fail to keep an eye out for traps end up falling prey to them. In addition, a serverless interface must be carefully monitored because of the multiple risks it presents~\cite{glikson2017deviceless}.

\item Switching vendors: Lock-in circumstances can be seen in two different ways, depending on who is looking at the event. In their opinion, the problem with serverless computing today is not the programme itself. Another contributing cause is serverless computing, a new and fragmented technology that now has a younger audience that is overly reliant on it~\cite{baresi2019towards}. Businesses should only agree to a platform after conducting comprehensive research and receiving multiple offers from rival service providers. A single cloud may be all you need instead of hurrying to implement a slew of different ones.

\item Less transparency: As the backend infrastructure is handled by an outside company, there is less transparency about how things actually function. The program's inner workings may be obscured, especially if it interacts with other applications. Here are a few examples of how this may be put to good use: The security measures associated with an external PaaS service connected to your application, for example, are generally not well known~\cite{baresi2017empowering}.
 
\item Others are in-charge of infrastructure management: Because our infrastructure is in the hands of a third party, it is more difficult to have a comprehensive understanding of the entire system. Using end-user-targeted devices under this paradigm, infrastructure is under the control of another firm, malware can nonetheless infiltrate your environment~\cite{cicconetti2020decentralized}.

\item Sustainability: An increasing amount of data generated on the edge is being submitted to the cloud for ML training/inference; the overall transmission energy is around 30\% of the total energy requirements of datacenters globally and rising fast. Research is required to ensure the upcoming serverless paradigm is sustainable with a focus on power-off techniques, increased computational density via smart workload consolidation, submitting the kilobytes-long function to the data vs., submitting terrabytes of data to the function, and effectively combining serverless edge resources for multitenant clients---e.g., by sharing artificial neural network layers with an acceptable tradeoff in accuracy~\cite{patros2021toward}.
\end{itemize}

AI has revolutionised market research and customer behaviour. Customers' preferences are recorded and analysed by an AI model, which then shows their customised content. Serverless computing simplifies the AI development process by removing the need for a dedicated server. As a result, serverless architecture entails handing over the management and monitoring of the infrastructure to a third party. That is why it is a good idea to work with a reputable cloud service provider. To guarantee that the infrastructure works smoothly, the cloud provider must have handled multiple comparable projects and have expertise in hosting and handling AI/machine learning and serverless architectures.

\subsection{Quantum Computing} 
Quantum computing promise is to be the one technology that has the potential to fundamentally alter AI. This section introduces the capabilities of quantum computing and its potential influences on AI and the economy~\cite{gill2020quantum}. The consequences of this approach to computation could affect a considerable range of aspect of intellectual and economic activities on our societies. With the enormous influence is AI having, all across the world, the combination with Quantum computing may have a multiplier effect to trigger a revolutionary effect on AI.

Quantum computing utilises a novel approach to data and information processing: Information, encoded in the quantum states of quantum systems, is processed accordingly to the law of quantum mechanics opening up some opportunities that are not available to the classical way of processing information. For example, quantum superposition and quantum entanglement \cite{chuang}. Quantum entangled is the property of quantum systems of limiting the amount of information an observer may obtain on parts of a global quantum state, making it impossible to provide a complete description from the knowledge of only the component states. The term ``superposition'' refers to the possibility of \emph{combining quantum state} in order to produce another valid quantum state.

Previous features of quantum systems trigger, from one side, the power of quantum computing (if sufficiently shielded from interactions with environment), but represent also the main limitations that do not allow an efficient simulation of quantum systems by present computer systems, even AI-powered supercomputers. In fact, the scaling of the phase space within which composite quantum systems evolve growths exponentially with the number of component systems.

The unit of information used by Quantum computers is the qubit, which replaces the bit used in classical computers. The state of a qubit $|\psi\rangle$, which could be an atom, a photon, a circuit, etc., can be represented, mathematically, as a vector in the complex Hilbert space \cite{chuang}, with two mutually orthogonal basis states $\{|0\rangle$, $|1\rangle\}$ as follows
\begin{equation}\label{eq:qubit}
    |\psi\rangle = a|0\rangle+b|1\rangle,
\end{equation}
where $|a|^{2}+|b|^2=1$, and $a, b \in\mathbb{C}$ are Complex numbers. 
The exploitation of quantum superposition (See Eq.~(\ref{eq:qubit})), and quantum entanglement is what makes quantum computing considerably more powerful for certain tasks than classical counterpart ~\cite{chuang}.

The simulation of quantum systems has been the original scope motivating the endeavour to build a quantum computer ~\cite{feynman}, but it has only been after the discovery of quantum algorithms able to achieve practical goals that the interest in building this devices started to attract increasing attention.
After the seminal works to formalize the concept of quantum computer \cite{Deutsch}, several algorithms followed that allowed to achieve tasks that were considered hard for classical computers. The discovery of the Shor's algorithm~\cite{shor} provided an efficient solution for factoring large numbers, that had critical implications for crypto-analysis, boosting studies in both quantum computation and quantum cryptography. Running effectively the Shor algorithm on a working quantum hardware, however, it would require a level of accuracy in implementing register initialization, quantum operations on multiple qubits, and storage of quantum states that are not yet achieved by current state-of-the-art devices ~\cite{googleQ19}. It is also worth to mention that quantum computers have their own limitations. For example, it is not expected that they can efficiently solve NP-hard optimization problems ~\cite{BennetNP-hard} and, coming to searching, the speed-up offered by quantum computers scales quadratically with respect to the time needed by a classical computer (Grover’s algorithm~\cite{grover}).

Building a quantum computer is, in fact, not an easy task: as experimentalists know pretty well, the advantages of quantum computing, offered by features like quantum superposition and entanglement, tend to vanish exponentially faster with the size and complexity (i.e., the number of quantum systems involved) of the hardware.
Nevertheless, in recent years, we have seen a spectacular increase in the interest of major high-tech players (IBM, Microsoft, Google, Amazon, Intel, Honeywell), and a flourishing of many young companies aiming at proposing solutions for quantum computing, with various core technologies employed, ranging form superconducting devises~\cite{googleQ19}, to trapped ions~\cite{ionQ}, to integrated light circuits~\cite{psiquantum}. These are just some of the many companies that are today financing quantum initiatives and are interested in developing this technology.
Despite the difficult challenges ahead, Google AI group has made significant progress during the recent years~\cite{googleQ19}, achieving what is known as quantum advantage building a programmable quantum computer, named Sycamore. Similarly, IBM has recently announced the first quantum computer to pack more than 100 qubits in their hardware, \emph{Eagle} chip~\cite{EagleIBM}, representing only a first step of busy research and engineering program during which the tech-giant is planning to push figures to more than 1000 qubits by 2023.
However, as said, the challenges to preserve the delicate features of composite quantum states rely on the ability to shield these devices from the external environment to allow coherent quantum evolution to take place under the presence of even very small amount of noise. For this reason these devices need ultra-low temperature of fractions of Kelvin, which pose challenges also for designing the appropriate materials able to perform well at such low temperatures.

\subsubsection{Open Challenges}
While universal quantum computers remain the long term challenge of quantum computing, Noisy intermediate-Scale Quantum (NISQ) devices are a foreseeable target to achieve in the forthcoming years.
With such devices physicist may start to effectively simulate complex composite quantum systems, and should be able to study exotic quantum states that have not been accessible in physics laboratory yet.

For the next step, once NISQ devices will be reliable and well developed, we will need to overcome the limitations imposed by the presence of noise during computation, by supporting the main computing unit with effective quantum error correction (QEC) circuitry. That may open road towards fault-tolerant quantum computation that will need to involve thousands and more qubits. In fact, QEC requires a considerable cost in terms of the number of qubits and logic gates to be implemented.
The road towards the implementation of complex operations like those needed by the Shor factoring algorithm is still long but while the research is focused on improving the performance of quantum devices and the optimization of quantum operations, numerous entrepreneurs are also interested in producing quantum software solutions. Consequently, many investors are expected to invest in start-up companies that are revolving around quantum computing technologies, and in perspective the interest in quantum computing is likely to increase.

Pharmaceutical investors' interest in quantum computing has sparked. Many sectors can benefit from quantum computers and commercial solutions. The financial industry, healthcare, genetics, pharmacology, transportation, sustainability, and cybersecurity are all direct beneficiaries of quantum computing~\cite{padhi2021quantum}. Quantum computing's potential has been picked up by the banking industry. Financial analysts frequently make use of quantum computational models that incorporate probabilities and assumptions about how markets and portfolios operate. To do this, quantum computers might help by processing data faster, running better foresight models, and balancing conflicting options more accurately. They might also assist in the resolution of complex optimization problems, such as portfolio risk optimization and fraud detection~\cite{foulkes2001quantum}.

Quantum algorithms in IBM's Cloud Computing platforms outperform classical Monte Carlo simulations, according to a research the company just presented. There is a lot of potential for the healthcare business to profit from quantum solutions. Quantum computing might lead to improved approaches to personalised medicine by allowing quicker genomic analysis to create personalised treatment strategies specific to each patient~\cite{caputo2015pan}. Genealogy research generates a lot of data. As a result, analysing DNA information requires a significant amount of processing power~\cite{di2016decoding}. Currently, companies are reducing the costs and resources needed to sequence the human genome; but, a powerful quantum computer might sift through this data considerably faster, making genome sequencing more efficient and scalable~\cite{spector1999finding}.

Another area where drug development might benefit from quantum computing is in the field of protein folding~\cite{protein-folding}. This might help speed up drug discovery efforts by making it easier to predict the effects of pharmacological compounds~\cite{richards2013quantum}. 

A crucial aspect where quantum computers, and the promise to build one, have had considerable impact is in the field of security and cryptography. Public-key cryptosystems are the foundation of today's era of communication. Rivest-Shamir-Adleman (RSA) encryption is in fact the most common cryptosystems for securing transmission of data over networks, and its working mechanism and security requires factoring large prime numbers, beyond the capabilities of current classical computing limits. However, as mentioned above, quantum computing capabilities, exploiting Shor's factoring algorithm may renders such encryption models obsolete. This has led to increasing research, and investment, over the last decades to build safe cryptosystems in a quantum computing era, and the projections for the next years show this interest to grow, e.g., Toshiba's quantum cryptography revenue target is \$3 billion by 2030~\cite{Toshiba}. In the meantime, while the efforts to design and implement effective quantum-key distribution (QKD) protocols expand, the National Institute of Standards and Technology (NIST) has also issued recommendations for post-quantum cryptography standards~\cite{NIST}. It has begun a process to request, assess, and standardise one or more public-key cryptography algorithms that are resistant to eavesdropping performed by quantum hardware.

Quantum computers have also been proposed for environmental applications, in the hope that quantum computing may open up new avenues for dealing with climate crises, identifying and optimising process that may help cope with global warming and other climate change effects~\cite{singh2021quantum}. 

\subsubsection{Quantum Artificial Intelligence (QAI)}

\textcolor{black}{Quantum computing is more effective than classical computing in managing large amounts of data. 
A quantum algorithm is a mathematical algorithm that executes on a realistic model of quantum computation; the quantum circuitry of computation is the most often used model. The state of a quantum computing system can be seen as the information encoded in the physical quantum state supporting the specific implementation. Quantum information theory is based on these fundamental object, quantum bits (the unit of information), quantum gates (the devices that operates on quantum bits), and quantum channels that connect gates and circuitry to preserve quantum superposition and entanglement. Quantum computers can handle and process exponentially larger amount of data than conventional computers can, because they intrinsically incorporate and manage the tensor product structure of composite quantum systems, which the number of parameters needed to obtain a complete description scale as $2^{N}$, where $N$ is the number of qubits. So, for instance, if $N=100$ the vector space in which the quantum system evolves would have a dimension that is $2^{100}$. That means that while we may need a $100$ qubits on a quantum computer (assuming that we do not need error correction) to describe the evolution of composite system, the same described on a classical machine, would require $2^{100}$ parameters. It is then clear that quantum computers are intrinsically prepared in managing the evolution of systems described by a large number of parameters.}

\textcolor{black}{Quantum devices, coherently controlling quantum superposition and entanglement, can exploit \emph{quantum parallelism}, i.e., they can simultaneously \emph{explore} multiple evolution. For example, in classical decision the problem is often represented as a \emph{decision tree}, where the evolution of the decision path is determined by a binary choice and the state of initialization of the register. This method becomes less effective when the creation of branches form each decision split-point slows down because the time needed to obtain an outcome to decide the split is too long. Quantum devices, exploiting quantum superposition to initialize the input register, and quantum evolution (that coherently preserves superposition) may explore, simultaneously, the various possible branches of a decision, speeding up considerably the application of these kind of decision-tree approaches. It is still difficult to provide a certain foreseeable future for the development and to know precisely how and when we will have a complete and deep application of quantum advantage, however it is reasonable to say that the exponential speedup of quantum computing will involve all sort of problems with large amounts of data to manage, like pattern recognition, or training in machine learning models.} 

\textcolor{black}{Training of machine learning models, like in reinforcement learning base much of their effectiveness on the speed at which the \emph{agents} learn, interacting with the environment: They interact, obtain some feedback, and adapt (learning) their behaviour on the base of the feedback received. This approach has been confirmed, e.g., in a recent experiment \cite{saggioetal2021} where not only the training of agents has been improved and accelerated by using a quantum channel, but it has been also integrated to implement a hybrid quantum/classical scenario, using a very promising integrated nano-photonic processor,  where classical communication are used for tuning and to achieve the optimal control of the learning progress.}

\begin{figure}[t]
    \centering
    \includegraphics[width=0.75\linewidth]{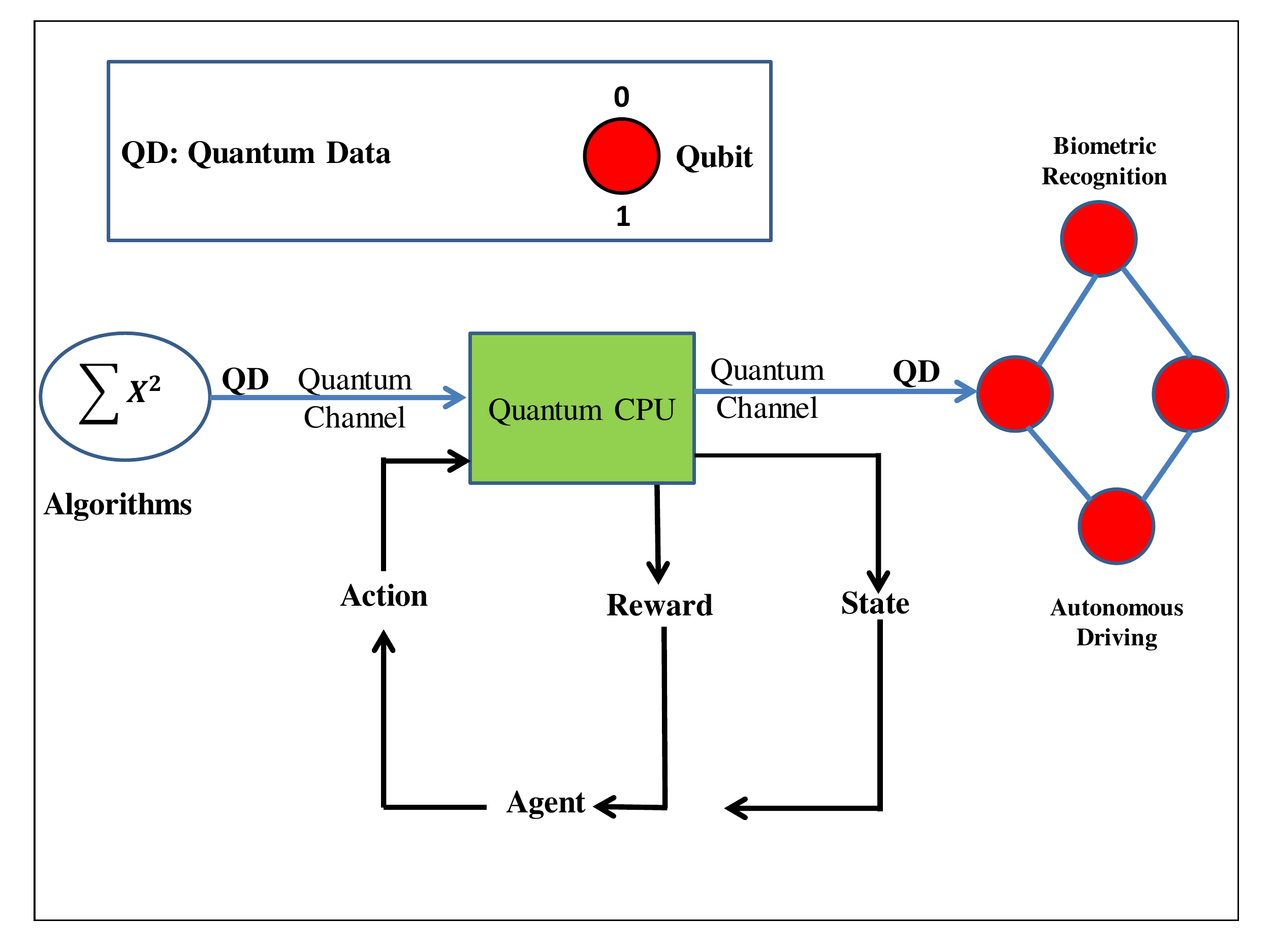}
    \caption{Overview of Quantum Artificial Intelligence}
    \label{fig:model5}
\end{figure}

\textcolor{black}{Clearly, if the computational speedup is exponential, as it is for quantum computing vs classical, when dealing with large amounts of data, such a speedup may have application on many different aspects of the development of AI.  Figure~\ref{fig:model5} provides a clear view of quantum computing with AI for modern applications.}

\textcolor{black}{Biometric recognition and autonomous driving are two critical examples that can utilize QAI for processing workloads. The fact that quantum computers can process more data in a shorter time than traditional computers has revealed the concept of QAI \cite{sgarbas2007road}. An example scheme for QAI is given in Figure~\ref{fig:model5}. QAI involves combination of Quantum Computing and AI to achieve superior performance results compared to classical AI \cite{biamonte2017quantum}. Reinforcement Learning (RL) is a well-established branch of ML that aims to maximize the reward by trial and error by means of an the agent \cite{schuler2021ai}. It is certain that combining RL with Quantum Computing will lead to great advances in computing systems. With quantum computers accelerating machine learning, the potential for impact is certain to be enormous\cite{huang_broughton_mohseni_babbush_boixo_neven_mcclean_2021}. Applications of quantum AI for quantum search, quantum game theory, quantum Algorithms for decision problems, algorithms for learning are shown in the Figure ~\ref{fig:model5}.}

\textcolor{black}{The ability of quantum computing to execute a task quickly, may be helpful for AI systems employed, e.g., in problems related to autonomous driving natural, natural language processing (NLP) algorithms~\cite{ayanzadeh2018quantum} and, in general, in tasks where classical approaches are extremely time-consuming and expensive. Characters and words are the basis for the current algorithms. The idea of becoming ``meaning aware'' is a goal of quantum algorithms~\cite{meichanetzidis2020quantum}. To build real-time speech patterns, these algorithms may use phrases and paragraphs. It is important to note that predictive analytics are a key AI application and commercial use case. Massive amounts of data can be used to train AI systems that are adept at machine learning and deep learning. However, complex and ambiguous issues such as stock market projections and climate change control systems require unique data created by quantum principles employing entanglement and superpositions~\cite{singh2021quantum}. New discoveries in Artificial Intelligence algorithms intended for quantum computers,, or Quantum Artificial Intelligence (QAI), are expected to deliver the critical breakthroughs required to advance the science of climate change. Improvements in weather and climate forecasting as a consequence of this research are predicted to have a cascading effect on a wide range of socioeconomic advantages. For example, NASA has established the Quantum Artificial Intelligence Laboratory (QuAIL), which is dedicated to investigating the possibilities of using quantum computers and algorithms to machine learning problems in NASA's missions.}

Nanotechnology and nanoscience may also be integrated into AI for very small, microscopic devices at molecular, atomic, and subatomic levels, thanks to quantum computing. Quantum physics finds use in nanotechnology. These are only a few examples of quantum computing's impact on AI and machine learning~\cite{sacha2013artificial}. Machine learning applications for quantum devices are already being developed, in the hope to employ quantum computing to speed up the training of machine learning models and produce more efficient algorithms for learning~\cite{biamonte2017quantum}. Machine learning and artificial intelligence are likely to benefit from improvements in quantum computing technology even before a comprehensive quantum computing solution is ready. \textcolor{black}{Hence, the field of Quantum Machine Learning (QML) is expected to pick up, followed by its autonomic expansion, Adaptive Quantum Machine Learning, which will be able to leverage quantum computing to adaptively achieve self-learning.}

Open-source quantum machine learning library TensorFlow Quantum (TFQ) is available from Google~\cite{broughton2020tensorflow}. Cirq is integrated with TensorFlow and provides high-level abstractions for the design and implementation of both discriminative and generative quantum-classical models by providing quantum computing primitives that are compatible with existing TensorFlow APIs, along with high-performance quantum circuit simulators.
Quantum computing has, indeed, the potential to transform AI in a number of ways. Constraint resolution, uncertainty handling, and constraint fulfilment will all be improved by quantum computing, as will adaptive machine learning and spatial and temporal reasoning~\cite{dunjko2018machine}.
Even if quantum computing is still in its infancy now, from a commercial and economic standpoint, it is an excellent moment for startups to join this path. The future of our economy will not be decided by cryptocurrencies, but rather by quantum computing solutions~\cite{holmes2021assessment}.

\section{Modern Autonomic Computing with Embedded Intelligence}
This section discusses the new research developments related to autonomic computing with embedded intelligence. New advances in intelligent edge, intelligent things and sensors as actuators are detailed here. 

\subsection{Intelligent Edge}
IoT is bringing billions of new gadgets online, creating an unprecedented amount of data that will be challenging to manage. Over 75 billion IoT devices are expected to be in use by the year 2025, according to research \cite{toldinas2019mqtt}. Businesses are developing and installing more and more things meant to improve the end-user experience and also generate massive amounts of data, such as linked automobiles, smart metres, and in-store sensors. This new data, meanwhile, requires real-time collection, management, and processing \cite{dick2019embedded}. Exactly how will this happen? One approach to go forward may be through the use of edge and fog computing \cite{guo2013internet}.

Edge computing is expected to get a huge amount of attention than fog computing in the next years, but what exactly is it? As opposed to typical cloud computing, in which data is stored and handled in a central location, edge computing processes data on-the-fly \cite{dai2019industrial}. The cloud and edge are not mutually exclusive in fog computing, which means that some computation can be done in the cloud, while other parts are handled by edge devices.

As a result, the data sent between linked devices might take too long, and edge computing utilises significantly less network capacity than traditional computing. Time can be saved by processing it locally on the device or inside a local network \cite{singh2021fog}. Edge computing, on the other hand, might provide cloud computing with much-needed assistance in dealing with the massive amounts of data generated by the IoT and other connected devices \cite{singh2021quantum}.

In both fog and edge, emerging IoT devices generate and transfer data, and the processing capacity of those devices is used to accomplish operations which might normally be carried out in the cloud. Both fog and edge allude to these new IoT device locations in the network. As a result, they help organisations lessen their dependence on the cloud by sending data to analytics platforms, where it can be analysed and turned into useful information. Corporations may cut down on network latency by using edge and fog technologies to reduce their reliance on cloud platforms for data analysis. As a result, you will be able to make data-driven decisions faster. In addition, because edge devices lack storage capacity, they must send data to the cloud when real-time processing is finished so that analytics may be conducted on it.

With today's cloud computing, bandwidth, and processing power, business communication network focuses primarily on supporting all of your distant applications, as well as offering infinite storage space. That will eventually change. In order to get the most out of data, it must be processed in real time, at the edge \cite{singh2021fog}. Looking for future, network infrastructures must be more adaptive and willing to manage a far greater number of smart devices than they currently are now. Having the decision-making process near to where the data is created is essential for real-time intelligence. For example, self-driving cars or self-maintaining smart manufacturing equipment can make fast judgments on the go \cite{vermesan2018next}. Real-time engine performance data generated by sensors installed in aeroplanes can be used to take preventative action before the aircraft returns from the sky. The savings might be substantial. A company's ability to deliver processing power and an intelligent environment will increase as it expands its network of corporate endpoints.

We will need quicker and more reliable data processing as our demand for data grows and billions of devices are linked to the network. Despite the benefits of cloud computing, the development of IoT and mobile devices has necessitated ever higher bandwidth requirements \cite{plastiras2018edge}. Cloud computing isn't required by every smart device, and avoiding transferring data back and forth over the cloud is a good idea in some circumstances. The edge may help enterprises become more nimble, decrease expenses, minimise latency, and better regulate network capacity \cite{gobieski2019intelligence}.

How to provide enough computing power for intelligent applications at the edge has become a serious challenge. Intelligent edge is a promising way that pushes intelligence to the Edges of Internet, which has played the role of intelligent decision-making in many aspects of edge computing, including task offloading, edge caching, and resource scheduling. Among them, edge offloading is a distributed computing paradigm that provides computing services for edge caching, edge training, and edge inference. By integrating methods such as Distributed Machine Learning (DML), Deep Reinforcement Learning (DRL) and Collaborative Machine Learning (CML) into the edge computing, it is beneficial to cope with the explosive growth of communication and computing of emerging IoT applications \cite{anwar2021recommender}, and achieve the energy-efficient and real-time processing~\cite{Xue2021EosDNN}. 

Instead, single pass AI techniques that can operate on resource-restricted environments have been proposed. For instance, an alternative to traditional machine learning is data-stream mining. This ML paradigm treats its datasets as individual datapoints coming in one at a time, while performing adapting learning with a finite memory budget. From an autonomics perspective, data-stream mining also leverages the notion of adaptive concept drifts, i.e., to save resources on the edge, the data-stream model is retrained only when its performance crosses bellow a threshold~\cite{gomes2017survey}.

\subsection{Intelligent Things}
Intelligence integrated in a technology relates to the capacity of a product to analyse and consider its own performance. In addition, it must be able to deal with the workload and its own working conditions \cite{chen2013internet}. As a result of this, the overall experience of the end user is enhanced. When building a new products/services, it is important to keep in mind the notion of self-evaluation of the product based on data from embedded sensors \cite{arsenio2014internet}. A business intelligence system must be at the heart of an integral model to product or service introduction. Employing an embedded intelligent system and a machine learning algorithm model is a major benefit \cite{nathani2017internet}. This may be assessed by looking at how well it performs in areas like launching smart product systems and setting up smart business services, to name just two. With the help of this cutting-edge infrastructure, it is possible to better understand and anticipate how the business landscape will shift in the future \cite{wazid2020tutorial}. Because of this, it is an area where human analysts typically fall short. As a result, machine intelligence capabilities based on sensors integrated in current gadgets and goods have become a need in today's corporate environment \cite{chen2019intelligent}.

The convergence of emerging applications and machine intelligence capabilities has led to an evolution in this process \cite{chen2013internet, arsenio2014internet, nathani2017internet, wazid2020tutorial, chen2019intelligent}. Embedded intelligence is increasingly being used to design the future IoT:

\begin{itemize}
\item AI: Human-machine intelligence synthesis is what this term denotes. In a specific gadget or service, the ability to make choices like the intellectual is possible.

\item Data Integrity: Device history may be tracked using blockchain, a business intelligence application or innovation. \textcolor{black}{Blockchain is made up of strings of interconnected block headers and blocks bodies \cite{9126779}. The block body consists of all transactions in the block. The block header, on the other hand, is generated using the hash value, timestamp of the previous block, and a Merkle root of the transactions it contains. Therefore, each block is created using the hash value of the previous block and linked to each other. Interfering with any block will change the hash value of that block and thus all blocks will be affected. This promises the availability of Blockchain technology in protecting the integrity of data \cite{8071359}. Blockchain technology, apart from cryptocurrency (e.g., Bitcoin), has been used to further enhance and optimise existing solutions such as cloud storage (e.g., \cite{doyleblockchainbus}), authentication (e.g., \cite{hammi2018bubbles}), health-care (e.g., \cite{abaid2019health}), and more. Integrating blockchain allows to have a clear audit trail of the data and models used to verify the machine decision process, which will lead to increasing the device's trustworthiness \cite{sisi2021blockchain}. The latter is of extreme importance in machine-to-machine communication. Furthermore, running AI code over Decentralized Autonomous Organization (DAO) with smart contracts attached to it limits catastrophic risk scenarios by limiting the action space.}

\item Smart Healthcare: Utilizing IoT in the healthcare business may pay out. As a result, healthcare would be even more widely available, and progress could proceed at a rapid pace.

\item Predictive Maintenance: Predictive maintenance is a notion that has emerged as a result of the growth of the Internet of Things. This means the practise of adding sensors to household appliances so that they may send out notifications when they need to be serviced.
\end{itemize}

From natural language processing, face recognition, bio-medicine to autonomous driving, more and more intelligent applications are being deployed on IoT devices \cite{IoTpi}. Due to the slow hardware development in small-sized equipment, the contradiction between the limited computing capacity of IoT devices and running complicated AI applications cannot be efficiently solved in a short time~\cite{Xue2021DDPQN}. In addition, there remains significant challenges in developing system-level, algorithm-level, architectural-level or infrastructure-level technologies for embedded intelligence, e.g., real-time decision making, energy-efficient Deep Neural Network (DNN) training and DNN inference, and security deployments.

\subsection{Things as Sensor-Actuator Network}
Cyber-Physical Systems (CPS) are the next generation of embedded Information and Communications Technology (ICT) systems that employ sensor-actuator networks to offer users with a wide range of smart applications and services by being aware of their physical surroundings \cite{sha2017empirical}. With the help of autonomous control loops, many IoT sensors are conceivable because of the inclusion of improved processing and analysis of data collected by sensors, as well as planning and executing plans utilizing actuators \cite{liu2003state}. Methods are needed to aid in the design and development of these systems because of their complexity \cite{cceltek2017internet}. In the context of CPS, the systems that are embedded or software integrated into physical things, networked, and offering residents and companies with a wide range of smart applications and services are referred to as ICT systems \cite{van1993sensor}. 

Transport systems, buildings, electricity grids, and water infrastructure are all examples of CPS \cite{deshmukh2018monitoring}. This type of CPS is meant to detect and react to the physical environment, allowing for quick, reliable autonomic control loops combining sensing and actuation, perhaps with linguistic and cognitive capabilities, as well \cite{joshi2020performance}. Using wireless sensor/actor networks, CPS can monitor and respond to the physical environment. Sensors and other alternative sources collect historical and real-time data, which is used to perform advanced analysis and processing in the type of autonomic control loops \cite{shi2018digs}. These loops then plan and execute actions in accordance with a set of goals or rules. Real-time or historical data is used to support this implementation \cite{bragarenco2020sensor}. There are several elements that make CPS systems difficult to manage, including the use of a wide range of sensors and actuators, the necessity for real-time processing of enormous amounts of data, and the implementation of plans for issue solving \cite{muralidhara2020air}. As a result of the system's complexity, engineers need tools and techniques to aid them in the designing process\textcolor{black}{; adaptive digital twins are poised to play a significant role in de-risking such complex engineering and cyber-physical projects.}

\section{Explainable AI (XAI) for Next Generation Computing}
Intelligent decisions are critical to the success of computing initiatives. Is a computing system stable and robust enough to execute workloads? Are the trained models black boxes or causally explainable? These are just a few examples of challenges that are faced before a computing system can be implemented \cite{linardatos2021explainable}. Inaccurate decision-making is expensive in terms of both cost and resource usage when it comes to these complex and advanced technologies \cite{gohel2021explainable}. There have been a number of AI/ML applications in computing systems to enhance decision-making for allocation of resources and energy efficiency. Nevertheless, these AI/ML models' predictions for computing systems are still not practical, explainable, or executable \cite{liao2021introduction}. AI/ML  models are frequently hampered by these constraints. Even if QoS continues to be a primary concern, some recent research have turned their attention to explaining how QoS is achieved \cite{souri2020hybrid}. Is there anything that researchers can do to further the advancement of the computing community? As a result, a thorough understanding of Explainable AI (XAI) and hands-on expertise with XAI tools and approaches \cite{zhang2021explainable} is needed in order to make informed resource management decisions (an example application of AI for Computing). These issues may be addressed by using Explainable AI approaches, such as formulating predictions about resource and energy usage and SLA deviations and then implementing timely, intelligent action to solve them. In order to make computing more feasible, explainable, and executable, XAI prediction models must be properly implemented.

\section{Potential Risks of AI-integrated Computing}
AI can save money, but it requires a highly-trained workforce, that can be expensive at the start. AI's other drawbacks in computing systems would include following ones:
\begin{itemize} 
\item Internet Connectivity Issues: ML/AI techniques based on autonomous computing are frequently hindered by slow internet connections. There seems to be a latency between transferring information to the cloud/fog/edge and receiving replies, even if autonomic computing is faster than traditional computing. ML methods for servers are prone to this issue since forecasting speed is among the most important considerations.
\item Privacy of Data: AI systems need an enormous quantity of data, which include information on customers and providers. Understanding who owns the data is far more beneficial than having private information that can't be attributed to a specific individual. Challenges about data security and compliance with regulations arise frequently when businesses make utilisation confidential material. Autonomic computing with AI necessitates privacy regulations and data security.
\item Possibility of Errors: Although AI may seem exciting, like with any experimentation, it isn't always effective in accomplishing its goals. During its search for solutions, the AI system generated many problematic statements on sensitive topics. With AI, there is a high risk of mistakes because to the many options. Before further usage of this innovation, trust and control must be established.
\item Over-reliance on AI Models: As it stands, AI/ML algorithms are only a small part of complex software-intensive systems. Software engineers leverage them for completing highly specialized tasks, while designing an large apparatus of more traditional algorithms around them, including sensor-data sanitization and filtering. Further, an AI model is only as good as the data it was trained on. Hence, strong data engineering processes are required to select appropriate/representative datasets and conduct various data engineering steps as well as thorough review processes need to be followed when transfer learning models are selected. Hence, from an autonomics perspective, all these steps need to be carefully handled by the MAPE-K loop, which will essentially need to act both as a software and a data engineer when conducting adaptations to the ML models.
\end{itemize}

\section{Emerging Trends and Future Directions}

\begin{figure}[t]
    \centering
    \includegraphics[width=1.1\linewidth]{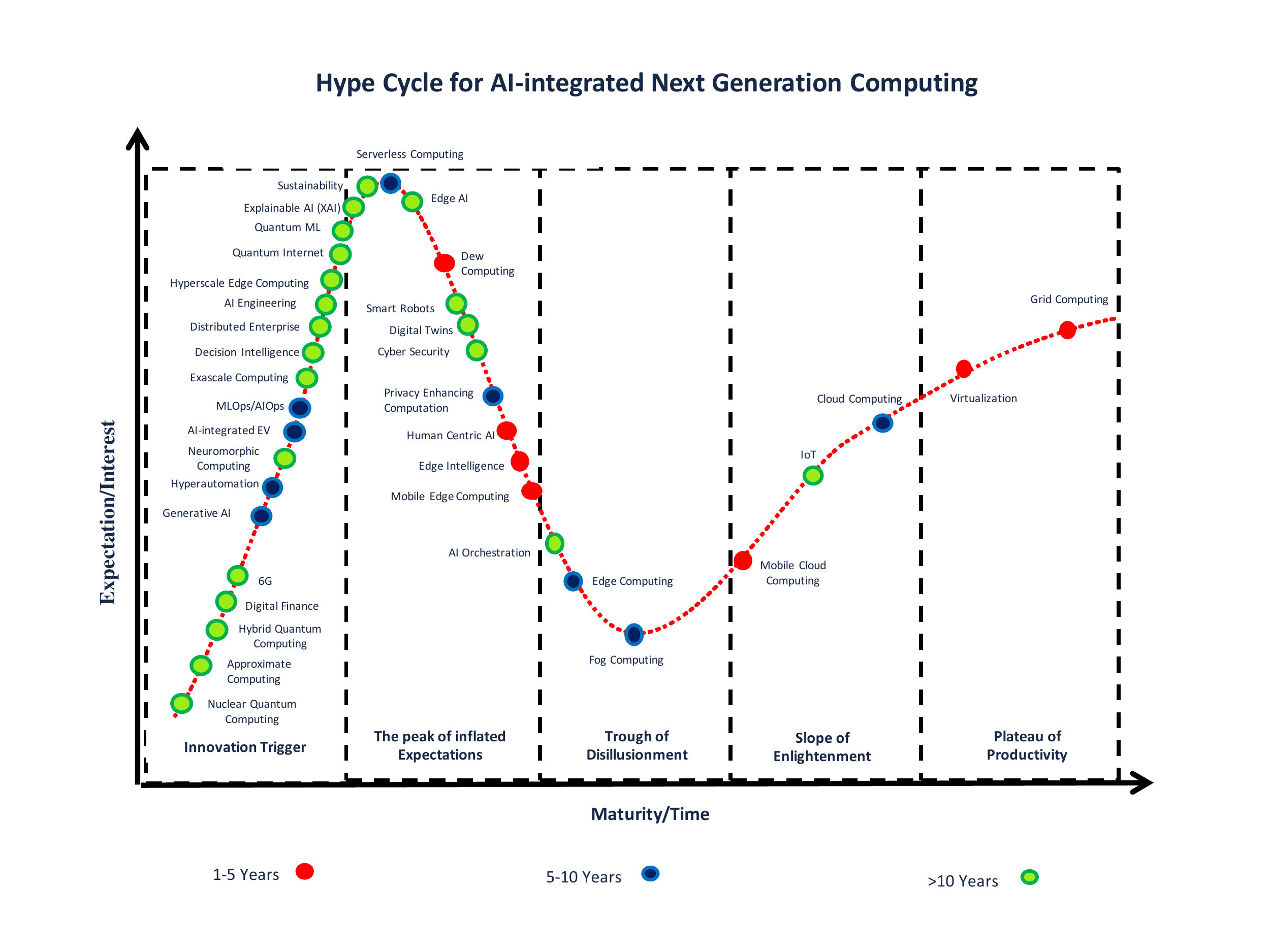}
    \caption{Hype Cycle for AI-integrated Next Generation Computing }
    \label{fig:model6}
\end{figure}

On the basis of current research, we've selected a number of computing study fields for three distinct maturity levels (5 years, 5 to 10 years, and more than ten years). We have identified a number of emerging technologies over the next decade, all of which have the potential to make efficient use of AI/ML-integrated next-generation computing~\cite{bermbach2021future, gartner, gartner1, ZDNET}. Figure~\ref{fig:model6} depicts the hype cycle for next-generation computing. There has been a lot of study done on cloud computing and IoT, but serverless computing is currently at its peak. Under the umbrella of computing, research fields including fog computing, edge computing, AI Orchestration, and mobile edge computing are only getting started. It might take up to ten years for the application of computing in these fields to reach maturity. However, the hype cycle for Explainable AI (XAI), AI engineering, hyperscale edge computing, distributed enterprise, sustainability, quantum Internet and quantum ML is projected to last more than ten years. The hype around smart robots, digital twins, cyber security, edge AI, human-centric AI, edge intelligence, dew computing and privacy enhancing computation is at an all-time high. It is anticipated that they will be fully developed in less than five years under the umbrella of autonomous computing. Expected to significantly evolve in the next five to ten years, MLOps, AIOps, AI-integrated electric vehicles (EV), decision intelligence and exascale computing have likewise hit their height of overblown expectations. Generative AI, hyperautomation, neuromorphic computing, hybrid quantum computing, digital finance, 6G and quantum computing all have a long way to go before they reach the heights of the hype cycle. Grid computing and virtualization have received a lot of attention in the past several years, and they might continue to do so over the next five years.

 In the following, we have highlighted a number of unsolved problems and research paths that require further investigation.

\subsection{Cloud Computing}
\begin{itemize}
\item New ensemble machine learning approaches for container management systems, such as Docker Swarm and Kubernetes, are needed to govern user-based QoS-based container clusters.
\item Cloud service dependability and QoS must be maintained through the use of advanced machine or deep learning techniques.
\item Network virtualization must be provided at a reasonable cost in an SDN-based cloud computing environment that uses AI/ML models to minimise energy consumption and boost dependability.
\item Using AI/ML, cloud-based Big Data analysis tools may find trends in client behaviour, make better decisions, and better understand their customers. It is a difficult challenge that has to be solved in order to ensure that scaling choices are executed or processed in a timely manner utilising AI/ML.
\item Thermal-aware task and resource scheduling can be improved by using new AI/ML-inspired methodologies.
\item AI/ML-based autonomic computing is becoming increasingly important as the IoT and scientific applications grow.

\end{itemize}
\subsection{Fog Computing}
\begin{itemize}
\item It is imperative that the newest AI and ML approaches be used in order to forecast security vulnerabilities in the fog layer and IoT devices because of their intrinsic decentralisation.
\item AI-based deep learning approaches are needed to estimate resource requirements in advance for different geographic resources for fog and cloud computing, which need new policies for provisioning and scheduling resources.
\item On diverse fog environments, state-of-the-art AI/ML approaches may be employed to schedule tasks.
\end{itemize}

\subsection{Edge Computing}
\begin{itemize}
\item Modern computing systems, which incorporate edge devices as component of datacenters, demand specific IoT-based apps to be created in order to provide for more encrypted transmission and to protect the privacy of data.
\item Due to the resource limitations of IoT edge devices, which can't run the robust security software and firewalls built for desktop PCs, Blockchain technology must be used to enhance security using AI/ML. Moreover, innovative software architectures such as that facilitate IoT devices patching and maintenance could be further enhanced by leveraging AI and ML. 
\item AI/ML-based automated decision-making, rather than human-encoded heuristics, presents a lucrative path for optimising edge systems with massive volumes of data through engineering speed and efficiency.
\item AI-based big data analytics methods are needed to handle edge device data in IoT applications at runtime.
\end{itemize}

\subsection{Serverless Computing}
\begin{itemize}
\item AI may be used to enhance the delay and reaction time of tasks in Serverless computing for IoT applications.
\item Automatic heart disease detection in IoT and Serverless computing contexts requires an ensemble deep learning-based intelligent healthcare system.
\item How can deep learning on IoT devices increase real-world performance in AI-based intelligent systems by leveraging Serverless Computing?
\item Serverless systems can benefit from threat mitigation strategies based on AI/ML, such as clustering model-based security analysis.
 
\end{itemize}
\subsection{Quantum Computing}
\begin{itemize}
\item To \emph{increase the size} of current quantum chips, keeping under control the amount of noise present during the evolution of the quantum states.
\item To enter, in full, within the era of Noisy Intermediate-Scale Quantum (NISQ) devices, which should allow us to simulate the dynamics of \emph{complex} quantum systems.
\item To integrate quantum chip with quantum error correction (QEC) that will allow to progress towards Fault-Tolerant Quantum Computation (FTQC). That will allow to simulate the design and behaviour of novel materials on general purpose quantum computers, opening up new possibilities in virtually all area of knowledge, from physics matter to the design of novel AI applications.
\item Develop cloud quantum computing infrastructures that, very likely, will be the way in which we will use quantum computer and simulators: as a booster supporting our local, classical devices.
\item To handle the massive amounts of data created by IoT devices, powerful AI and reinforcement learning may be used.

\item The most recent AI and ML-based methods may be utilised to dynamically discover and rectify faults to provide a valued and dependable service. Recent AI and ML approaches can enhance dependability, but they can also raise system complexity by increasing data processing, which results in higher training costs for AI and ML techniques as a side effect.
 
\end{itemize}

\subsection{Miscellaneous}
The advancements in cloud, fog, and edge in the context of IoT have lead to concepts such as Cloud-to-Things. The following are interesting research directions to explore:

\begin{itemize}
\item The ability to form Decentralized Autonomous Organizations (DAOs) is a concept fundamental to the Cloud-to-Things computing continuum.
\item Rules and agreements can take the form of Blockchain-enabled smart contracts, thus providing for trust among the actors and organizing computing across the continuum \cite{KOCHOVSKI2019747}.
\end{itemize}

AI/ML can also open up new research directions when used in conjunction with blockchain technology:

\begin{itemize}
\item Smart Oracles can be used to establish decentralised monitoring of vertically and horizontally distributed heterogeneous infrastructures.
\item AI/ML and knowledge management methods can be incorporated both within functions of Blockchain-enabled smart contracts and Smart Oracles.
\end{itemize}

\section{Conclusions and Summary}
\textcolor{black}{Computing systems have advanced computer science in the past couple of decades and are now the heart of the corporate world, providing services based on Cloud, Fog, Edge, Serverless, and Quantum Computing.  Many real-world problems that require low latency and response time have been solved due to modern computing systems. This has helped young talents around the globe to launch start-ups allowing large computing capacity for solving challenging problems to speed up scientific progress. }

\textcolor{black}{Artificial Intelligence (AI), Machine Learning (ML) and Deep Learning (DL) have become increasingly popular in recent years because of the advances in accuracy pioneered by them in areas such as computer vision, natural language processing and other allied applications. For training these models, massive amounts of data has been collected in the previous few years in addition to the development in state-of-the-art computing hardware such as the Graphics Processing Unit (GPU), Google's Tensor Processing Unit (TPU) and AI Tesla's Dojo Processing Unit (DPU). Computational researchers and practitioners should be aware of AI/ML/DL algorithms and models. With AI/ML/DL, modern computing may profit from more efficient resource management, while computing is a vital platform for hosting AI/ML/DL services because of its huge computational capacity. This means that both sides gain from the other. Large-scale computing power and external data sources are needed for many AI/ML/DL techniques, which may be more easily obtained via computing systems. This is especially important now that methods for training sophisticated AI, ML, and DL models can be implemented in parallel and in large quantities. To that end, it is foreseen that continued interest in AI/ML/DL applications will spur new research into well-established data centre resource management issues including VM provisioning, consolidation, and load balancing, while also making it easier to cope with scale-out challenges. Innovative research on Explainable AI (XAI) might pave the way for more widespread use of AI in modern computer systems.}

\begin{figure}[t]
    \centering
    \includegraphics[width=1.1\linewidth]{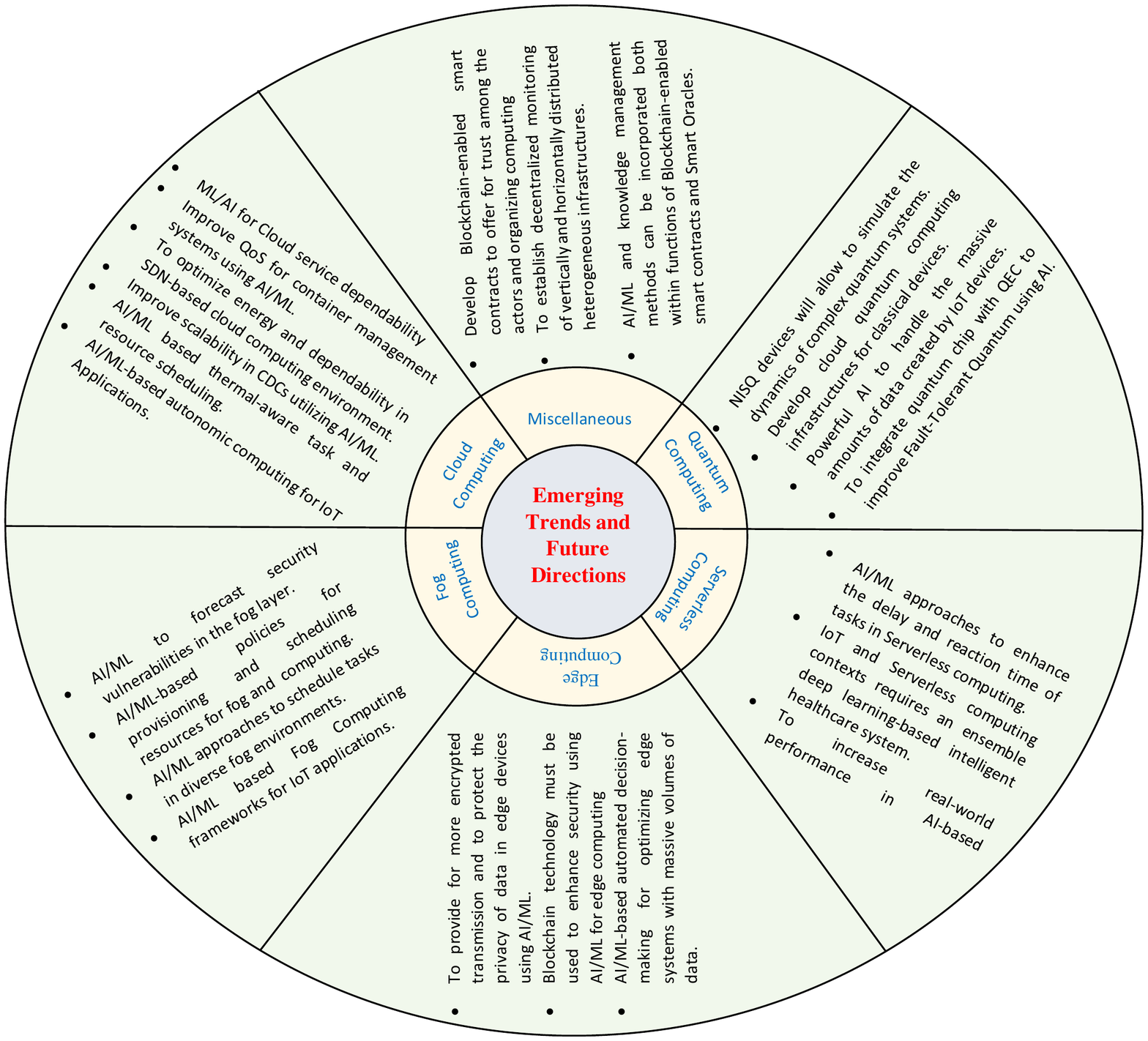}
    \caption{Summary of Emerging Trends and Future Directions for AI-integrated Next Generation Computing}
    \label{fig:model7}
\end{figure}

\textcolor{black}{AI and ML are bringing important necessary demands for computing systems in the upcoming years, from large-scale heterogeneous IoT and sensor networks generating extremely huge data streams to store, maintain, and investigate to QoS-aware (latency, energy, cost, response time) customised computing service adapting for an array of hardware devices while maximising for multicriteria including software-level QoS constraints and financial restraints. As a result of these needs, new methodologies and research strategies are needed to harness the AI and ML models in order to overcome the challenges such as latency and scalability as well as resource and security management. As a cost-effective technique to increase computing application performance, scaling and flexibility are functional abilities that are yet to completely utilise AI and ML models. AI and ML may be strategically used in resource management and scheduling techniques to maximise QoS to improve modern computing. At this time, there are no comprehensive models of service resilience, autonomous methods for managing availability and reliability, and provisioning algorithms that are cognizant of failures in the current research. Next-generation or futuristic computing could be established with the help of AI/ML techniques, which can handle these problems quickly and effectively. The implementation of AI/ML-based resource management policies can help data to automatically adjust their own energy usage and deliver QoS without affecting the system's reliability. AI and ML techniques may also be used to predict the demand for energy usage in advance by combining renewable and non-renewable energy sources. Further, AI/ML modes can be used to analyse Big Data for security breaching. Figure~\ref{fig:model7} shows the summary of new trends and future directions for AI-integrated next generation computing.}

\textcolor{black}{In this article, we have given our vision and explored numerous new trends in AI and ML for cloud, fog, edge, serverless and quantum computing, as well as for other computing platforms and technologies. This is a holistic futuristic research article that has drawn together breakthroughs and highlighted the obstacles remaining to be solved in implementing the use of AI/ML for modern computing. We have also developed a conceptual framework for integrating cutting-edge technology in the future to provide effective computing services. New research developments related to autonomic computing with embedded intelligence have been discussed. In addition, various potential risks of AI-integrated next generation computing have been presented. This work recognised recent significant challenges in AI/ML-embedded next generation computing and has summarised research findings with limitations. Additionally, this futuristic work has examined how current computing issues would be affected by new trends. In this visionary work, potential research directions for AI/ML-based next-generation or modern computing are highlighted. It is clear that AI and ML can be used to solve complicated issues in the future, and this forward-thinking strategy inspires other scholars and researchers to follow suit in a similar fashion. We expect that this visionary research will be useful to practitioners, scientists, engineers and researchers who are interested in conducting research in any area of AI/ML-integrated next generation computing in the future.}

\section*{Declaration of Competing Interests}
{The authors declare that they have no known competing financial interests or personal relationships that could have appeared to influence the work reported in this paper.}

\section*{Acknowledgments}
\textcolor{black}{We thank Prof. Fatos Xhafa (Editor-in-Chief) and anonymous reviewers for their constructive suggestions and guidance on improving the content and quality of this paper.} We also thank Prof. Rajkumar Buyya (The University of Melbourne) and Dr Felix Cuadrado (Technical University of Madrid) for their comments and suggestions for improving the paper. Regarding funding, Minxain Xu has been supported by the National Natural Science Foundation of China (62102408).

% \newpage
\printcredits

\bibliographystyle{elsarticle-num}
% \bibliographystyle{cas-model2-names}

\bibliography{ms}

\begin{thebibliography}{100}
\expandafter\ifx\csname url\endcsname\relax
  \def\url#1{\texttt{#1}}\fi
\expandafter\ifx\csname urlprefix\endcsname\relax\def\urlprefix{URL }\fi
\expandafter\ifx\csname href\endcsname\relax
  \def\href#1#2{#2} \def\path#1{#1}\fi

\bibitem{kephart2003vision}
J.~O. Kephart, D.~M. Chess, The vision of autonomic computing, Computer 36~(1)
  (2003) 41--50.

\bibitem{singh2017star}
S.~Singh, I.~Chana, R.~Buyya, Star: Sla-aware autonomic management of cloud
  resources, IEEE Transactions on Cloud Computing (2017).

\bibitem{parashar2018autonomic}
M.~Parashar, S.~Hariri, Autonomic computing: concepts, infrastructure, and
  applications, CRC press, 2018.

\bibitem{puviani2013self}
M.~Puviani, R.~Frei, Self-management for cloud computing, in: 2013 Science and
  Information Conference, IEEE, 2013, pp. 940--946.

\bibitem{huebscher2008survey}
M.~C. Huebscher, J.~A. McCann, A survey of autonomic computing—degrees,
  models, and applications, ACM Computing Surveys (CSUR) 40~(3) (2008) 1--28.

\bibitem{elmroth2011self}
E.~Elmroth, J.~Tordsson, F.~Hern{\'a}ndez, A.~Ali-Eldin, P.~Sv{\"a}rd,
  M.~Sedaghat, W.~Li, Self-management challenges for multi-cloud architectures,
  in: European Conference on a Service-Based Internet, Springer, 2011, pp.
  38--49.

\bibitem{kephart2007achieving}
J.~O. Kephart, R.~Das, Achieving self-management via utility functions, IEEE
  Internet Computing 11~(1) (2007) 40--48.

\bibitem{singh2015qos}
S.~Singh, I.~Chana, Qos-aware autonomic resource management in cloud computing:
  a systematic review, ACM Computing Surveys (CSUR) 48~(3) (2015) 1--46.

\bibitem{gill2018chopper}
S.~S. Gill, I.~Chana, M.~Singh, R.~Buyya, Chopper: an intelligent qos-aware
  autonomic resource management approach for cloud computing, Cluster Computing
  21~(2) (2018) 1203--1241.

\bibitem{gill2019resource}
S.~S. Gill, R.~Buyya, Resource provisioning based scheduling framework for
  execution of heterogeneous and clustered workloads in clouds: from
  fundamental to autonomic offering, Journal of Grid Computing 17~(3) (2019)
  385--417.

\bibitem{derbel2009anema}
H.~Derbel, N.~Agoulmine, M.~Sala{\"u}n, Anema: Autonomic network management
  architecture to support self-configuration and self-optimization in ip
  networks, Computer Networks 53~(3) (2009) 418--430.

\bibitem{herrmann2005self}
K.~Herrmann, G.~Muhl, K.~Geihs, Self management: the solution to complexity or
  just another problem?, IEEE distributed systems online 6~(1) (2005).

\bibitem{kephart2015symbiotic}
J.~O. Kephart, J.~Lenchner, A symbiotic cognitive computing perspective on
  autonomic computing, in: 2015 IEEE International Conference on Autonomic
  Computing, IEEE, 2015, pp. 109--114.

\bibitem{kephart2004artificial}
J.~O. Kephart, W.~E. Walsh, An artificial intelligence perspective on autonomic
  computing policies, in: Proceedings. Fifth IEEE International Workshop on
  Policies for Distributed Systems and Networks, 2004. POLICY 2004., IEEE,
  2004, pp. 3--12.

\bibitem{anderson2021self}
C.~Anderson, T.~G. Walmsley, P.~Patros, A self-learning architecture for
  digital twins with self-protection, in: 2021 IEEE International Conference on
  Autonomic Computing and Self-Organizing Systems Companion (ACSOS-C), IEEE,
  2021, pp. 291--292.

\bibitem{rutten2017feedback}
E.~Rutten, N.~Marchand, D.~Simon, Feedback control as mape-k loop in autonomic
  computing, Software engineering for self-adaptive systems iii. assurances
  (2017) 349--373.

\bibitem{donepudi2018application}
P.~K. Donepudi, Application of artificial intelligence in automation industry,
  Asian Journal of Applied Science and Engineering 7~(1) (2018) 7--20.

\bibitem{ullah2020applications}
Z.~Ullah, F.~Al-Turjman, L.~Mostarda, R.~Gagliardi, Applications of artificial
  intelligence and machine learning in smart cities, Computer Communications
  154 (2020) 313--323.

\bibitem{gill2019transformative}
S.~S. Gill, S.~Tuli, M.~Xu, I.~Singh, K.~V. Singh, D.~Lindsay, S.~Tuli,
  D.~Smirnova, M.~Singh, U.~Jain, et~al., Transformative effects of iot,
  blockchain and artificial intelligence on cloud computing: Evolution, vision,
  trends and open challenges, Internet of Things 8 (2019) 100118.

\bibitem{buyya2018manifesto}
R.~Buyya, S.~N. Srirama, G.~Casale, R.~Calheiros, Y.~Simmhan, B.~Varghese,
  E.~Gelenbe, B.~Javadi, L.~M. Vaquero, M.~A. Netto, et~al., A manifesto for
  future generation cloud computing: Research directions for the next decade,
  ACM computing surveys (CSUR) 51~(5) (2018) 1--38.

\bibitem{kettimuthu2018towards}
R.~Kettimuthu, Z.~Liu, I.~Foster, P.~H. Beckman, A.~Sim, K.~Wu, W.-k. Liao,
  Q.~Kang, A.~Agrawal, A.~Choudhary, Towards autonomic science infrastructure:
  architecture, limitations, and open issues, in: Proceedings of the 1st
  International Workshop on Autonomous Infrastructure for Science, 2018, pp.
  1--9.

\bibitem{harman2012role}
M.~Harman, The role of artificial intelligence in software engineering, in:
  2012 First International Workshop on Realizing AI Synergies in Software
  Engineering (RAISE), IEEE, 2012, pp. 1--6.

\bibitem{lopez2019self}
S.~L{\'o}pez-Huguet, A.~P{\'e}rez, A.~Calatrava, C.~de~Alfonso, M.~Caballer,
  G.~Molt{\'o}, I.~Blanquer, A self-managed mesos cluster for data analytics
  with qos guarantees, Future Generation Computer Systems 96 (2019) 449--461.

\bibitem{salehie2005autonomic}
M.~Salehie, L.~Tahvildari, Autonomic computing: emerging trends and open
  problems, ACM SIGSOFT Software Engineering Notes 30~(4) (2005) 1--7.

\bibitem{nami2007survey}
M.~R. Nami, K.~Bertels, A survey of autonomic computing systems, in: Third
  international conference on autonomic and autonomous systems (ICAS'07), IEEE,
  2007, pp. 26--26.

\bibitem{tarbell2019ai}
M.~A. Tarbell, Ai and the transcendence of true autonomy, in: Micro-and
  Nanotechnology Sensors, Systems, and Applications XI, Vol. 10982,
  International Society for Optics and Photonics, 2019, p. 1098223.

\bibitem{psaier2011survey}
H.~Psaier, S.~Dustdar, A survey on self-healing systems: approaches and
  systems, Computing 91~(1) (2011) 43--73.

\bibitem{lynn2018toward}
T.~Lynn, P.~Rosati, P.~T. Endo, Toward the intelligent internet of everything:
  Observations on multidisciplinary challenges in intelligent systems research,
  Technology, Science, and Culture: A Global Vision (2018) 52.

\bibitem{ganek2004response}
A.~G. Ganek, C.~P. Hilkner, J.~W. Sweitzer, B.~Miller, J.~L. Hellerstein, The
  response to it complexity: autonomic computing, in: Third IEEE International
  Symposium on Network Computing and Applications, 2004.(NCA 2004).
  Proceedings., IEEE, 2004, pp. 151--157.

\bibitem{chaurasia2021comprehensive}
N.~Chaurasia, M.~Kumar, R.~Chaudhry, O.~P. Verma, Comprehensive survey on
  energy-aware server consolidation techniques in cloud computing, The Journal
  of Supercomputing 77~(10) (2021) 11682--11737.

\bibitem{zhou2019distributing}
L.~Zhou, H.~Wen, R.~Teodorescu, D.~H. Du, Distributing deep neural networks
  with containerized partitions at the edge, in: 2nd $\{$USENIX$\}$ Workshop on
  Hot Topics in Edge Computing (HotEdge 19), 2019, pp. 1--5.

\bibitem{varghese2018next}
B.~Varghese, R.~Buyya, Next generation cloud computing: New trends and research
  directions, Future Generation Computer Systems 79 (2018) 849--861.

\bibitem{abdulkareem2019review}
K.~H. Abdulkareem, M.~A. Mohammed, S.~S. Gunasekaran, M.~N. Al-Mhiqani, A.~A.
  Mutlag, S.~A. Mostafa, N.~S. Ali, D.~A. Ibrahim, A review of fog computing
  and machine learning: concepts, applications, challenges, and open issues,
  IEEE Access 7 (2019) 153123--153140.

\bibitem{merenda2020edge}
M.~Merenda, C.~Porcaro, D.~Iero, Edge machine learning for ai-enabled iot
  devices: A review, Sensors 20~(9) (2020) 2533.

\bibitem{kumar2021survey}
A.~Kumar, S.~Bhatia, K.~Kaushik, S.~M. Gandhi, S.~G. Devi, A.~D.~J. Diego,
  A.~Mashat, Survey of promising technologies for quantum drones and networks,
  IEEE Access 9 (2021) 125868--125911.

\bibitem{li2020quantum}
Y.~Li, M.~Tian, G.~Liu, C.~Peng, L.~Jiao, Quantum optimization and quantum
  learning: A survey, Ieee Access 8 (2020) 23568--23593.

\bibitem{hassan2021survey}
H.~B. Hassan, S.~A. Barakat, Q.~I. Sarhan, Survey on serverless computing,
  Journal of Cloud Computing 10~(1) (2021) 1--29.

\bibitem{goel2021review}
S.~S. Goel, A.~Goel, M.~Kumar, G.~Molt{\'o}, A review of internet of things:
  qualifying technologies and boundless horizon, Journal of Reliable
  Intelligent Environments 7~(1) (2021) 23--33.

\bibitem{desai2021healthcloud}
F.~Desai, et~al., Healthcloud: A system for monitoring health status of heart
  patients using machine learning and cloud computing, Internet of Things 17
  (2022) 100485.

\bibitem{gill2021quantum}
S.~S. Gill, Quantum and blockchain based serverless edge computing: A vision,
  model, new trends and future directions, Internet Technology Letters (2021)
  e275.

\bibitem{XuTSUSC2021}
M.~Xu, A.~N. Toosi, R.~Buyya, A self-adaptive approach for managing
  applications and harnessing renewable energy for sustainable cloud computing,
  IEEE Transactions on Sustainable Computing 6~(4) (2021) 544--558.
\newblock \href {https://doi.org/10.1109/TSUSC.2020.3014943}
  {\path{doi:10.1109/TSUSC.2020.3014943}}.

\bibitem{pusztai2021slo}
T.~Pusztai, A.~Morichetta, V.~C. Pujol, S.~Dustdar, S.~Nastic, X.~Ding, D.~Vij,
  Y.~Xiong, Slo script: A novel language for implementing complex cloud-native
  elasticity-driven slos, in: 2021 IEEE International Conference on Web
  Services (ICWS), IEEE, 2021, pp. 21--31.

\bibitem{tuli2022hunter}
S.~Tuli, et~al., Hunter: Ai based holistic resource management for sustainable
  cloud computing, Journal of Systems and Software 184 (2022) 111124.

\bibitem{abdelaziz2018machine}
A.~Abdelaziz, M.~Elhoseny, A.~S. Salama, A.~Riad, A machine learning model for
  improving healthcare services on cloud computing environment, Measurement 119
  (2018) 117--128.

\bibitem{ulrich2020top}
L.~Ulrich, Top 10 tech cars: The scramble for electric dominance has begun,
  IEEE Spectrum 57~(4) (2020) 30--39.

\bibitem{masood2019aiops}
A.~Masood, A.~Hashmi, Aiops: Predictive analytics \& machine learning in
  operations, in: Cognitive Computing Recipes, Springer, 2019, pp. 359--382.

\bibitem{dang2019aiops}
Y.~Dang, Q.~Lin, P.~Huang, Aiops: real-world challenges and research
  innovations, in: 2019 IEEE/ACM 41st International Conference on Software
  Engineering: Companion Proceedings (ICSE-Companion), IEEE, 2019, pp. 4--5.

\bibitem{pwc}
\href{https://www.pwc.com/us/en/tech-effect/cloud/covid19-cloud-infrastructure.html}{Can
  you meet customer demand for cloud-based computing?} (2019).
\newline\urlprefix\url{https://www.pwc.com/us/en/tech-effect/cloud/covid19-cloud-infrastructure.html}

\bibitem{nastic2020sloc}
S.~Nastic, A.~Morichetta, T.~Pusztai, S.~Dustdar, X.~Ding, D.~Vij, Y.~Xiong,
  Sloc: Service level objectives for next generation cloud computing, IEEE
  Internet Computing 24~(3) (2020) 39--50.

\bibitem{younas2020elicitation}
M.~Younas, D.~N.~A. Jawawi, M.~A. Shah, A.~Mustafa, M.~Awais, M.~K. Ishfaq,
  K.~Wakil, Elicitation of nonfunctional requirements in agile development
  using cloud computing environment, IEEE Access 8 (2020) 209153--209162.

\bibitem{breiter2009life}
G.~Breiter, M.~Behrendt, Life cycle and characteristics of services in the
  world of cloud computing, IBM Journal of Research and Development 53~(4)
  (2009) 3--1.

\bibitem{pop2016machine}
D.~Pop, Machine learning and cloud computing: Survey of distributed and saas
  solutions, arXiv preprint arXiv:1603.08767 (2016).

\bibitem{elger2020ai}
P.~Elger, E.~Shanaghy, AI as a Service: Serverless machine learning with AWS,
  Manning Publications, 2020.

\bibitem{XuContainerSurvey}
Z.~Zhong, M.~Xu, M.~A. Rodriguez, C.~Xu, R.~Buyya, Machine learning-based
  orchestration of containers: A taxonomy and future directions, ACM Computing
  Surveys (2022).

\bibitem{chaudhary2018optimized}
R.~Chaudhary, G.~S. Aujla, N.~Kumar, J.~J. Rodrigues, Optimized big data
  management across multi-cloud data centers: Software-defined-network-based
  analysis, IEEE Communications Magazine 56~(2) (2018) 118--126.

\bibitem{rajeswari2022ai}
S.~Rajeswari, V.~Ponnusamy, Ai-based iot analytics on the cloud for diabetic
  data management system, in: Integrating AI in IoT Analytics on the Cloud for
  Healthcare Applications, IGI Global, 2022, pp. 143--161.

\bibitem{surya2018streamlining}
L.~Surya, Streamlining cloud application with ai technology, International
  Journal of Innovations in Engineering Research and Technology [IJIERT] ISSN
  (2018) 2394--3696.

\bibitem{marshall2018cloud}
T.~E. Marshall, S.~L. Lambert, Cloud-based intelligent accounting applications:
  accounting task automation using ibm watson cognitive computing, Journal of
  Emerging Technologies in Accounting 15~(1) (2018) 199--215.

\bibitem{jha2021combining}
N.~Jha, D.~Prashar, A.~Nagpal, Combining artificial intelligence with robotic
  process automation—an intelligent automation approach, in: Deep Learning
  and Big Data for Intelligent Transportation, Springer, 2021, pp. 245--264.

\bibitem{robertson2021cloud}
J.~Robertson, J.~M. Fossaceca, K.~W. Bennett, A cloud-based computing framework
  for artificial intelligence innovation in support of multidomain operations,
  IEEE Transactions on Engineering Management (2021).

\bibitem{horn2019cost}
G.~Horn, P.~Skrzypek, K.~Materka, T.~Prze{\'z}dziek, Cost benefits of
  multi-cloud deployment of dynamic computational intelligence applications,
  in: Workshops of the International Conference on Advanced Information
  Networking and Applications, Springer, 2019, pp. 1041--1054.

\bibitem{gonzalez2019biomedical}
G.~Gonz{\'a}lez, C.~L. Evans, Biomedical image processing with containers and
  deep learning: An automated analysis pipeline: Data architecture, artificial
  intelligence, automated processing, containerization, and clusters
  orchestration ease the transition from data acquisition to insights in
  medium-to-large datasets, BioEssays 41~(6) (2019) 1900004.

\bibitem{chatterjee2020adoption}
S.~Chatterjee, S.~K. Ghosh, R.~Chaudhuri, S.~Chaudhuri, Adoption of
  ai-integrated crm system by indian industry: from security and privacy
  perspective, Information \& Computer Security (2020).

\bibitem{carvalho2019off}
A.~Carvalho, A.~Levitt, S.~Levitt, E.~Khaddam, J.~Benamati, Off-the-shelf
  artificial intelligence technologies for sentiment and emotion analysis: a
  tutorial on using ibm natural language processing, Communications of the
  Association for Information Systems 44~(1) (2019) 43.

\bibitem{blohm2019towards}
M.~Blohm, C.~Dukino, M.~Kintz, M.~Kochanowski, F.~Koetter, T.~Renner, Towards a
  privacy compliant cloud architecture for natural language processing
  platforms., in: ICEIS (1), 2019, pp. 454--461.

\bibitem{singh2021fog}
J.~Singh, et~al., Fog computing: A taxonomy, systematic review, current trends
  and research challenges, Journal of Parallel and Distributed Computing 157
  (2021) 56--85.

\bibitem{gubbi2013internet}
J.~Gubbi, R.~Buyya, S.~Marusic, M.~Palaniswami, Internet of things (iot): A
  vision, architectural elements, and future directions, Future generation
  computer systems 29~(7) (2013) 1645--1660.

\bibitem{mahmud2019quality}
R.~Mahmud, S.~N. Srirama, K.~Ramamohanarao, R.~Buyya, Quality of experience
  (qoe)-aware placement of applications in fog computing environments, Journal
  of Parallel and Distributed Computing 132 (2019) 190--203.

\bibitem{tong2019artificial}
W.~Tong, A.~Hussain, W.~X. Bo, S.~Maharjan, Artificial intelligence for
  vehicle-to-everything: A survey, IEEE Access 7 (2019) 10823--10843.

\bibitem{gill2020thermosim}
S.~S. Gill, S.~Tuli, A.~N. Toosi, F.~Cuadrado, P.~Garraghan, R.~Bahsoon,
  H.~Lutfiyya, R.~Sakellariou, O.~Rana, S.~Dustdar, et~al., Thermosim: Deep
  learning based framework for modeling and simulation of thermal-aware
  resource management for cloud computing environments, Journal of Systems and
  Software 166 (2020) 110596.

\bibitem{maroof2019plar}
U.~Maroof, A.~Shaghaghi, S.~Jha, Plar: Towards a pluggable software
  architecture for securing iot devices, in: Proceedings of the 2nd
  International ACM Workshop on Security and Privacy for the
  Internet-of-Things, 2019, pp. 50--57.

\bibitem{zou2019edge}
Z.~Zou, Y.~Jin, P.~Nevalainen, Y.~Huan, J.~Heikkonen, T.~Westerlund, Edge and
  fog computing enabled ai for iot-an overview, in: 2019 IEEE International
  Conference on Artificial Intelligence Circuits and Systems (AICAS), IEEE,
  2019, pp. 51--56.

\bibitem{firouzi2021convergence}
F.~Firouzi, B.~Farahani, A.~Marin{\v{s}}ek, The convergence and interplay of
  edge, fog, and cloud in the ai-driven internet of things (iot), Information
  Systems (2021) 101840.

\bibitem{teoh2021iot}
Y.~K. Teoh, et~al., Iot and fog computing based predictive maintenance model
  for effective asset management in industry 4.0 using machine learning, IEEE
  Internet of Things Journal (2021).

\bibitem{czentye2019optimizing}
J.~Czentye, I.~Pelle, A.~Kern, B.~P. Gero, L.~Toka, B.~Sonkoly, Optimizing
  latency sensitive applications for amazon's public cloud platform, in: 2019
  IEEE Global Communications Conference (GLOBECOM), IEEE, 2019, pp. 1--7.

\bibitem{satyanarayanan2015brief}
M.~Satyanarayanan, A brief history of cloud offload: A personal journey from
  odyssey through cyber foraging to cloudlets, GetMobile: Mobile Computing and
  Communications 18~(4) (2015) 19--23.

\bibitem{tuli2020ithermofog}
S.~Tuli, et~al., ithermofog: Iot-fog based automatic thermal profile creation
  for cloud data centers using artificial intelligence techniques, Internet
  Technology Letters 3~(5) (2020) e198.

\bibitem{skouby2014smart}
K.~E. Skouby, P.~Lynggaard, Smart home and smart city solutions enabled by 5g,
  iot, aai and cot services, in: 2014 International Conference on Contemporary
  Computing and Informatics (IC3I), IEEE, 2014, pp. 874--878.

\bibitem{gill2022manifesto}
S.~S. Gill, A manifesto for modern fog and edge computing: Vision, new
  paradigms, opportunities, and future directions, in: Operationalizing
  Multi-Cloud Environments, Springer, 2022, pp. 237--253.

\bibitem{bonomi2012fog}
F.~Bonomi, R.~Milito, J.~Zhu, S.~Addepalli, Fog computing and its role in the
  internet of things, in: Proceedings of the first edition of the MCC workshop
  on Mobile cloud computing, 2012, pp. 13--16.

\bibitem{naha2018fog}
R.~K. Naha, S.~Garg, D.~Georgakopoulos, P.~P. Jayaraman, L.~Gao, Y.~Xiang,
  R.~Ranjan, Fog computing: Survey of trends, architectures, requirements, and
  research directions, IEEE access 6 (2018) 47980--48009.

\bibitem{goudarzi2020application}
M.~Goudarzi, H.~Wu, M.~Palaniswami, R.~Buyya, An application placement
  technique for concurrent iot applications in edge and fog computing
  environments, IEEE Transactions on Mobile Computing 20~(4) (2020) 1298--1311.

\bibitem{luong2020machine}
N.~C. Luong, Y.~Jiao, P.~Wang, D.~Niyato, D.~I. Kim, Z.~Han, A
  machine-learning-based auction for resource trading in fog computing, IEEE
  Communications Magazine 58~(3) (2020) 82--88.

\bibitem{naveen2018search}
S.~Naveen, M.~R. Kounte, In search of the future technologies: Fusion of
  machine learning, fog and edge computing in the internet of things, in:
  International conference on Computer Networks, Big data and IoT, Springer,
  2018, pp. 278--285.

\bibitem{lin2020fog}
S.-Y. Lin, Y.~Du, P.-C. Ko, T.-J. Wu, P.-T. Ho, V.~Sivakumar, et~al., Fog
  computing based hybrid deep learning framework in effective inspection system
  for smart manufacturing, Computer Communications 160 (2020) 636--642.

\bibitem{wang2019mobility}
D.~Wang, Z.~Liu, X.~Wang, Y.~Lan, Mobility-aware task offloading and migration
  schemes in fog computing networks, IEEE Access 7 (2019) 43356--43368.

\bibitem{waqas2018mobility}
M.~Waqas, Y.~Niu, M.~Ahmed, Y.~Li, D.~Jin, Z.~Han, Mobility-aware fog computing
  in dynamic environments: Understandings and implementation, IEEE Access 7
  (2018) 38867--38879.

\bibitem{zhao2017reinforcement}
T.~Zhao, W.~Zhang, H.~Zhao, Z.~Jin, A reinforcement learning-based framework
  for the generation and evolution of adaptation rules, in: 2017 IEEE
  International Conference on Autonomic Computing (ICAC), IEEE, 2017, pp.
  103--112.

\bibitem{wang2022optimization}
J.~Wang, H.~Ke, X.~Liu, H.~Wang, Optimization for computational offloading in
  multi-access edge computing: A deep reinforcement learning scheme, Computer
  Networks (2022) 108690.

\bibitem{wang2020reinforcement}
D.~Wang, X.~Tian, H.~Cui, Z.~Liu, Reinforcement learning-based joint task
  offloading and migration schemes optimization in mobility-aware mec network,
  China Communications 17~(8) (2020) 31--44.

\bibitem{rusman2019fog}
J.~Rusman, Z.~Tahir, A.~E.~U. Salam, Fog computing concept implementation in
  work error detection system of the industrial machine using support vector
  machine (svm), in: 2019 International Seminar on Research of Information
  Technology and Intelligent Systems (ISRITI), IEEE, 2019, pp. 160--164.

\bibitem{majumdar2018kyasanur}
A.~Majumdar, T.~Debnath, S.~K. Sood, K.~L. Baishnab, Kyasanur forest disease
  classification framework using novel extremal optimization tuned neural
  network in fog computing environment, Journal of medical systems 42~(10)
  (2018) 1--16.

\bibitem{ray2018compassionate}
A.~Ray, Compassionate Artificial Intelligence: Frameworks and Algorithms,
  Compassionate AI Lab (An Imprint of Inner Light Publishers), 2018.

\bibitem{hussain2020machine}
F.~Hussain, R.~Hussain, S.~A. Hassan, E.~Hossain, Machine learning in iot
  security: Current solutions and future challenges, IEEE Communications
  Surveys \& Tutorials 22~(3) (2020) 1686--1721.

\bibitem{queralta2019edge}
J.~P. Queralta, T.~N. Gia, H.~Tenhunen, T.~Westerlund, Edge-ai in lora-based
  health monitoring: Fall detection system with fog computing and lstm
  recurrent neural networks, in: 2019 42nd international conference on
  telecommunications and signal processing (TSP), IEEE, 2019, pp. 601--604.

\bibitem{rihan2020deep}
M.~Rihan, M.~Elwekeil, Y.~Yang, L.~Huang, C.~Xu, M.~M. Selim, Deep-vfog: When
  artificial intelligence meets fog computing in v2x, IEEE Systems Journal
  (2020).

\bibitem{tanwani2019fog}
A.~K. Tanwani, N.~Mor, J.~Kubiatowicz, J.~E. Gonzalez, K.~Goldberg, A fog
  robotics approach to deep robot learning: Application to object recognition
  and grasp planning in surface decluttering, in: 2019 International Conference
  on Robotics and Automation (ICRA), IEEE, 2019, pp. 4559--4566.

\bibitem{tian2019fog}
N.~Tian, A.~K. Tanwani, J.~Chen, M.~Ma, R.~Zhang, B.~Huang, K.~Goldberg,
  S.~Sojoudi, A fog robotic system for dynamic visual servoing, in: 2019
  International Conference on Robotics and Automation (ICRA), IEEE, 2019, pp.
  1982--1988.

\bibitem{li2018deep}
L.~Li, K.~Ota, M.~Dong, Deep learning for smart industry: Efficient manufacture
  inspection system with fog computing, IEEE Transactions on Industrial
  Informatics 14~(10) (2018) 4665--4673.

\bibitem{ng2020anomaly}
B.~A. NG, S.~Selvakumar, Anomaly detection framework for internet of things
  traffic using vector convolutional deep learning approach in fog environment,
  Future Generation Computer Systems 113 (2020) 255--265.

\bibitem{pusztai2021pogonip}
T.~Pusztai, F.~Rossi, S.~Dustdar, Pogonip: Scheduling asynchronous applications
  on the edge, in: 2021 IEEE 14th International Conference on Cloud Computing
  (CLOUD), IEEE, 2021, pp. 660--670.

\bibitem{nastic2021polaris}
S.~Nastic, T.~Pusztai, A.~Morichetta, V.~C. Pujol, S.~Dustdar, D.~Vii,
  Y.~Xiong, Polaris scheduler: Edge sensitive and slo aware workload scheduling
  in cloud-edge-iot clusters, in: 2021 IEEE 14th International Conference on
  Cloud Computing (CLOUD), IEEE, 2021, pp. 206--216.

\bibitem{XuSPE2022}
M.~Xu, Q.~Zhou, H.~Wu, W.~Lin, K.~Ye, C.~Xu,
  \href{https://onlinelibrary.wiley.com/doi/abs/10.1002/spe.3014}{Pdma:
  Probabilistic service migration approach for delay-aware and mobility-aware
  mobile edge computing}, Software: Practice and Experience 52~(2) (2022)
  394--414.
\newblock \href
  {http://arxiv.org/abs/https://onlinelibrary.wiley.com/doi/pdf/10.1002/spe.3014}
  {\path{arXiv:https://onlinelibrary.wiley.com/doi/pdf/10.1002/spe.3014}},
  \href {https://doi.org/https://doi.org/10.1002/spe.3014}
  {\path{doi:https://doi.org/10.1002/spe.3014}}.
\newline\urlprefix\url{https://onlinelibrary.wiley.com/doi/abs/10.1002/spe.3014}

\bibitem{yu2017survey}
W.~Yu, F.~Liang, X.~He, W.~G. Hatcher, C.~Lu, J.~Lin, X.~Yang, A survey on the
  edge computing for the internet of things, IEEE access 6 (2017) 6900--6919.

\bibitem{edgeai}
\href{https://towardsdatascience.com/edge-ai-is-the-next-wave-of-ai-a3e98b77c2d7}{Edge
  ai is the next wave of ai} (2020).
\newline\urlprefix\url{https://towardsdatascience.com/edge-ai-is-the-next-wave-of-ai-a3e98b77c2d7}

\bibitem{pusztai2021novel}
T.~Pusztai, A.~Morichetta, V.~C. Pujol, S.~Dustdar, S.~Nastic, X.~Ding, D.~Vij,
  Y.~Xiong, A novel middleware for efficiently implementing complex
  cloud-native slos, in: 2021 IEEE 14th International Conference on Cloud
  Computing (CLOUD), IEEE, 2021, pp. 410--420.

\bibitem{sun2019ai}
W.~Sun, J.~Liu, Y.~Yue, Ai-enhanced offloading in edge computing: When machine
  learning meets industrial iot, IEEE Network 33~(5) (2019) 68--74.

\bibitem{lv2021intelligent}
Z.~Lv, D.~Chen, R.~Lou, Q.~Wang, Intelligent edge computing based on machine
  learning for smart city, Future Generation Computer Systems 115 (2021)
  90--99.

\bibitem{hao2018edge}
T.~Hao, Y.~Huang, X.~Wen, W.~Gao, F.~Zhang, C.~Zheng, L.~Wang, H.~Ye, K.~Hwang,
  Z.~Ren, et~al., Edge aibench: towards comprehensive end-to-end edge computing
  benchmarking, in: International Symposium on Benchmarking, Measuring and
  Optimization, Springer, 2018, pp. 23--30.

\bibitem{calo2017edge}
S.~B. Calo, M.~Touna, D.~C. Verma, A.~Cullen, Edge computing architecture for
  applying ai to iot, in: 2017 IEEE International Conference on Big Data (Big
  Data), IEEE, 2017, pp. 3012--3016.

\bibitem{shakarami2020survey}
A.~Shakarami, M.~Ghobaei-Arani, A.~Shahidinejad, A survey on the computation
  offloading approaches in mobile edge computing: A machine learning-based
  perspective, Computer Networks (2020) 107496.

\bibitem{deng2020edge}
S.~Deng, H.~Zhao, W.~Fang, J.~Yin, S.~Dustdar, A.~Y. Zomaya, Edge intelligence:
  The confluence of edge computing and artificial intelligence, IEEE Internet
  of Things Journal 7~(8) (2020) 7457--7469.

\bibitem{yang2019mobile}
B.~Yang, X.~Cao, X.~Li, Q.~Zhang, L.~Qian, Mobile-edge-computing-based
  hierarchical machine learning tasks distribution for iiot, IEEE Internet of
  Things Journal 7~(3) (2019) 2169--2180.

\bibitem{zhu2020toward}
G.~Zhu, D.~Liu, Y.~Du, C.~You, J.~Zhang, K.~Huang, Toward an intelligent edge:
  Wireless communication meets machine learning, IEEE communications magazine
  58~(1) (2020) 19--25.

\bibitem{zhou2019edge}
Z.~Zhou, X.~Chen, E.~Li, L.~Zeng, K.~Luo, J.~Zhang, Edge intelligence: Paving
  the last mile of artificial intelligence with edge computing, Proceedings of
  the IEEE 107~(8) (2019) 1738--1762.

\bibitem{huh2019understanding}
J.-H. Huh, Y.-S. Seo, Understanding edge computing: Engineering evolution with
  artificial intelligence, IEEE Access 7 (2019) 164229--164245.

\bibitem{ranjan2014streaming}
R.~Ranjan, Streaming big data processing in datacenter clouds, IEEE Cloud
  Computing 1~(1) (2014) 78--83.

\bibitem{chen2019artificial}
Z.~Chen, Q.~He, L.~Liu, D.~Lan, H.-M. Chung, Z.~Mao, An artificial intelligence
  perspective on mobile edge computing, in: 2019 IEEE International Conference
  on Smart Internet of Things (SmartIoT), IEEE, 2019, pp. 100--106.

\bibitem{xu2020artificial}
Z.~Xu, W.~Liu, J.~Huang, C.~Yang, J.~Lu, H.~Tan, Artificial intelligence for
  securing iot services in edge computing: a survey, Security and Communication
  Networks 2020 (2020).

\bibitem{tuli2020healthfog}
S.~Tuli, et~al., Healthfog: An ensemble deep learning based smart healthcare
  system for automatic diagnosis of heart diseases in integrated iot and fog
  computing environments, Future Generation Computer Systems 104 (2020)
  187--200.

\bibitem{dhillon2020iotpulse}
A.~Dhillon, et~al., Iotpulse: machine learning-based enterprise health
  information system to predict alcohol addiction in punjab (india) using iot
  and fog computing, Enterprise Information Systems (2020) 1--33.

\bibitem{bermbach2021future}
D.~Bermbach, A.~Chandra, C.~Krintz, A.~Gokhale, A.~Slominski, L.~Thamsen,
  E.~Cavalcante, T.~Guo, I.~Brandic, R.~Wolski, On the future of cloud
  engineering, in: 2021 IEEE International Conference on Cloud Engineering
  (IC2E), IEEE, 2021, pp. 264--275.

\bibitem{kounev2021toward}
S.~Kounev, C.~Abad, I.~Foster, N.~Herbst, A.~Iosup, S.~Al-Kiswany, A.~A.-E.
  Hassan, B.~Balis, A.~Bauer, A.~Bondi, et~al., Toward a definition for
  serverless computing, Report from Dagstuhl Seminar 21201 (2021).

\bibitem{castro2019rise}
P.~Castro, V.~Ishakian, V.~Muthusamy, A.~Slominski, The rise of serverless
  computing, Communications of the ACM 62~(12) (2019) 44--54.

\bibitem{fox2017status}
G.~C. Fox, V.~Ishakian, V.~Muthusamy, A.~Slominski, Status of serverless
  computing and function-as-a-service (faas) in industry and research, arXiv
  preprint arXiv:1708.08028 (2017).

\bibitem{akkus2018sand}
I.~E. Akkus, R.~Chen, I.~Rimac, M.~Stein, K.~Satzke, A.~Beck, P.~Aditya,
  V.~Hilt, $\{$SAND$\}$: Towards high-performance serverless computing, in:
  2018 Annual Technical Conference, 2018, pp. 923--935.

\bibitem{lee2018evaluation}
H.~Lee, K.~Satyam, G.~Fox, Evaluation of production serverless computing
  environments, in: 2018 IEEE 11th International Conference on Cloud Computing
  (CLOUD), IEEE, 2018, pp. 442--450.

\bibitem{jangda2019formal}
A.~Jangda, D.~Pinckney, Y.~Brun, A.~Guha, Formal foundations of serverless
  computing, Proceedings of the ACM on Programming Languages 3~(OOPSLA) (2019)
  1--26.

\bibitem{lloyd2018serverless}
W.~Lloyd, S.~Ramesh, S.~Chinthalapati, L.~Ly, S.~Pallickara, Serverless
  computing: An investigation of factors influencing microservice performance,
  in: 2018 IEEE International Conference on Cloud Engineering (IC2E), IEEE,
  2018, pp. 159--169.

\bibitem{al2018making}
Z.~Al-Ali, S.~Goodarzy, E.~Hunter, S.~Ha, R.~Han, E.~Keller, E.~Rozner, Making
  serverless computing more serverless, in: 2018 IEEE 11th International
  Conference on Cloud Computing (CLOUD), IEEE, 2018, pp. 456--459.

\bibitem{mohanty2018evaluation}
S.~K. Mohanty, G.~Premsankar, M.~Di~Francesco, et~al., An evaluation of open
  source serverless computing frameworks., in: CloudCom, 2018, pp. 115--120.

\bibitem{feng2018exploring}
L.~Feng, P.~Kudva, D.~Da~Silva, J.~Hu, Exploring serverless computing for
  neural network training, in: 2018 IEEE 11th International Conference on Cloud
  Computing (CLOUD), IEEE, 2018, pp. 334--341.

\bibitem{perez2018serverless}
A.~P{\'e}rez, G.~Molt{\'o}, M.~Caballer, A.~Calatrava, Serverless computing for
  container-based architectures, Future Generation Computer Systems 83 (2018)
  50--59.

\bibitem{baldini2017serverless}
I.~Baldini, P.~Castro, K.~Chang, P.~Cheng, S.~Fink, V.~Ishakian, N.~Mitchell,
  V.~Muthusamy, R.~Rabbah, A.~Slominski, et~al., Serverless computing: Current
  trends and open problems, in: Research advances in cloud computing, Springer,
  2017, pp. 1--20.

\bibitem{golec2021ifaasbus}
M.~Golec, et~al., ifaasbus: A security and privacy based lightweight framework
  for serverless computing using iot and machine learning, IEEE Transactions on
  Industrial Informatics 18~(5) (2022) 3522--3529.
\newblock \href {https://doi.org/10.1109/TII.2021.3095466}
  {\path{doi:10.1109/TII.2021.3095466}}.

\bibitem{eivy2017wary}
A.~Eivy, J.~Weinman, Be wary of the economics of" serverless" cloud computing,
  IEEE Cloud Computing 4~(2) (2017) 6--12.

\bibitem{wang2019distributed}
H.~Wang, D.~Niu, B.~Li, Distributed machine learning with a serverless
  architecture, in: IEEE INFOCOM 2019-IEEE Conference on Computer
  Communications, IEEE, 2019, pp. 1288--1296.

\bibitem{gupta2020utility}
V.~Gupta, S.~Phade, T.~Courtade, K.~Ramchandran, Utility-based resource
  allocation and pricing for serverless computing, arXiv preprint
  arXiv:2008.07793 (2020).

\bibitem{schuler2021ai}
L.~Schuler, S.~Jamil, N.~K{\"u}hl, Ai-based resource allocation: Reinforcement
  learning for adaptive auto-scaling in serverless environments, in: 2021
  IEEE/ACM 21st International Symposium on Cluster, Cloud and Internet
  Computing (CCGrid), IEEE, 2021, pp. 804--811.

\bibitem{tu2018pay}
Z.~Tu, M.~Li, J.~Lin, Pay-per-request deployment of neural network models using
  serverless architectures, in: Proceedings of the 2018 Conference of the North
  American Chapter of the Association for Computational Linguistics:
  Demonstrations, 2018, pp. 6--10.

\bibitem{bilal2021great}
M.~Bilal, M.~Canini, R.~Fonseca, R.~Rodrigues, With great freedom comes great
  opportunity: Rethinking resource allocation for serverless functions, arXiv
  preprint arXiv:2105.14845 (2021).

\bibitem{douceur2001optimizing}
J.~R. Douceur, R.~P. Wattenhofer, Optimizing file availability in a secure
  serverless distributed file system, in: Proceedings 20th IEEE Symposium on
  Reliable Distributed Systems, IEEE, 2001, pp. 4--13.

\bibitem{nastic2017serverless}
S.~Nastic, T.~Rausch, O.~Scekic, S.~Dustdar, M.~Gusev, B.~Koteska, M.~Kostoska,
  B.~Jakimovski, S.~Ristov, R.~Prodan, A serverless real-time data analytics
  platform for edge computing, IEEE Internet Computing 21~(4) (2017) 64--71.

\bibitem{glikson2017deviceless}
A.~Glikson, S.~Nastic, S.~Dustdar, Deviceless edge computing: extending
  serverless computing to the edge of the network, in: Proceedings of the 10th
  ACM International Systems and Storage Conference, 2017, pp. 1--1.

\bibitem{baresi2019towards}
L.~Baresi, D.~F. Mendon{\c{c}}a, Towards a serverless platform for edge
  computing, in: 2019 IEEE International Conference on Fog Computing (ICFC),
  IEEE, 2019, pp. 1--10.

\bibitem{baresi2017empowering}
L.~Baresi, D.~F. Mendon{\c{c}}a, M.~Garriga, Empowering low-latency
  applications through a serverless edge computing architecture, in: European
  Conference on Service-Oriented and Cloud Computing, Springer, 2017, pp.
  196--210.

\bibitem{cicconetti2020decentralized}
C.~Cicconetti, M.~Conti, A.~Passarella, A decentralized framework for
  serverless edge computing in the internet of things, IEEE Transactions on
  Network and Service Management 18~(2) (2020) 2166--2180.

\bibitem{patros2021toward}
P.~Patros, J.~Spillner, A.~V. Papadopoulos, B.~Varghese, O.~Rana, S.~Dustdar,
  Toward sustainable serverless computing, IEEE Internet Computing 25~(6)
  (2021) 42--50.

\bibitem{gill2020quantum}
S.~S. Gill, A.~Kumar, H.~Singh, M.~Singh, K.~Kaur, M.~Usman, R.~Buyya, Quantum
  computing: A taxonomy, systematic review and future directions, Software:
  Practice and Experience 52~(1) (2022) 66--114.

\bibitem{chuang}
M.~Nielsen, I.~Chuang, Quantum Computation and Quantum Information, Cambridge
  University Press, 2010.

\bibitem{feynman}
R.~Feynman, Simulating physics with computers, International Journal of
  Theoretical Physics 21 (1981) 467.

\bibitem{Deutsch}
D.~Deutsch, Quantum theory, the church–turing principle and the universal
  quantum computer, Proceedings of the Royal Society of London. A. Mathematical
  and Physical Sciences 400~(1818) (1985) 97.

\bibitem{shor}
P.~Shor, Polynomial-time algorithms for prime factorization and discrete loga-
  rithms on a quantum computer, SIAM Review 41~(2) (1999) 303--332.

\bibitem{googleQ19}
F.~Arute, K.~Arya, R.~Babbush, D.~Bacon, J.~C. Bardin, R.~Barends, R.~Biswas,
  S.~Boixo, F.~G. Brandao, D.~A. Buell, et~al., Quantum supremacy using a
  programmable superconducting processor, Nature 574~(7779) (2019) 505--510.

\bibitem{BennetNP-hard}
C.~H. Bennett, E.~Bernstein, G.~Brassard, U.~Vazirani, Strengths and weaknesses
  of quantum computing, SIAM journal on Computing 26~(5) (1997) 1510--1523.

\bibitem{grover}
L.~K. Grover., Quantum mechanics helps in searching for a needle in a haystack,
  Physical Review Letters 79~(2) (1997) 325.

\bibitem{ionQ}
J.~M. e.~a. Pino, Demonstration of the trapped-ion quantum ccd computer
  architecture, Nature 592 (2021) 209–213.

\bibitem{psiquantum}
E.~Gibney, Quantum gold rush: the private funding pouring into quantum
  start-ups, Nature 574 (2019) 22--24.

\bibitem{EagleIBM}
P.~Ball, First quantum computer to pack 100 qubits enters crowded race, Nature
  599 (2021) 542.

\bibitem{padhi2021quantum}
P.~K. Padhi, F.~Charrua-Santos, Quantum biotech and internet of virus things:
  Towards a theoretical framework, Applied System Innovation 4~(2) (2021) 27.

\bibitem{foulkes2001quantum}
W.~Foulkes, L.~Mitas, R.~Needs, G.~Rajagopal, Quantum monte carlo simulations
  of solids, Reviews of Modern Physics 73~(1) (2001) 33.

\bibitem{caputo2015pan}
A.~Caputo, V.~Merhej, K.~Georgiades, P.-E. Fournier, O.~Croce, C.~Robert,
  D.~Raoult, Pan-genomic analysis to redefine species and subspecies based on
  quantum discontinuous variation: the klebsiella paradigm, Biology Direct
  10~(1) (2015) 1--12.

\bibitem{di2016decoding}
M.~Di~Ventra, M.~Taniguchi, Decoding dna, rna and peptides with quantum
  tunnelling, Nature nanotechnology 11~(2) (2016) 117--126.

\bibitem{spector1999finding}
L.~Spector, H.~Barnum, H.~J. Bernstein, N.~Swamy, Finding a
  better-than-classical quantum and/or algorithm using genetic programming, in:
  Proceedings of the 1999 Congress on Evolutionary Computation-CEC99 (Cat. No.
  99TH8406), Vol.~3, IEEE, 1999, pp. 2239--2246.

\bibitem{protein-folding}
A.~e.~a. Robert, Resource-efficient quantum algorithm for protein folding, npj
  Quantum information 7~(38) (2021).

\bibitem{richards2013quantum}
W.~G. Richards, Quantum pharmacology, Elsevier, 2013.

\bibitem{Toshiba}
\href{https://www.reuters.com/article/us-toshiba-cyber-idUSKBN2730KW}{Toshiba
  targets \$3 billion revenue in quantum cryptography by 2030} (2020).
\newline\urlprefix\url{https://www.reuters.com/article/us-toshiba-cyber-idUSKBN2730KW}

\bibitem{NIST}
\href{https://csrc.nist.gov/projects/post-quantum-cryptography}{Post-quantum
  cryptography} (2020).
\newline\urlprefix\url{https://csrc.nist.gov/projects/post-quantum-cryptography}

\bibitem{singh2021quantum}
M.~Singh, et~al., Quantum artificial intelligence for the science of climate
  change, arXiv preprint arXiv:2108.10855 (2021).

\bibitem{saggioetal2021}
V.~Saggio, B.~E. Asenbeck, A.~Hamann, T.~Strömberg, P.~Schiansky, V.~Dunjko,
  N.~Friis, N.~C. Harris, M.~Hochberg, D.~Englund, S.~Wölk, H.~J. Briegel,
  P.~Walther, Experimental quantum speed-up in reinforcement learning agents,
  Nature 591 (2021) 229--236.

\bibitem{sgarbas2007road}
K.~N. Sgarbas, The road to quantum artificial intelligence, arXiv preprint
  arXiv:0705.3360 (2007).

\bibitem{biamonte2017quantum}
J.~Biamonte, P.~Wittek, N.~Pancotti, P.~Rebentrost, N.~Wiebe, S.~Lloyd, Quantum
  machine learning, Nature 549~(7671) (2017) 195--202.

\bibitem{huang_broughton_mohseni_babbush_boixo_neven_mcclean_2021}
H.-Y. Huang, M.~Broughton, M.~Mohseni, R.~Babbush, S.~Boixo, H.~Neven, J.~R.
  McClean, Power of data in quantum machine learning, Nature Communications
  12~(1) (2021).
\newblock \href {https://doi.org/10.1038/s41467-021-22539-9}
  {\path{doi:10.1038/s41467-021-22539-9}}.

\bibitem{ayanzadeh2018quantum}
R.~Ayanzadeh, Quantum artificial intelligence for natural language processing
  applications, in: Proceedings of the 49th ACM Technical Symposium on Computer
  Science Education, 2018, pp. 273--273.

\bibitem{meichanetzidis2020quantum}
K.~Meichanetzidis, S.~Gogioso, G.~De~Felice, N.~Chiappori, A.~Toumi, B.~Coecke,
  Quantum natural language processing on near-term quantum computers, arXiv
  preprint arXiv:2005.04147 (2020).

\bibitem{sacha2013artificial}
G.~M. Sacha, P.~Varona, Artificial intelligence in nanotechnology,
  Nanotechnology 24~(45) (2013) 452002.

\bibitem{broughton2020tensorflow}
M.~Broughton, G.~Verdon, T.~McCourt, A.~J. Martinez, J.~H. Yoo, S.~V. Isakov,
  P.~Massey, R.~Halavati, M.~Y. Niu, A.~Zlokapa, et~al., Tensorflow quantum: A
  software framework for quantum machine learning, arXiv preprint
  arXiv:2003.02989 (2020).

\bibitem{dunjko2018machine}
V.~Dunjko, H.~J. Briegel, Machine learning \& artificial intelligence in the
  quantum domain: a review of recent progress, Reports on Progress in Physics
  81~(7) (2018) 074001.

\bibitem{holmes2021assessment}
S.~Holmes, L.~Chen, Assessment of quantum threat to bitcoin and derived
  cryptocurrencies, Cryptology ePrint Archive (2021).

\bibitem{toldinas2019mqtt}
J.~Toldinas, B.~Lozinskis, E.~Baranauskas, A.~Dobrovolskis, Mqtt quality of
  service versus energy consumption, in: 2019 23rd International Conference
  Electronics, IEEE, 2019, pp. 1--4.

\bibitem{dick2019embedded}
R.~P. Dick, L.~Shang, M.~Wolf, S.-W. Yang, Embedded intelligence in the
  internet-of-things, IEEE Design \& Test 37~(1) (2019) 7--27.

\bibitem{guo2013internet}
B.~Guo, D.~Zhang, Z.~Yu, Y.~Liang, Z.~Wang, X.~Zhou, From the internet of
  things to embedded intelligence, World Wide Web 16~(4) (2013) 399--420.

\bibitem{dai2019industrial}
W.~Dai, H.~Nishi, V.~Vyatkin, V.~Huang, Y.~Shi, X.~Guan, Industrial edge
  computing: Enabling embedded intelligence, IEEE Industrial Electronics
  Magazine 13~(4) (2019) 48--56.

\bibitem{vermesan2018next}
O.~Vermesan, M.~EisenHauer, M.~Serrano, P.~Guillemin, H.~Sundmaeker, E.~Z.
  Tragos, J.~Valino, B.~Copigneaux, M.~Presser, A.~Aagaard, et~al., The next
  generation internet of things--hyperconnectivity and embedded intelligence at
  the edge, Next Generation Internet of Things. Distributed Intelligence at the
  Edge and Human Machine-to-Machine Cooperation (2018).

\bibitem{plastiras2018edge}
G.~Plastiras, M.~Terzi, C.~Kyrkou, T.~Theocharidcs, Edge intelligence:
  Challenges and opportunities of near-sensor machine learning applications,
  in: 2018 ieee 29th international conference on application-specific systems,
  architectures and processors (asap), IEEE, 2018, pp. 1--7.

\bibitem{gobieski2019intelligence}
G.~Gobieski, B.~Lucia, N.~Beckmann, Intelligence beyond the edge: Inference on
  intermittent embedded systems, in: Proceedings of the Twenty-Fourth
  International Conference on Architectural Support for Programming Languages
  and Operating Systems, 2019, pp. 199--213.

\bibitem{anwar2021recommender}
M.~H. Anwar, et~al., Recommender system for optimal distributed deep learning
  in cloud datacenters, Wireless Personal Communications (2021) 1--25.

\bibitem{Xue2021EosDNN}
M.~Xue, H.~Wu, R.~Li, M.~Xu, P.~Jiao, Eosdnn: An efficient offloading scheme
  for dnn inference acceleration in local-edge-cloud collaborative
  environments, IEEE Transactions on Green Communications and Networking (2021)
  1--1\href {https://doi.org/10.1109/TGCN.2021.3111731}
  {\path{doi:10.1109/TGCN.2021.3111731}}.

\bibitem{gomes2017survey}
H.~M. Gomes, J.~P. Barddal, F.~Enembreck, A.~Bifet, A survey on ensemble
  learning for data stream classification, ACM Computing Surveys (CSUR) 50~(2)
  (2017) 1--36.

\bibitem{chen2013internet}
Y.~Chen, H.~Hu, Internet of intelligent things and robot as a service,
  Simulation Modelling Practice and Theory 34 (2013) 159--171.

\bibitem{arsenio2014internet}
A.~Ars{\'e}nio, H.~Serra, R.~Francisco, F.~Nabais, J.~Andrade, E.~Serrano,
  Internet of intelligent things: Bringing artificial intelligence into things
  and communication networks, in: Inter-cooperative collective intelligence:
  Techniques and applications, Springer, 2014, pp. 1--37.

\bibitem{nathani2017internet}
B.~Nathani, R.~Vijayvergia, The internet of intelligent things: An overview,
  in: 2017 International Conference on Intelligent Communication and
  Computational Techniques (ICCT), IEEE, 2017, pp. 119--122.

\bibitem{wazid2020tutorial}
M.~Wazid, A.~K. Das, S.~Shetty, M.~Jo, A tutorial and future research for
  building a blockchain-based secure communication scheme for internet of
  intelligent things, IEEE Access 8 (2020) 88700--88716.

\bibitem{chen2019intelligent}
N.~Chen, T.~Qiu, X.~Zhou, K.~Li, M.~Atiquzzaman, An intelligent robust
  networking mechanism for the internet of things, IEEE Communications Magazine
  57~(11) (2019) 91--95.

\bibitem{9126779}
M.~K. Choi, C.~Y. Yeun, P.~H. Seong, A novel monitoring system for the data
  integrity of reactor protection system using blockchain technology, IEEE
  Access 8 (2020) 118732--118740.
\newblock \href {https://doi.org/10.1109/ACCESS.2020.3005134}
  {\path{doi:10.1109/ACCESS.2020.3005134}}.

\bibitem{8071359}
I.~Zikratov, A.~Kuzmin, V.~Akimenko, V.~Niculichev, L.~Yalansky, Ensuring data
  integrity using blockchain technology, in: 2017 20th Conference of Open
  Innovations Association (FRUCT), 2017, pp. 534--539.
\newblock \href {https://doi.org/10.23919/FRUCT.2017.8071359}
  {\path{doi:10.23919/FRUCT.2017.8071359}}.

\bibitem{doyleblockchainbus}
J.~Doyle, et~al., Blockchainbus: A lightweight framework for secure virtual
  machine migration in cloud federations using blockchain, Security and Privacy
  (2021) e197.

\bibitem{hammi2018bubbles}
M.~T. Hammi, B.~Hammi, P.~Bellot, A.~Serhrouchni, Bubbles of trust: A
  decentralized blockchain-based authentication system for iot, Computers \&
  Security 78 (2018) 126--142.

\bibitem{abaid2019health}
Z.~Abaid, A.~Shaghaghi, R.~Gunawardena, S.~Seneviratne, A.~Seneviratne, S.~Jha,
  Health access broker: Secure, patient-controlled management of personal
  health records in the cloud, in: Computational Intelligence in Security for
  Information Systems Conference, Springer, 2019, pp. 111--121.

\bibitem{sisi2021blockchain}
Z.~Sisi, A.~Souri, Blockchain technology for energy-aware mobile crowd sensing
  approaches in internet of things, Transactions on Emerging Telecommunications
  Technologies (2021) e4217.

\bibitem{IoTpi}
T.~Shao, et~al.,
  \href{https://onlinelibrary.wiley.com/doi/abs/10.1002/itl2.355}{Iot-pi: A
  machine learning-based lightweight framework for cost-effective distributed
  computing using iot}, Internet Technology Letters (2022) e355\href
  {http://arxiv.org/abs/https://onlinelibrary.wiley.com/doi/pdf/10.1002/itl2.355}
  {\path{arXiv:https://onlinelibrary.wiley.com/doi/pdf/10.1002/itl2.355}},
  \href {https://doi.org/https://doi.org/10.1002/itl2.355}
  {\path{doi:https://doi.org/10.1002/itl2.355}}.
\newline\urlprefix\url{https://onlinelibrary.wiley.com/doi/abs/10.1002/itl2.355}

\bibitem{Xue2021DDPQN}
M.~Xue, H.~Wu, G.~Peng, K.~Wolter, Ddpqn: An efficient dnn offloading strategy
  in local-edge-cloud collaborative environments, IEEE Transactions on Services
  Computing (2021) 1--1\href {https://doi.org/10.1109/TSC.2021.3116597}
  {\path{doi:10.1109/TSC.2021.3116597}}.

\bibitem{sha2017empirical}
M.~Sha, D.~Gunatilaka, C.~Wu, C.~Lu, Empirical study and enhancements of
  industrial wireless sensor--actuator network protocols, IEEE Internet of
  Things Journal 4~(3) (2017) 696--704.

\bibitem{liu2003state}
J.~Liu, M.~Chu, J.~Reich, F.~Zhao, State-centric programming for
  sensor-actuator network systems, IEEE Pervasive Computing 2~(4) (2003)
  50--62.

\bibitem{cceltek2017internet}
S.~A. {\c{C}}eltek, M.~Durgun, H.~Soy, Internet of things based smart home
  system design through wireless sensor/actuator networks, in: 2017 2nd
  International Conference on Advanced Information and Communication
  Technologies (AICT), IEEE, 2017, pp. 15--18.

\bibitem{van1993sensor}
M.~Van~de Panne, E.~Fiume, Sensor-actuator networks, in: Proceedings of the
  20th annual conference on Computer graphics and interactive techniques, 1993,
  pp. 335--342.

\bibitem{deshmukh2018monitoring}
P.~V.~M. Deshmukh, Monitoring and control of gas leakages of industrial sector
  using pic 18f4550, zigbee and wireless sensor actuator network, i-Manager’s
  Journal on Electronics Engineering 8~(3) (2018) 5.

\bibitem{joshi2020performance}
T.~Joshi, K.~Nagiya, M.~Ram, Performance evaluation of a wireless sensor
  actuator network under reliability approach., Mathematics in Engineering,
  Science \& Aerospace (MESA) 11~(1) (2020).

\bibitem{shi2018digs}
J.~Shi, M.~Sha, Z.~Yang, Digs: Distributed graph routing and scheduling for
  industrial wireless sensor-actuator networks, in: 2018 IEEE 38th
  International Conference on Distributed Computing Systems (ICDCS), IEEE,
  2018, pp. 354--364.

\bibitem{bragarenco2020sensor}
A.~Bragarenco, Sensor-actuator software component stack for industrial internet
  of things applications, in: 2020 24th International Conference on System
  Theory, Control and Computing (ICSTCC), IEEE, 2020, pp. 540--545.

\bibitem{muralidhara2020air}
S.~Muralidhara, N.~Hegde, Air quality monitoring and gas leakage detection with
  automatic shut-off using wireless sensor-actuator networks, Internet
  Technology Letters 3~(5) (2020) e185.

\bibitem{linardatos2021explainable}
P.~Linardatos, V.~Papastefanopoulos, S.~Kotsiantis, Explainable ai: A review of
  machine learning interpretability methods, Entropy 23~(1) (2021) 18.

\bibitem{gohel2021explainable}
P.~Gohel, P.~Singh, M.~Mohanty, Explainable ai: current status and future
  directions, arXiv preprint arXiv:2107.07045 (2021).

\bibitem{liao2021introduction}
Q.~V. Liao, M.~Singh, Y.~Zhang, R.~Bellamy, Introduction to explainable ai, in:
  Extended Abstracts of the 2021 CHI Conference on Human Factors in Computing
  Systems, 2021, pp. 1--3.

\bibitem{souri2020hybrid}
A.~Souri, A.~M. Rahmani, N.~J. Navimipour, R.~Rezaei, A hybrid formal
  verification approach for qos-aware multi-cloud service composition, Cluster
  Computing 23~(4) (2020) 2453--2470.

\bibitem{zhang2021explainable}
K.~Zhang, J.~Zhang, P.-D. Xu, T.~Gao, D.~W. Gao, Explainable ai in deep
  reinforcement learning models for power system emergency control, IEEE
  Transactions on Computational Social Systems (2021).

\bibitem{gartner}
\href{https://www.gartner.com/en/articles/5-impactful-technologies-from-the-gartner-emerging-technologies-and-trends-impact-radar-for-2022}{5
  impactful technologies from the gartner emerging technologies and trends
  impact radar for 2022} (2021).
\newline\urlprefix\url{https://www.gartner.com/en/articles/5-impactful-technologies-from-the-gartner-emerging-technologies-and-trends-impact-radar-for-2022}

\bibitem{gartner1}
\href{https://www.gartner.com/en/newsroom/press-releases/2021-10-18-gartner-identifies-the-top-strategic-technology-trends-for-2022}{Gartner
  identifies the top strategic technology trends for 2022} (2021).
\newline\urlprefix\url{https://www.gartner.com/en/newsroom/press-releases/2021-10-18-gartner-identifies-the-top-strategic-technology-trends-for-2022}

\bibitem{ZDNET}
\href{https://www.zdnet.com/article/generative-ai-autonomic-systems-hyperautomation-and-more-top-gartner-list-of-top-tech-trends-in-2022/}{Generative
  ai, autonomic systems, hyperautomation and more top gartner list of top tech
  trends in 2022} (2021).
\newline\urlprefix\url{https://www.zdnet.com/article/generative-ai-autonomic-systems-hyperautomation-and-more-top-gartner-list-of-top-tech-trends-in-2022/}

\bibitem{KOCHOVSKI2019747}
P.~Kochovski, S.~Gec, V.~Stankovski, M.~Bajec, P.~D. Drobintsev,
  \href{https://www.sciencedirect.com/science/article/pii/S0167739X19301281}{Trust
  management in a blockchain based fog computing platform with trustless smart
  oracles}, Future Generation Computer Systems 101 (2019) 747--759.
\newblock \href {https://doi.org/https://doi.org/10.1016/j.future.2019.07.030}
  {\path{doi:https://doi.org/10.1016/j.future.2019.07.030}}.
\newline\urlprefix\url{https://www.sciencedirect.com/science/article/pii/S0167739X19301281}

\end{thebibliography}

% \newpage
\section*{Authors Biography}

\parpic{\includegraphics[width=0.8in,clip,keepaspectratio]{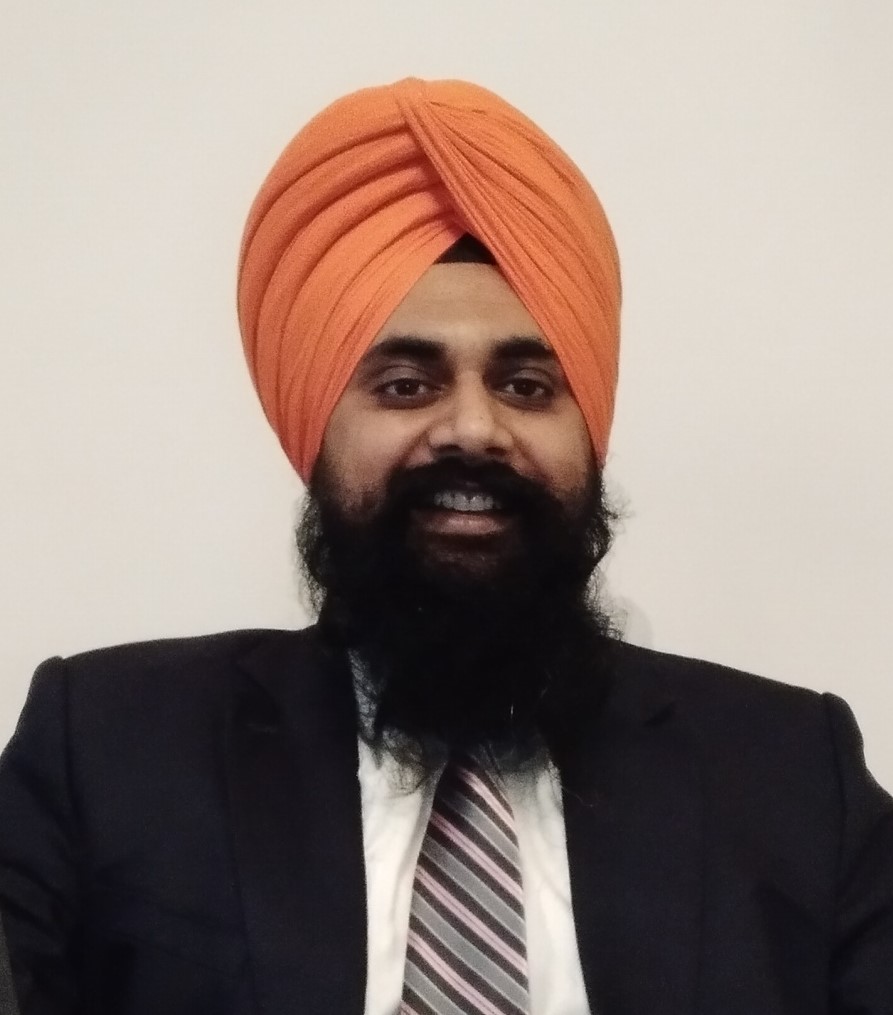}}
\noindent {\footnotesize \sffamily{\bf Sukhpal Singh Gill} is a Lecturer (Assistant Professor) in Cloud Computing at the School of Electronic Engineering and Computer Science, Queen Mary University of London, UK. Prior to his present stint, Dr. Gill has held positions as a Research Associate at the School of Computing and Communications, Lancaster University, UK and also as a Postdoctoral Research Fellow at CLOUDS Laboratory, The University of Melbourne, Australia. Dr. Gill received his Bachelor’s degree in Computer Science and Engineering from Punjab Technical University with Distinction in 2010. Then, he obtained the Degree of Master of Engineering in Software Engineering (Gold Medalist), as well as a Doctoral Degree specialization in Autonomic Cloud Computing from Thapar University. He was a DST (Department of Science \& Technology) Inspire Fellow during his doctoral studies and also worked as a Senior Research Fellow (Professional) on a DST Project sponsored by the Government of India. One of his research paper has been nominated and selected for the ACM's 21st annual Best of Computing Notable Books and Articles as one of the notable items published in computing – 2016.  Dr. Gill was a research visitor at Monash University, University of Manitoba, University of Manchester and Imperial College London. Dr. Gill is serving as an Associate Editor in Wiley ETT and IET Networks Journal. He has co-authored 70+ peer-reviewed papers (with H-index 30+) and has published in prominent international journals and conferences such as IEEE TCC, IEEE TSC, IEEE TII, IEEE IoT Journal, Elsevier JSS and IEEE CCGRID. He has received several awards, including the Distinguished Reviewer Award from SPE (Wiley), 2018, Best Paper Award AusPDC at ACSW 2021 and has also served as the PC member for venues such as PerCom, UCC, CCGRID, CLOUDS, ICFEC, AusPDC. Dr. Gill served as a Guest Editor for SPE (Wiley) and JCC Springer Journal. He is a regular reviewer for IEEE TPDS, IEEE TSC, IEEE TNSE, IEEE TSC, ACM CSUR and Wiley SPE.  He has edited a research book for Elsevier. Dr. Gill has reviewed 370+ research articles of high ranked journals and prestigious conferences according to the data from Publons. His research interests include Cloud Computing, Fog Computing, Software Engineering, Internet of Things and Energy Efficiency. For further information, please visit \url{http://www.ssgill.me}}. \\

\parpic{\includegraphics[width=0.8in,clip,keepaspectratio]{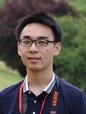}}
\noindent {\footnotesize \sffamily{\bf Minxian Xu} is currently an associate professor at Shenzhen Institute of Advanced Technology, Chinese Academy of Sciences. He received the B.Sc. degree and the M.Sc. degree, both in software engineering from University of Electronic Science and Technology of China. He obtained his Ph.D. degree from the University of Melbourne in 2019. His research interests include resource scheduling and optimization in cloud computing. He has co-authored 30+ peer-reviewed papers published in prominent international journals and conferences, such as ACM Computing Surveys, IEEE Transactions on Sustainable Computing, IEEE Transactions on Cloud Computing, Journal of Parallel and Distributed Computing, Software: Practice and Experience, International Conference on Service-Oriented Computing. His Ph.D. Thesis was awarded the 2019 IEEE TCSC Outstanding Ph.D. Dissertation Award. He is member of CCF and IEEE. More information can be found at: \url{http://www.minxianxu.info}}. \\

\parpic{\includegraphics[width=0.8in,clip,keepaspectratio]{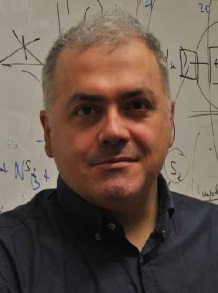}}
\noindent {\footnotesize \sffamily{\bf Carlo Ottaviani} is Associate Lecturer at Department of Computer Science and Research Fellow at the York Centre for Quantum Technologies (YCQT) of the University of York, UK. His research activity is in Quantum Information and Quantum Technologies. He worked on theoretical proposals to design quantum-phase gates, quantum memories and quantum repeaters using non-linear quantum atom-optics systems. In recent years he focused on security, quantum-key distribution, and the realisation of quantum relays and quantum networks based on quantum continuous-variable systems.} \\

\parpic{\includegraphics[width=0.8in,clip,keepaspectratio]{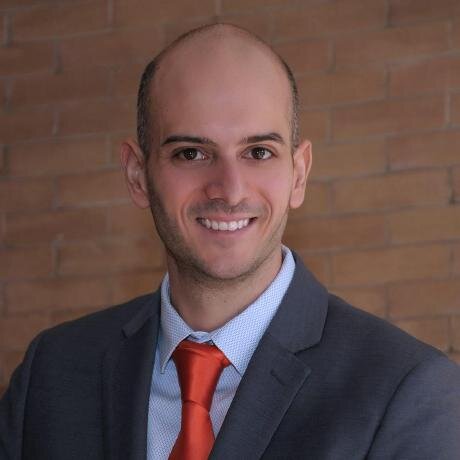}}
\noindent {\footnotesize \sffamily{\bf Panos Patros} is a Senior Lecturer with the Department of Software Engineering, University of Waikato, Aotearoa New Zealand. He leads the Cloud and Adaptive Systems (Ohu Rangahau Kapua Aunoa) ORKA Lab, is a member of the Ahuora Smart Energy Systems Centre and focuses on self-adaptive computer systems and digital twins. He received the Ph.D. degree in computer science from the University of New Brunswick, Fredericton, NB, Canada, on multitenancy, performance, and modeling of cloud systems. Contact him at panos.patros@waikato.ac.nz.} \\

\parpic{\includegraphics[width=0.8in,clip,keepaspectratio]{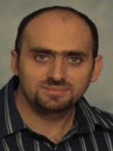}}
\noindent {\footnotesize \sffamily{\bf Rami Bahsoon} is is a Reader at the School of Computer Science, University of Birmingham, UK. Bahsoon’s research is in the area of software architecture, cloud and services software engineering, self-aware software architectures, self-adaptive and managed software engineering, economics-driven software engineering and technical debt management in software. He co-edited four books on Software Architecture, including Economics-Driven Software Architecture; Software Architecture for Big Data and the Cloud; Aligning Enterprise, System, and Software Architecture. He was a Visiting Scientist at the Software Engineering Institute (SEI), Carnegie Mellon University, USA (June–August 2018) and was the 2018 Melbourne School of Engineering (MSE) Visiting Fellow of The School of Computing and Information Systems, the University of Melbourne (August to Nov 2018). He holds a Ph.D. in Software Engineering from University College London (2006) and was MBA Fellow in Technology at London Business School (2003–2005). He is a fellow of the Royal Society of Arts and Associate Editor of IEEE Software - Software Economies}. \\

\parpic{\includegraphics[width=0.8in,clip,keepaspectratio]{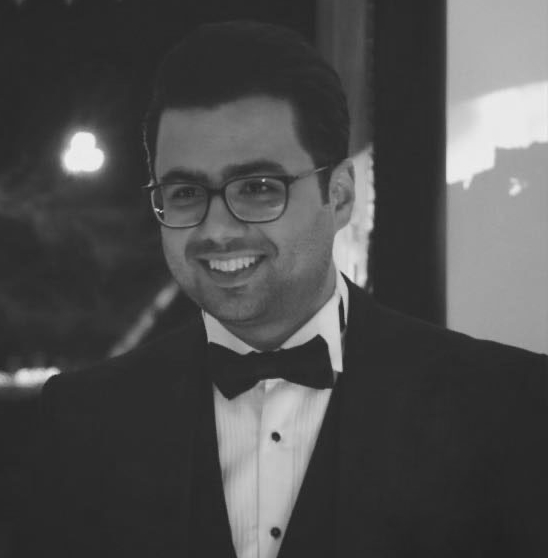}}
\noindent {\footnotesize \sffamily{\bf Arash Shaghaghi} received the B.Sc. degree from Heriot-Watt University, the M.Sc. degree in information security from University College London (UCL), and the PhD degree in Computer Science and Engineering from The University of New South Wales (UNSW Sydney), Australia. He is currently a Senior Lecturer in Cyber Security at RMIT University, Melbourne, Australia. He is also a Visiting Fellow at the School of Computer Science and Engineering, UNSW Sydney. He has previously been affiliated with Deakin University, UNSW Sydney, Data61 CSIRO, The University of Melbourne, and The University of Texas at Dallas. He is a multi-award winner cyber security educator and researcher with a track record of publications at competitive international conferences and journals. To this date, he has received a total funding of more than 300,000 AUD (as PI and CI combined) for his cyber security research from various internal and external sources, including the Australian Government. Arash currently serves as an Associate Editor for Ad Hoc Networks journal and has had roles (TPC member, organizing member, and reviewer) at prestigious security and networking journals and conferences. He is a member of IEEE and the Australian Information Security Association. Visit \url{http://www.arashshaghaghi.com} for more information}. \\

\parpic{\includegraphics[width=0.8in,clip,keepaspectratio]{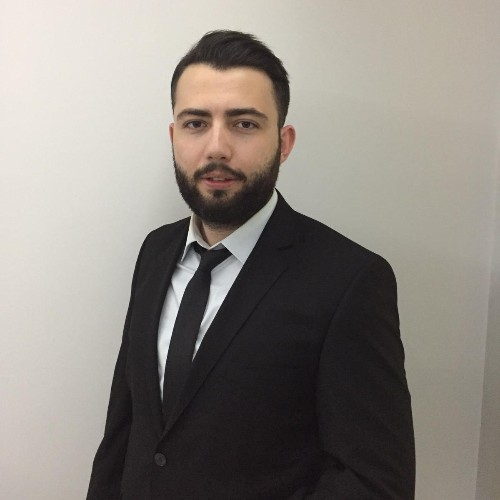}}
\noindent {\footnotesize \sffamily{\bf Muhammed Golec} is a PhD student in Computer Science at Queen Mary University. After his undergraduate graduation, he was awarded the Ministry of Education Scholarship, one of the most prestigious scholarships in his country. Within the scope of this scholarship, he graduated from Queen Mary University of London Computer Science with a high degree (Distinction). His master thesis was found successful and published in IEEE Consumer Electronics Magazine. He worked at Sisecam Company as an Electrical and Electronics Maintenance Engineer for one year to consolidate his academic skills in the private sector. His research interests include AI, Cloud Computing, and Security and Privacy. For further information, please visit \url{https://www.linkedin.com/in/muhammed-golec-b55756119/}.} \\

\parpic{\includegraphics[width=0.8in,clip,keepaspectratio]{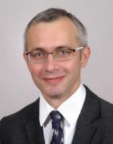}}
\noindent {\footnotesize \sffamily{\bf Vlado Stankovski} is a Full Professor of Computer and Information Science at the University of Ljubljana. Vlado Stankovski was awarded his Eng. Comp. Sc., M.Sc. and Ph.D. degrees in computer science from the University of Ljubljana in 1995, 2000 and 2009, respectively. He began his career in 1995 as consultant and later as project manager with the Fujitsu-ICL Corporation in Prague. From 1998-2002 he worked as researcher at the University Medical Centre in Ljubljana. In the period 2003-2019, he was with the Department of Construction Informatics at the University of Ljubljana. From 2020 he lectures at all levels (undergraduate, postgraduate) at the Faculty of Computer and Information Science. Vlado Stankovski's research interests are in semantic and distributed-computing technologies. He is currently the scientific and technical coordinator of the Horizon Europe Next Generation Internet project ONTOCHAIN.} \\

\parpic{\includegraphics[width=0.8in,clip,keepaspectratio]{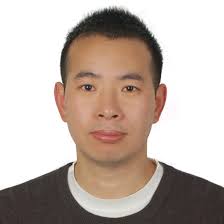}}
\noindent {\footnotesize \sffamily{\bf Huaming Wu} received the BE and MS degrees from the Harbin Institute of Technology, China, in 2009 and 2011, respectively, both in electrical engineering, and the PhD degree with the highest honor in computer science at Freie Universit\"at Berlin, Germany, in 2015. He is currently an associate professor with the Center for Applied Mathematics, Tianjin University. His research interests include model-based evaluation, wireless and mobile network systems, mobile cloud computing and deep learning. He is a member of IEEE and ACM.} \\

\parpic{\includegraphics[width=0.8in,clip,keepaspectratio]{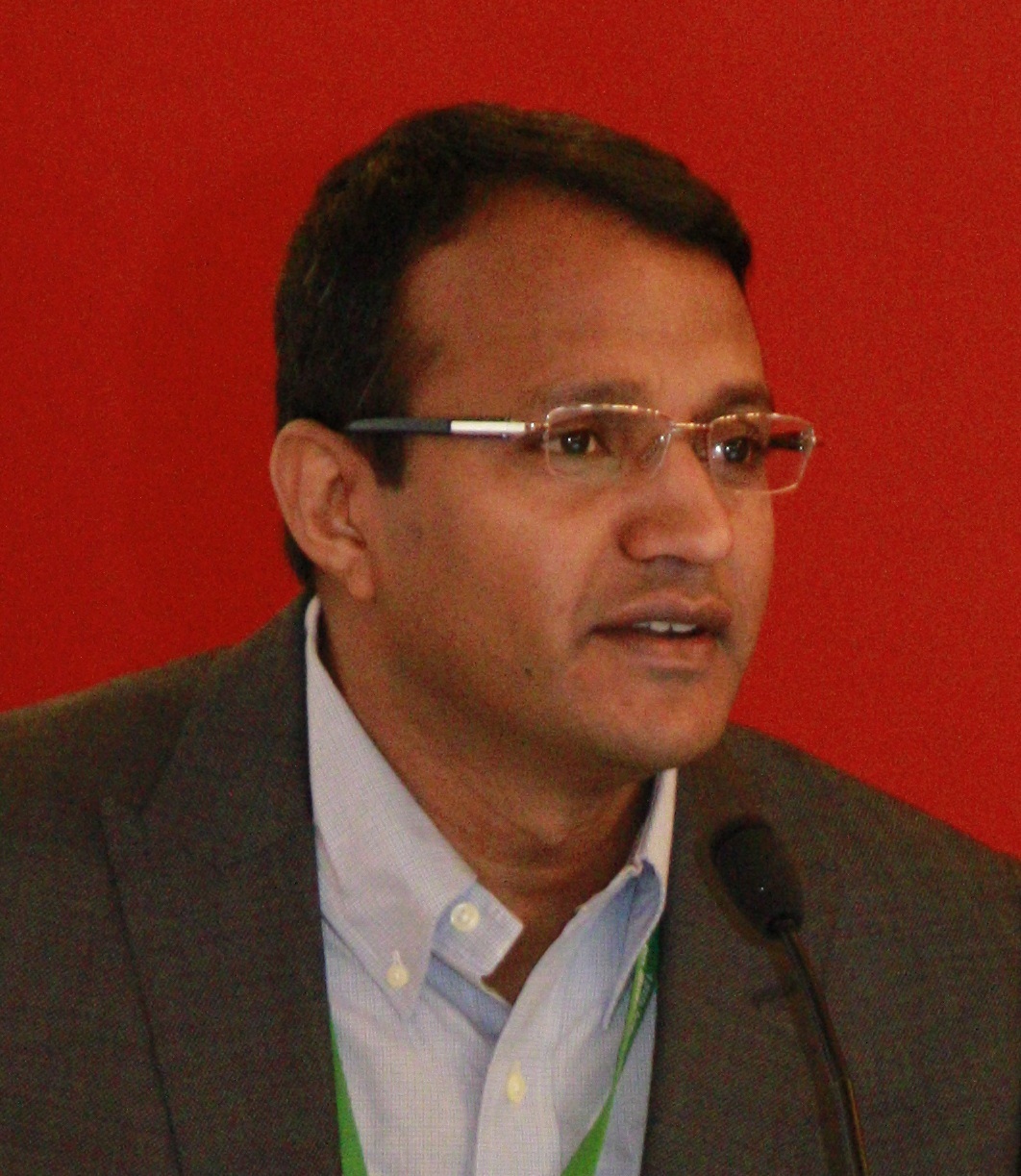}}
\noindent {\footnotesize \sffamily{\bf Ajith Abraham} received the M.Sc. degree from Nanyang Technological University, Singapore, in 1998, and the Ph.D. degree in computer science from Monash University, Melbourne, Australia, in 2001. He is currently the Director of the Machine Intelligence Research Labs (MIR Labs), a Not-for-Profit Scientific Network for Innovation and Research Excellence connecting Industry and Academia. The Network with HQ in Seattle, USA, has currently more than 1000 scientific members from more than 100 countries. As an Investigator/Co-Investigator, he has won research grants worth more than U.S.\$100 million from Australia, USA, EU, Italy, Czech Republic, France, Malaysia, and China. He works in a multi-disciplinary environment involving machine intelligence, cyber-physical systems, the Internet of Things, network security, sensor networks, web intelligence, web services, data mining, and applied to various real-world problems. In these areas, he has authored/coauthored more than 1400 research publications out of which there are more than 100 books covering various aspects of computer science. He has more than 46,000 academic citations (H-index of 100 as per Google Scholar). He is also the Editor-in-Chief of Engineering Applications of Artificial Intelligence (EAAI) and serves/served the editorial board of more than 15 international journals indexed by Thomson ISI. More information can be available at \url{http://www.softcomputing.net}.} \\

\parpic{\includegraphics[width=0.8in,clip,keepaspectratio]{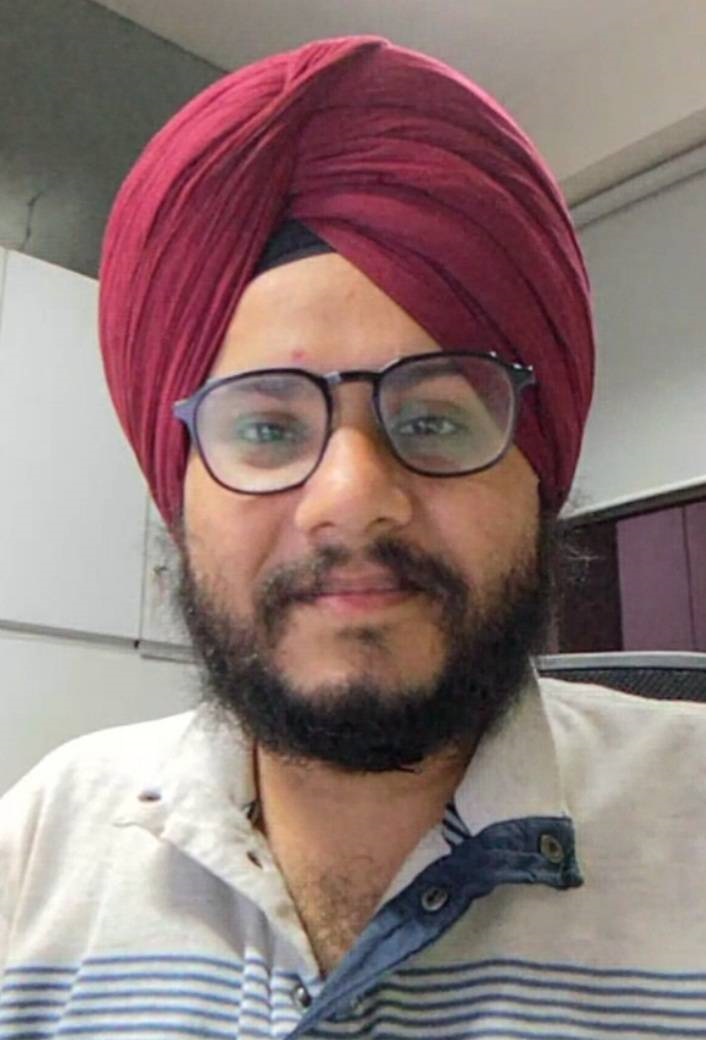}}
\noindent {\footnotesize \sffamily{\bf Manmeet Singh} is Scientist C at the Centre for Climate Change Research, Indian Institute of Tropical Meteorology, Pune since 2013. He was a Fulbright-Kalam fellow at the Jackson School of Geosciences, The University of Texas at Austin in 2021. He is specially interested in AI/ML techniques, causal approaches, recurrence plots, complex networks and non-linear time series analysis for solving grand challenges in Earth System Science. He is an experienced climate modeller having contributed to the IITM Earth System Model simulations towards the IPCC AR6 report. Together with his PhD co-advisor Dr Ayantika Dey Choudhury, he developed and coupled the aerosol module of IITM-ESM. Recently, his work has focussed on the impacts of the proposals suggesting volcanic eruptions as an analogue of solar geoengineering to halt climate change. His international collaborations include researchers from PIK Potsdam, UT Austin, Penn State University, University of Tübingen, Magdeburg-Stendal University of Applied Sciences, Germany.} \\

\parpic{\includegraphics[width=0.8in,clip,keepaspectratio]{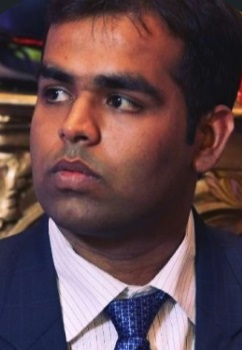}}
\noindent {\footnotesize \sffamily{\bf Harshit Mehta} is Principal Data Scientist with Dell Technologies. He has led many Data Science projects from ideation to production.He has extensive experience in different domains of AI like Natural language Processing, Fraud Detection and Anomaly Detection. Harshit graduated from University of Texas at Austin with a Masters in Operation Research. } \\  \\  \\ \\

\parpic{\includegraphics[width=0.8in,clip,keepaspectratio]{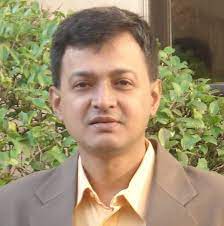}}
\noindent {\footnotesize \sffamily{\bf Soumya K Ghosh} received the M.Tech. and Ph.D. degrees from the Department of Computer Science and Engineering, Indian Institute of Technology (IIT) Kharagpur, India. He is currently a Professor with the Department of Computer Science and Engineering, IIT Kharagpur. He was with the Indian Space Research Organization, India. He has authored or coauthored more than 300 research papers in reputed journals and conference proceedings. His current research interests include spatial data science, spatial web services, and cloud computing. He is the recipient of National Geospatial Chair Professor award from Department of Science and Technology, Govt. of India. He is senior member of IEEE and member of ACM.} \\

\parpic{\includegraphics[width=0.8in,clip,keepaspectratio]{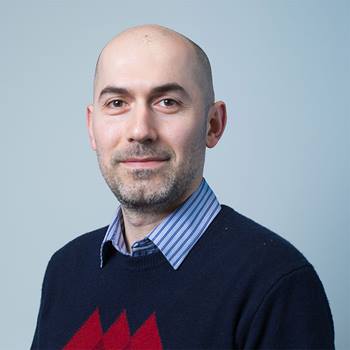}}
\noindent {\footnotesize \sffamily{\bf Thar Baker} is Associate Professor in the Department of Computer Science at The University of Sharjah (UoS) in UAE. Before joining UoS, Thar was Reader in Cloud Engineering and Head of Applied Computing Research Group (ACRG) in the Faculty of Engineering and Technology at Liverpool John Moores University (LJMU, UK). He received his PhD in Autonomic Cloud Applications from LJMU in 2010 and became a Senior Fellow of Higher Education Academy (SFHEA) in 2018. Dr Baker has published numerous refereed research papers in multidisciplinary research areas including parallel and distributed computing, autonomic computing, federated learning, and IoT.} \\

\parpic{\includegraphics[width=0.8in,clip,keepaspectratio]{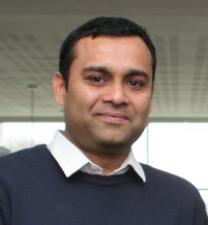}}
\noindent {\footnotesize \sffamily{\bf Ajith Kumar Parlikad} is Professor of Asset Management at Cambridge University Engineering Department. He is based at the Institute for Manufacturing, where he is the Head of the Asset Management research group. He is a Fellow and Tutor at Hughes Hall. Ajith leads research activities on engineering asset management and maintenance. His particular focus is examining how asset information can be used to improve asset performance through effective decision-making. He actively engages with industry through research and consulting projects. He is currently the Scientific Secretary of the IFAC TC5.1 Working Group on `Advanced Maintenance Engineering, Services and Technology' and sits on the steering committee of the UK Digital Twin Hub. Ajith's current research focuses on the development and exploitation of digital twins of complex asset systems bringing together data from disparate sources to improve asset management. Ajith joined Cambridge University to read for his PhD degree, which he successfully completed in August 2006. For his PhD, he developed a methodology for quantifying the benefits of improving product information availability and quality on the effectiveness of product recovery processes.} \\

\parpic{\includegraphics[width=0.8in,clip,keepaspectratio]{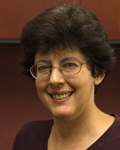}}
\noindent {\footnotesize \sffamily{\bf Hanan Lutfiyya} is a Professor and the Chair of the Department of Computer Science, University of Western Ontario (UWO), Canada. Her research interests include Internet of Things, software engineering, self-adaptive and self-managing systems, autonomic computing, monitoring and diagnostics, mobile systems, policies, and clouds. She was a recipient of the UWO Faculty Scholar Award in 2006. She is a Past Member of the Natural Science and Engineering Research Council of Canada (NSERC) Discovery Grant Committee, and a Past Member and the Chair of an NSERC Strategic Grants Committee. She was a member of the Computer Science Accreditation Council. She is currently an Associate Editor of the IEEE Transactions on Network and Service Management and has recently served as the Program Co-Chair for the IEEE/IFIP Network Operations and Management Symposium and the IEEE International Conference on Network and Service Management. She is currently on the steering committee for the Ontario Celebration of Women in Computing Conference.} \\

\parpic{\includegraphics[width=0.8in,clip,keepaspectratio]{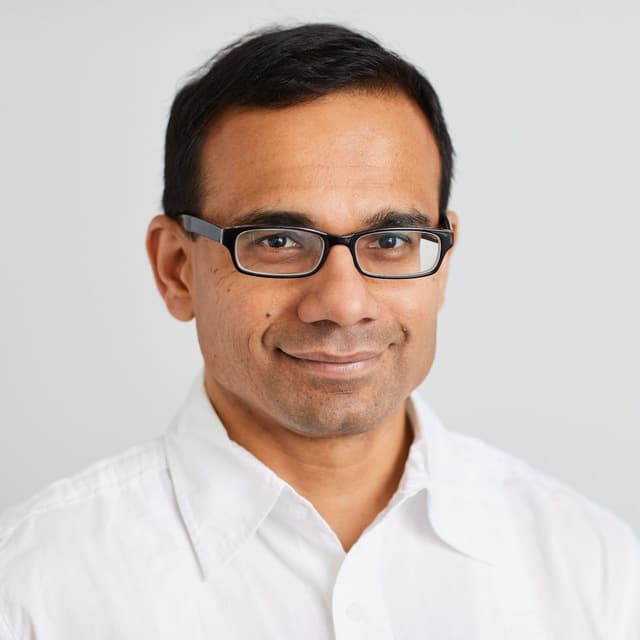}}
\noindent {\footnotesize \sffamily{\bf Salil S. Kanhere} received the MS and PhD degrees from Drexel University, Philadelphia, USA. He is a Professor of Computer Science and Engineering at UNSW Sydney, Australia. His research interests include the Internet of Things, cyber-physical systems, blockchain, pervasive computing, cybersecurity, and applied machine learning. Salil is also affiliated with CISRO’s Data61 and the Cybersecurity Cooperative Research Centre. He is a Senior Member of the IEEE and ACM, an ACM Distinguished Speaker and an IEEE Computer Society Distinguished Visitor. He has received the Friedrich Wilhelm Bessel Research Award (2020) and the Humboldt Research Fellowship (2014), both from the Alexander von Humboldt Foundation in Germany. He has held visiting positions at I2R Singapore, Technical University Darmstadt, University of Zurich and Graz University of Technology. He serves as the Editor in Chief of the Ad Hoc Networks Journal and as an Associate Editor of IEEE Transactions On Network and Service Management, Computer Communications, and Pervasive and Mobile Computing. He has served on the organising committee of several IEEE/ACM international conferences including IEEE PerCom, IEEE/ACM IPSN, IEEE ICBC, IEEE WoWMoM, ACM MSWiM, etc. He has co-authored a book titled Blockchain for Cyberphysical Systems published by Artech House in 2020. Further details are at: \url{https://salilkanhere.net/}.} \\

\parpic{\includegraphics[width=0.8in,clip,keepaspectratio]{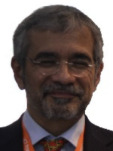}}
\noindent {\footnotesize \sffamily{\bf Rizos Sakellariou} obtained his Ph.D. from the University of Manchester in 1997. Since then he held positions with Rice University and the University of Cyprus, while currently he is Professor with the University of Manchester leading a laboratory that carries out research in High-Performance, Parallel and Distributed systems. He has carried out research on a number of topics related to parallel and distributed computing, with an emphasis on problems stemming from efficient resource utilization and workload allocation and scheduling issues. He has published over 160 research papers, His research has been supported by several UK and EU projects and has been on the Program Committee of over 180 conferences and workshops. He values collaboration and a strong work ethic.} \\

\parpic{\includegraphics[width=0.8in,clip,keepaspectratio]{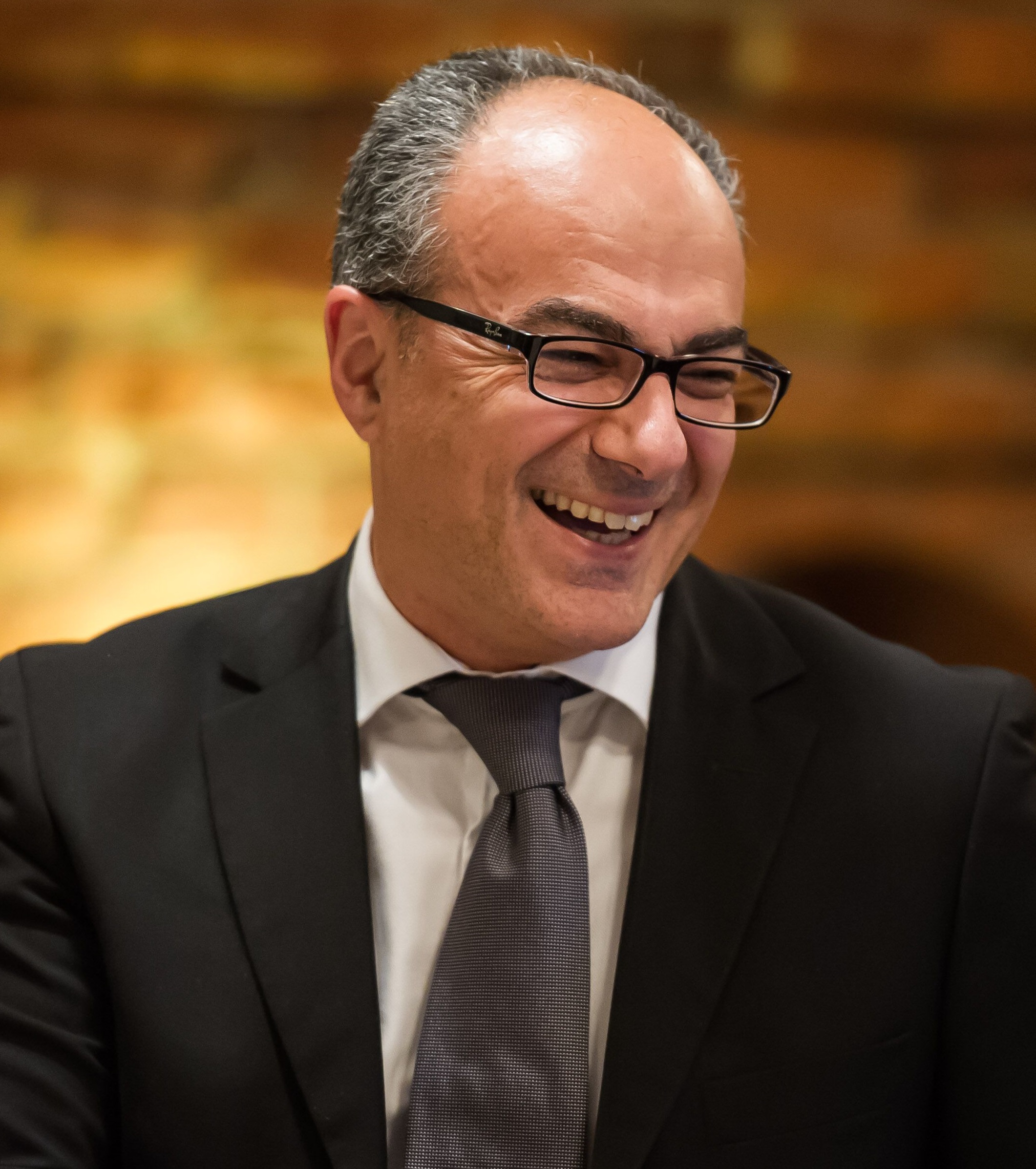}}
\noindent {\footnotesize \sffamily{\bf Schahram Dustdar} is Full Professor of computer science heading the Research Division of Distributed Systems at the TU Wien, Austria. He is founding Co-Editor-in-Chief of the new ACM Transactions on Internet of Things (ACM TIoT) as well as Editor-in-Chief of Computing (Springer). He is an Associate Editor of IEEE Transactions on Services Computing, IEEE Transactions on Cloud Computing, ACM Transactions on the Web, and ACM Transactions on Internet Technology, as well as on the editorial board of IEEE Internet Computing and IEEE Computer. Dustdar is IEEE Fellow (2016), recipient of the ACM Distinguished Scientist Award (2009), the ACM Distinguished Speaker ward (2021), the IBM Faculty Award (2012), an Elected Member of the Academia Europaea: The Academy of Europe, where he is Chairman of the Informatics Section. In 2021 Dustdar was elected EAI Fellow as well as Fellow and President for the Asia-Pacific Artificial Intelligence Association (AAIA).} \\

\parpic{\includegraphics[width=0.8in,clip,keepaspectratio]{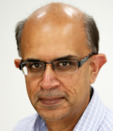}}
\noindent {\footnotesize \sffamily{\bf Omer Rana} is a Professor of Performance Engineering in School of Computer Science \& Informatics at Cardiff University and Deputy Director of the Welsh e-Science Centre. He holds a Ph.D. from Imperial College. His research interests extend to three main areas within computer science: problem solving environments, high performance agent systems and novel algorithms for data analysis and management. Moreover, he leads the Complex Systems research group in the School of Computer Science \& Informatics and is director of the `Internet of Things' Lab, at Cardiff University. He has published over 310 papers in peer-reviewed international conferences and journals. He serves on the Editorial Board of IEEE Transactions on Parallel and Distributed Systems, ACM Transactions on Internet Technology, and ACM Transactions on Autonomous and Adaptive Systems. He has served as a Co-Editor for a number of journals, including Concurrency: Practice and Experience (John Wiley), IEEE Transactions on System, Man, and Cybernetics: Systems, and IEEE Transactions on Cloud Computing.} \\

\parpic{\includegraphics[width=0.8in,clip,keepaspectratio]{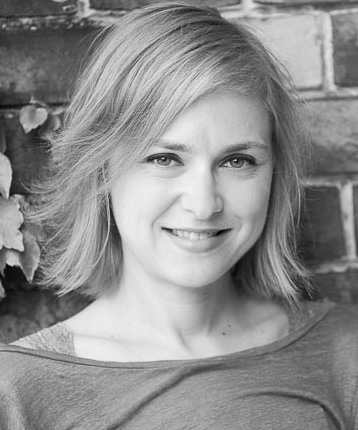}}
\noindent {\footnotesize \sffamily{\bf Ivona Brandic}  is University Professor for HPC Systems at the Institute of Information Systems Engineering, Vienna University of Technology (TU Wien) where she leads High Performance Computing Systems Research Group. In 2015, she was awarded the FWF START prize, the highest Austrian award for early career researchers. Since 2016 she has been a member of the Young Academy of the Austrian Academy of Sciences. She received her PhD degree in 2007 and her venia docendi for practical computer science in 2013, both from Vienna University of Technology. Brandic was on the Editorial Board of IEEE Magazine on Cloud Computing, IEEE Transactions on Parallel and Distributed Systems and IEEE Transactions on Cloud Computing. In 2011 she received the Distinguished Young Scientist Award from the Vienna University of Technology for her project on the Holistic Energy Efficient Hybrid Clouds. Her interests comprise virtualized HPC systems, energy efficient ultra-scale distributed systems, massive-scale data analytics, Cloud \& workflow Quality of Service (QoS), and service-oriented distributed systems. In 2019 she chaired the CHIST-ERA panel (ANR) on Smart Distribution of Computing in Dynamic Networks (SDCDN). } \\

\parpic{\includegraphics[width=0.8in,clip,keepaspectratio]{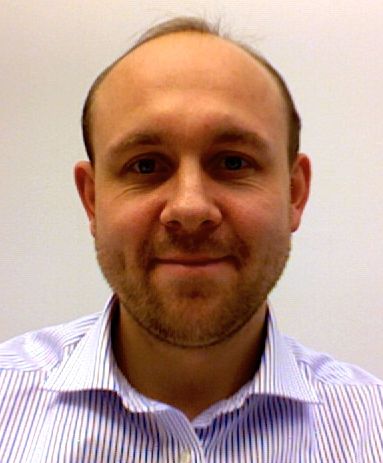}}
\noindent {\footnotesize \sffamily{\bf Steve Uhlig} obtained a Ph.D. degree in Applied Sciences from the University of Louvain, Belgium, in 2004. From 2004 to 2006, he was a Postdoctoral Fellow of the Belgian National Fund for Scientific Research (F.N.R.S.). His thesis won the annual IBM Belgium/F.N.R.S. Computer Science Prize 2005. Between 2004 and 2006, he was a visiting scientist at Intel Research Cambridge, UK, and at the Applied Mathematics Department of University of Adelaide, Australia. Between 2006 and 2008, he was with Delft University of Technology, the Netherlands. Prior to joining Queen Mary University of London, he was a Senior Research Scientist with Technische Universität Berlin/Deutsche Telekom Laboratories, Berlin, Germany. Since January 2012, he has been the Professor of Networks and Head of the Networks Research group at Queen Mary, University of London. Between 2012 and 2016, he was a guest professor at the Institute of Computing Technology, Chinese Academy of Sciences, Beijing, China. With expertise in network monitoring, large-scale network measurements and analysis, and network engineering, during his career he has been published in over 100 peer-reviewed journals, and awarded over £3million in grant funding. Awarded a Turing Fellow, Steve is also the Principal Investigator on a new project funded by the Alan Turing Institute: 'Learning-based reactive Internet Engineering' (LIME). He is currently the Editor in Chief of ACM SIGCOMM Computer Communication Review, the newsletter of the ACM SIGCOMM SIG on data communications. Since December 2020, Steve has also held the position of Head of School of Electronic Engineering and Computer Science. Current Research interests: Internet measurements, software-defined networking, content delivery.} \\

\end{document}